\documentclass[a4paper]{article}
\usepackage[table]{xcolor}
\usepackage[english]{babel}
\usepackage[utf8x]{inputenc}
\usepackage[T1]{fontenc}
\usepackage[a4paper,top=3cm,bottom=2cm,left=3cm,right=3cm,marginparwidth=1.75cm]{geometry}

\usepackage{amsmath}
\usepackage{setspace}
\usepackage{graphicx}
\usepackage[colorinlistoftodos]{todonotes}
\usepackage[colorlinks=true, allcolors=blue]{hyperref}
\usepackage{amssymb}
\usepackage{multirow}
\usepackage{dsfont}
\usepackage{amsthm}
\usepackage{siunitx}
\usepackage{eurosym}
\usepackage{caption}
\usepackage{booktabs}
\usepackage[authoryear]{natbib}
\usepackage{graphicx}
\usepackage{subcaption}
\usepackage{xcolor}
\newcommand{\customcite}[1]{\citeauthor{#1} \citeyear{#1}}
\usepackage{array}
\usepackage{booktabs}
\usepackage{tabularx}
\usepackage{bm}
\usepackage{amsmath}
\usepackage{comment}
\definecolor{hotpink}{RGB}{255, 105, 180}
\usepackage{multirow}
\usepackage{colortbl}
\usepackage{graphicx} 
\usepackage{booktabs} 
\usepackage{graphicx} 
\definecolor{SteelBlue1}{rgb}{0.39, 0.72, 1.0}
\usepackage{float}
\usepackage{amsmath} 
\usepackage{geometry}
\usepackage{tabularx}

\definecolor{bluegreen}{RGB}{0, 149, 139} 

\title{Advanced Models for Hourly Marginal CO$_2$ Emission Factor Estimation: A Synergy between Fundamental and Statistical Approaches}

\author{
Souhir Ben Amor\footnote{Brandenburgische Technische Universität Cottbus-Senftenberg, Germany. Email: benamor@b-tu.de}
\and Smaranda Sgarciu\footnote{Brandenburgische Technische Universität Cottbus-Senftenberg, Germany. Email: sgarcsma@b-tu.de}
\and Taimyra BatzLineiro\footnote{Brandenburgische Technische Universität Cottbus-Senftenberg, Germany. Email: batzltai@b-tu.de}
\and Felix Muesgens\footnote{Brandenburgische Technische Universität Cottbus-Senftenberg, Germany. Email: muesgens@b-tu.de}
}

\begin{document}
\maketitle
\begin{abstract}
\noindent 
Global warming is caused by increasing concentrations of greenhouse gases, particularly carbon dioxide (CO$_2$). A metric used to quantify the change in CO$_2$ emissions is the marginal emission factor, defined as the marginal change in CO$_2$ emissions resulting from a marginal change in electricity demand over a specified period of time.
This paper aims to present two methodologies to estimate the marginal emission factor in a decarbonized electricity system with high temporal resolution. First, we present an energy systems model that incrementally calculates the marginal emission factors. This calculation is computationally intensive when the time resolution is high, but is very accurate because it considers relevant market factors on both the supply and demand sides and emulates the electricity market dynamics. Second, we examine a Markov Switching Dynamic Regression model, a statistical model designed to estimate marginal emission factors faster, and use an incremental marginal emission factor as a benchmark to assess its precision. For the German electricity market, we estimate the marginal emissions factor time series both historically (2019, 2020) using Agora Energiewende and for the future (2025, 2030, and 2040) using estimated energy system data.
The results indicate that the Markov Switching Dynamic Regression model is more accurate in estimating marginal emission factors than the Dynamic Linear Regression models, which are frequently used in the literature. Hence, the Markov Switching Dynamic Regression model is a simpler alternative to the computationally intensive incremental marginal emissions factor, especially when short-term marginal emissions factor estimation is needed.
The results of the marginal emission factor estimation are applied to an exemplary low-emission vehicle charging scenario to estimate CO$_2$ savings by shifting the charge hours to those corresponding to the lower marginal emissions factor. By implementing this emission-minimized charging approach, an average reduction of 31\% in the marginal emission factor was achieved over the 5-year period.

\end{abstract}

\textbf{JEL classification}: 

\newpage
\section{Introduction}
Global warming is caused by continuous increases in greenhouse gas (GHG) concentrations, especially carbon dioxide (CO$_2$) (\customcite{IPCC2021}). The energy industry contributes significantly to global GHG emissions, accounting for more than three-quarters of total GHG emissions globally (\customcite{iea2023}). In 2021, CO$_2$ emissions from energy combustion and electricity generation accounted for almost 89\% of the GHG emissions in the energy sector (\customcite{iea2022}). Consequently, stakeholders have developed various techniques to reduce CO$_2$ emissions in the electricity sector. On the supply side, sustainable electricity generation technologies, with a particular emphasis on renewable energy sources such as wind and solar power, are ready for mass deployment. On the demand side, energy efficiency and demand response techniques can be used to reduce and shift electricity consumption (\customcite{HittingerAzevedo2015}). However, efficient implementation of such techniques requires computation of the associated change in GHG emissions, in both the short and long term.

The most widely used approach to compute changes in CO$_2$ emissions is the system-average emissions factor (AEF). The AEF is defined as the average emissions (typically expressed in tons of CO$_2$-equivalents per MWh of electricity) in a regional electricity system (often a country) over a certain period (often one year).
AEFs are easy to compute, historical values are available, and AEFs are well suited for system-level analysis. Based on AEFs, we can, for example, determine whether an electricity system from one country is more emissions-intensive than another. Also, AEFs are well suited to comparing emissions intensity over time.\\
However, AEFs are not optimal for assessing emissions changes in other contexts. For example, think about short-term flexibility measures on the demand side: If a consumer considers shifting consumption from one hour to another to reduce the CO$_2$ footprint, this would not be reflected in the AEF (which is the same number for both hours).
Furthermore, a change in demand or supply does not influence all aspects of the electricity system proportionally. However, this is a fundamental assumption of the AEF. Therefore, relying solely on the AEF is misleading in many cases.

Instead, it is recommended to use marginal emission factors (MEFs)(\customcite{ryan2016comparative}).\footnote{Note that the focus at this point is more on minimizing GHG emissions than on minimizing financial costs. Hence, the approach is most relevant for actors who are concerned with the environmental impact of their electricity consumption. And the environmental impact of a MWh consumed is context specific and varies over time.} MEFs are defined as the marginal change in CO$_2$ emissions resulting from a marginal change in electricity demand in a given period, e.g. an hour. Thus, the MEF is the partial derivative of emissions with respect to electricity demand. The literature argues that MEFs are a more appropriate metric for evaluating emissions reductions resulting from policy or technology interventions (\customcite{elenes2022well}, \customcite{thind2017marginal}). The key reason is that an additional MWh of electricity consumption changes the system in different ways. For example, during an hour where renewable energy sources are curtailed due to exceedingly high infeed of solar and wind power, a marginal increase in demand does not cause any CO$_2$ emissions. Instead, carbon-neutral renewables would produce more. Even a few hours later, conventional capacity may be needed in the system and additional consumption would be served by e.g. hard coal plants ramping up. Furthermore, compared to AEFs, MEFs react more sensitively and precisely to changes in power grids and demand (\customcite{kamjou2021comparison}). Hence, using AEFs for demand changes can be problematic because they often overlook the impact on specific generators, such as "peaker" fossil fuel plants, which are adjusted for demand fluctuations, while "baseload" hydropower or nuclear plants remain unchanged. These marginal generators have emissions characteristics different from the average. Policies that influence electricity consumption based on AEFs tend to underestimate their environmental impacts. To illustrate, \customcite{thind2017marginal} found that the actual environmental benefits of an energy efficiency program are nearly 20\% lower than initially anticipated when AEFs are employed. For a simple building lighting intervention, \customcite{donti2019much} found that using AEFs underestimated damages by 45\%. In contrast, MEFs by definition consider emissions characteristics of plants "on the margin," providing a more accurate estimate of CO$_2$ emissions changes in such scenarios (\customcite{elenes2022well}). 

In light of these advantages of MEFs, scientific efforts have been made to estimate them. Two primary approaches, namely energy system models and statistical models based on empirical data, have been utilized. While complex energy system models can be useful for estimating MEFs, they are time- and resource-intensive to implement. In contrast, statistical models offer advantages such as reflecting historical performance and reducing calculations and data requirements, but they have limitations including reliance on historical data, an inability to estimate future MEFs that might be different from the past, and limited flexibility to account for future changes in the electricity grid system (\customcite{ryan2016comparative}). \\
Interestingly, it is hard to assess the accuracy of empirical MEF estimates because MEFs are inherently unobservable “what-if” variables.\footnote{The MEF is considered a "what-if" variable because it addresses hypothetical questions: "What would be the variation in CO$_2$ emissions if we increased energy demand by 1 MWh?" Hence, the MEF represents model output rather than a real-world variable.} In this context, \customcite{elenes2022well} compared approximation methods in an analytical setting and found that an estimate of MEFs based on complex energy system modeling was the most accurate and could be adopted as a reference for comparison to evaluate the performance of different emissions factor methods. At the same time, they indicated that a linear regression model was a faster computing alternative to estimating MEFs. \\
Existing literature primarily focuses on either estimating empirical MEFs through statistical models or projecting MEFs for specific future years, using energy system models. Moreover, there is a notable gap in research addressing hourly MEFs, as the available studies have neglected temporal changes and their impact on emissions, hindering a comprehensive understanding of how MEFs evolve over time, especially in the context of climate mitigation measures. However, hourly MEF time series are required to properly assess the climate impact of flexibility, e.g. on the demand side. When centering only on seasonal or annual averages of MEFs, the peaks and troughs in the time series are smoothed out, which leads to missing critical periods throughout the year when MEFs are significantly high or low.\\ 
Moreover, hourly MEFs offer a more realistic depiction of the emissions related to the production of electricity at various times of the day. The hourly resolution grants a better understanding of how load shifts impact emissions, as the emissions per MWh can vary significantly over time, based on the marginal setting plant at any given hour. The development of more efficient carbon pricing schemes can benefit from the application of hourly MEFs. By implementing dynamic pricing schemes that accurately reflect the environmental costs associated with electricity generation, it is possible to encourage users to shift their usage to periods of lower emissions by taking advantage of the day-to-day variations in MEFs. MEFs facilitate the shift to low-carbon energy systems by offering a more precise illustration of how demand response might impact emissions \cite{fleschutz2021effect}.\
Following recent trends in energy systems of high storage shares and sector coupling, hourly MEFs can also provide insights into how different system parts affect emissions over time
\cite{ripp2018first}. \
 
An additional deficiency in prior research lies in the lack of a benchmark for evaluating MEF estimates derived with various statistical methods. The absence of such a benchmark limits the capacity to thoroughly evaluate the accuracy and effectiveness of diverse MEF estimation approaches.

The purpose of this paper is to present two methodologies to estimate MEFs in a decarbonized electricity system at a high temporal resolution. The first model calculates MEF benchmarks with an incremental methodology in a complex energy system model. The second methodology combines an energy system model and a statistical model to estimate MEFs. The MEF is parameterized with empirical data for the German electricity market, both historically (2019, 2020) and for the future (2025, 2030, and 2040). Furthermore, we apply the results to an exemplary emission-minimized vehicle charging scenario. This scenario serves as an example of demand-side flexibility and allows us to draw additional policy recommendations from the findings.

Our contribution can be divided into three methodological and three empirical key aspects. \\
In terms of \textit{methodology}, \\
(1) we present a complex energy system model (ESM) to calculate a benchmark for MEFs at an hourly resolution for selected years.
Building on the methodology established by \customcite{elenes2022well}, our MEFs are determined incrementally by calculating the difference between two model runs, each reflecting a one-MWh variation in demand at a given time step. This approach is computationally intensive, particularly at high temporal resolutions, so it is previously tested in the literature as a proof of concept. The novelty of our work lies in applying this methodology within an empirical model framework at a large European scale, offering insights that extend beyond previous studies.
The ESM used in our analysis considers international power trade, must-run feed-in (due to heat constraints from CHP or for reasons of system stability), and inter-temporal effects (start-up decisions and storage optimization).\\
(2) We present a statistical model for the estimation of MEFs. This approach is much faster and simpler to generate. We propose a Markov Switching Dynamic Regression model (MSDR). In contrast to the linear regression model adopted in previous studies (\customcite{hawkes2010estimating}), MSDR is a nonlinear model, which allows parameters to switch between different regimes according to a Markov process.\\
(3) We establish a connection between the ESM model and the statistical model by utilizing the ESM's output (CO$_2$ emissions and electricity generation) as input for the statistical model. This approach helps us overcome the limitation of the statistical model in estimating only empirical MEFs when historical data on CO$_2$ emissions and conventional generation are available. Additionally, it allows us to leverage the simplicity of the statistical model in estimating prospective hourly MEFs in the future.\\
The \textit{empirical} contribution is threefold. \\
(1) We apply both methodologies to data from historical years (2019 and 2020), as well as future years (2025, 2030, 2040). This provides an extended time series of hourly MEFs for the German market. To our knowledge, the literature up to now has covered only historical MEFs. Our findings thus extend the analysis to future years.\\
(2) We present an example of demand-side flexibility to showcase the improvements achieved using MEFs when the aim is to minimize GHG emissions caused by a particular process. To this end, we develop an emission-minimized vehicle charging scenario that quantifies the reduction in emissions resulting from shifting charging hours to periods with lower MEFs.
(3) Our methodology is easy to implement and can be applied to datasets with similar characteristics, such as those from other energy markets. Our code is open-source and can be found on GitHub: https://github.com/BTU-EnerEcon/Modeling-Hourly-Marginal-CO2-Factors. \\
As a result of our work, consumers and policymakers can make more informed decisions regarding energy use and emissions reduction strategies. They are of particular value to all parties interested in minimizing the CO$_2$ footprint of a certain amount of electricity taken from the public grid.\\ 
The structure of this paper is organized as follows. In Section \ref{literature}, we review the relevant literature on estimating MEFs. Section \ref{methodology} presents our methodology, which includes a two-step ESM model, the development of statistical models, and the exemplary  emission-minimized vehicle charging scenario applied to estimate CO$_2$ savings. Section \ref{data} defines the data used for the ESM and the statistical model. In Section \ref{results}, we present the empirical specification and results, focusing on the estimation of MEFs for the German power supply. In Section \ref{smartchargingsection}, we develop an emission-minimized vehicle charging application. Lastly, in Section \ref{conclusionandpolicy}, we conclude our research, highlighting the main contributions and providing recommendations for utilizing MEFs in various applications and finally suggest avenues for future research. \\

\section{Literature Review}
\label{literature}

Estimating the impact of CO$_2$ emissions caused by changes in the electricity system has been a widely discussed topic for several years. This is significant because the choice of the methodology used to calculate the emissions rate has substantial implications for determining which technology is most effective in reducing CO$_2$ emissions (\customcite{rees2014carbon}, \customcite{hawkes2014long}).\\
Current approaches for estimating MEFs vary in terms of the models employed, including merit order, energy system models, and statistical models. They also differ in terms of the time horizon considered and the temporal resolution (\customcite{boing2019hourly}). \\
In the current section, we will examine existing research and provide an overview of the time resolution and methodology used to estimate MEFs. We will analyze the strengths and weaknesses of each methodology. Following that, we will identify the gaps in the existing literature and outline how our paper addresses and fills these gaps.

In the literature, MEFs have been used to study load management (\customcite{wang2014locational}), environmental pricing (\customcite{holland2008real}), and community energy management (\customcite{yamaguchi2007transition}). MEFs for power generation have also been applied to investigate the environmental impacts of demand-related policies in power-consuming sectors, such as industrial motor replacement (\customcite{hasanuzzaman2011energy}). Furthermore, in the transportation sector, MEFs for electric power generation have been employed to estimate the environmental impacts of transitioning vehicle fuel technology from petroleum to electricity (\customcite{ma2012new}). \

In the context of MEFs, understanding the implications for the electricity system requires considering different time resolutions, such as annual (\customcite{hawkes2014long}) and hourly time resolutions (\customcite{fleschutz2021effect}, \customcite{beltrami2020did}, and \customcite{hawkes2010estimating}).\\  
While yearly resolutions may serve to estimate a long-term trend or establish emissions reduction targets, they may not provide enough information for investment incentives. Hence, the challenge lies in identifying time-varying MEFs. More precisely, hourly MEFs can be used to estimate the carbon intensity of demand changes by taking into account the carbon intensity of the marginal power plant \footnote{The power plant can be either a base-load plant or a marginal plant. Base-load power plants continuously generate electricity for customers and cannot react quickly to demand changes. They are unable to meet flexible demand. Conversely, marginal power plants can adjust their output quickly to meet fluctuating demand (\customcite{zheng2015assessment}).} for each time step (\customcite{hawkes2010estimating}).\\ 
\customcite{roder2020design} indicated that using hourly, rather than constant, emission factors of electricity generation is critical for the development of low-emissions local energy systems in Germany. \\
Despite an increasing body of literature recognizing the need for hourly MEFs (\customcite{ripp2018first}, \customcite{bazaraa2011linear}), calculating hourly MEFs requires detailed and comprehensive databases. For some countries, the lack of publicly available power plant-specific electricity generation data makes it challenging to empirically identify the marginal power plant (\customcite{doucette2011modeling}).
Our study focuses on determining the time series of hourly MEFs. Hourly resolution allows us to identify the changes in CO$_2$ emissions due to supplying one more unit of demand. Understanding hourly emission patterns is essential for future market design, as well as for supporting real-time decision-making.

In examining the time horizons, previous research on emission factors in Germany has mostly relied on retrospective analysis. However, there are a few relevant studies that have provided insight into future emission factors. As an example, by considering capacity factors and planned installed capacity from two scenarios, \customcite{maennel2018comparison} calculated Germany's future emission factors until 2030. To provide a broad overview, annual averages were used in the calculations. 
Similarly, \customcite{boing2019hourly} used a linear optimization model to forecast hourly AEFs and MEFs until 2050. They offered a more detailed temporal resolution compared to \customcite{maennel2018comparison}.
Another study, \customcite{jochem2015assessing}, used an energy model to explicitly calculate hourly AEFs and annual MEFs for the years 2020 and 2030, taking into account fluctuations in demand. While the study was limited to a few years, it provided useful insights into the temporal dynamics of emission factors. 
\customcite{regett2018emission} also used an energy model to determine hourly AEFs and MEFs for the year 2030. However, the computational time necessary for such calculations was deemed impractical, implying scalability restrictions. 
\customcite{buyle2019analysis} and \customcite{sengupta2022current} used a prospective method to determine MEFs using the linear regression method.\\
Despite recent studies on MEFs, there is still a literature gap concerning prospective emission factors with high temporal resolution for future power supply in Germany (\customcite{seckinger2021dynamic}). Thus, the objective of this study is to compute prospective MEFs on an hourly basis until 2040, to fill this gap and to provide more detailed information on future CO$_2$ emission factors in Germany.

Concerning the MEF estimation, existing methodologies vary in terms of input data and models employed, including empirical data with statistical relationship models vs. optimization models for power systems (\customcite{ryan2016comparative}, and \customcite{boing2019hourly}). \\
Three main approaches have been extensively applied to estimate MEFs: the merit order model, economic dispatch models, and statistical models. 

Numerous studies compute MEFs through a merit order-based method (\customcite{fleschutz2021effect}, \customcite{zheng2013assessment}, and \customcite{schram2019use}). Merit order refers to the sequential dispatch of generators based on their operational costs, where the cheapest generators are dispatched first until system demand is satisfied within a specific period. By combining this approach with a load duration curve, researchers can identify the generator and fuel type that is next in line (referred to as the "margin") to be brought online or taken offline for a given system load level. The CO$_2$ intensity of this marginal generator represents the MEF for that particular system load level and follows a step function pattern. 
However, this static merit order approach determines the dispatch based on the variable cost of operation at optimal utilization levels. This approach neglects several factors: international power trade, grid restrictions, must-run feed-in (due to heat constraints from CHP or for reasons of system stability), and especially intertemporal effects (start-up decisions and storage optimization). Neglecting these factors leads to inaccuracies in the estimation of MEFs.

The second method for estimating MEFs involves the use of energy system models or economic dispatch models that are built using a combination of theoretical and empirical assumptions (\customcite{hawkes2014long}, \customcite{zheng2015assessment}). Dispatch models are simulation optimization models that are frequently used to assess the impact of load changes, allowing the investigation of future emission factors while considering the overall effects on the power generation system; this includes long-term changes such as the construction and dispatch of new power plants. Using a dispatch model, the net emissions effect can be determined by assessing the operation and emissions of plants with and without the new technology, thus calculating the change in carbon emissions. These models are especially suited to forecast the environmental consequences of future demand-side management schemes.\\
For instance, \customcite{regett2018emission} suggested an emissions estimation approach that determined hourly MEFs for different energy carriers in multi-energy systems. This approach was also adopted by \customcite{boing2019hourly}. Besides this, \customcite{baumgartner2019design} applied an economic dispatch model based on German data from 2016, yielding hourly MEFs that were later used in a multi-objective synthesis problem for a low-carbon utility system. Moreover, \customcite{elenes2022well} used the electricity system dispatch model as a controlled environment to compare various approaches to emission factors. \\
However, developing robust energy system models necessitates defining a large number of parameters and input data, as well as making several assumptions to achieve reproducible results. This process entails substantial computational requirements and workload, which may not be feasible or practical depending on the complexity of the investigation (\customcite{seckinger2021dynamic}, \customcite{vandepaer2019integration}). \\
Recognizing the need for simpler approaches to estimate short-term emissions changes resulting from small technology additions, several efforts have been made in this direction. Three motivations underpin these efforts. Firstly, it is not evident whether the complexity of a dispatch model is necessary, as a simpler approach may yield a reasonable approximation. Secondly, a simpler approach allows individuals without access to fundamental data or expertise in dispatch models to study emissions effects, broadening the range of people capable of conducting such analyses. Lastly, a lack of publicly available and empirically validated dispatch models contributes to the motivation to find simpler alternatives. While some publicly shared dispatch models do exist, they are not yet considered standard practice (\customcite{elenes2022well}).


An alternative and simpler approach to estimating MEFs is through the use of statistical models, avoiding the explicit modeling of the merit order and the complexity of the energy system models. The linear regression model is a widely used statistical model that employs regression analysis to estimate marginal emissions changes. More precisely, it estimates the ratio of emissions change to demand change resulting from a small shift during normal operation (\customcite{hawkes2010estimating}, \customcite{siler2012marginal}, \customcite{donti2019much}, \customcite{thind2017marginal}). This model calculates the difference in emissions and conventional power generation between two successive hours. By performing a linear regression analysis on these differences, the gradient of the fitted regression line provides a reasonable approximation of the average MEF. 
\customcite{hawkes2010estimating} proposed a regression-based methodology for calculating MEFs in Great Britain, determining the average MEF by calculating the slope of the regression line using half-hourly data for system CO$_2$ emissions and system load. This approach takes into account trading decisions, operational constraints of power plants, and transmission and distribution restrictions, providing advantages over purely merit-order-based methods. Hawkes' methodology has been applied in various subsequent studies (\customcite{staffell2017measuring}, \customcite{thomas2012us}). \customcite{siler2012marginal}, \customcite{li2017marginal}, \customcite{holland2022marginal}, and \customcite{thind2017marginal} analyzed the American electricity markets. Other authors have adopted the regression model to analyze the effects of smart charging (\customcite{holland2015environmental}). For example, \customcite{pareschi2017assessment} adopted the methodology to analyze the effects of smart charging for the German and Swiss electricity markets. \customcite{huber2021carbon} utilized a regression model to derive MEFs for Germany in 2017 and incorporated a multilayer perception (MLP) for short-term predictions of MEFs, enabling valuable feedback for CO$_2$-efficient smart charging. Another robust empirical methodology for MEF estimation was proposed by \customcite{beltrami2020did}, based on Autoregressive Integrated Moving Average (ARIMA) models, using half-hourly data on carbon emissions and generation in Great Britain. \\
In addition to reducing complexity, the statistical approaches are also computationally efficient and transparent and provide real-time feedback, making them well-suited for short-term analysis (\customcite{yang2013fuel}). However, they only rely on empirical data (\customcite{elenes2022well}, \customcite{ryan2016comparative}). Estimating MEFs using historical data is more suitable for short-term considerations rather than long-term studies, as long-term analyses need to account for potential structural changes in the infrastructure that may not be adequately captured by the derived emission factors (\customcite{thind2017marginal}. To overcome this limitation, we have built a bridge between the energy system model and the statistical model. This bridge is formed by using CO$_2$ emissions and electricity generation data from the energy system model as inputs for the statistical model. This combination enables us to estimate the MEFs for both historical years (2019 and 2020) and future years (2025, 2030, and 2040). This approach goes beyond relying solely on empirical or historical MEFs, providing a means to generate future projections with a simple statistical method.
Some authors have used both energy system and statistical approaches, and then compared the modeled and empirical MEFs (\customcite{deetjen2019reduced}, \customcite{braeuer2020comparing}, and \customcite{elenes2022well}). 

The extensive literature has several drawbacks:\\
First, the literature concentrates on either estimating empirical MEFs using statistical models or providing an estimation of MEFs until a specific future year, such as 2050, using energy system models. On one hand, the economic dispatch model offers a comprehensive understanding of the entire power generation system and enables long-term assessments. However, its complexity and computational time pose challenges, especially when quick decisions are required. For instance, in real-time applications where rapid adjustments are crucial, the time-consuming nature of the economic dispatch model may hinder its adoption. On the other hand, statistical models, particularly linear regression models, have been widely used for estimating MEFs due to their simplicity. However, these models are limited to empirical applications (historical MEFs). Moreover, assuming a linear relationship between electricity generation and CO$_2$ emissions may oversimplify the complex dynamics of the power system. Non-linear relationships can better capture the intricate interplay between electricity generation sources and CO$_2$ emissions. Furthermore, existing statistical models used for estimating MEFs lack enhancements, such as the incorporation of dynamic merit order considerations, which can improve the accuracy and reliability of the results. To overcome these limitations, our methodology takes advantage of the strengths of both the economic dispatch model and an advanced statistical model.\\ 
Second, a crucial aspect lacking in the previous literature is the absence of a benchmark for evaluating MEFs estimated through different methodologies. While previous studies focus on estimating MEFs using a single approach, they often compare the MEF with the AEF. However, AEFs and MEFs are distinct ratios and cannot be estimated using the same methodology. Thus, the absence of a benchmark limits the ability to evaluate the accuracy and effectiveness of various MEF estimation methodologies comprehensively. To overcome this limitation, the ESM model is used to calculate a benchmark MEF, which, based on the literature, is the most accurate way of doing it (\customcite{elenes2022well}).

Third, a significant limitation in the literature is the unavailability of MEF data, since MEFs are unobservable, and the researchers who estimate them rarely make their MEF data publicly accessible. Hence, policymakers and researchers who could greatly benefit from MEFs for climate mitigation efforts face challenges in accessing the necessary data. This limited availability obstructs the development of effective strategies and policies aimed at reducing emissions based on accurate MEF information. We emphasize the importance of making estimated MEFs publicly available, providing access to our estimated hourly MEF data. Moreover, we offer a policy implication section that highlights the application of MEFs to emission-minimized vehicle charging, promoting effective strategies for reducing emissions based on reliable MEF information.

\section{Methodology}
\label{methodology}
Our research utilizes two different methodologies to estimate marginal emission factors: an energy system model and a statistical model. This section will first present the ESM, which is set up in two steps. In the first step, we employ the EM.POWER Invest model to compute installed capacities in future years. In the second step, we implement the EM.POWER Dispatch model, which is a dispatch model that requires exogenous capacities. The determination of marginal emission factors with the ESM is described in Section \ref{ESM mod}. Second, we present two statistical models that estimate MEFs econometrically, Markov Switching Dynamic Regression (MSDR) and Dynamic Linear Regression (DLR). These models are presented in Subsection \ref{stat mod}.

\subsection{The Energy System Model}
\label{ESM mod}
\begin{figure}[H]
  \centering
  \includegraphics[width=1\textwidth]{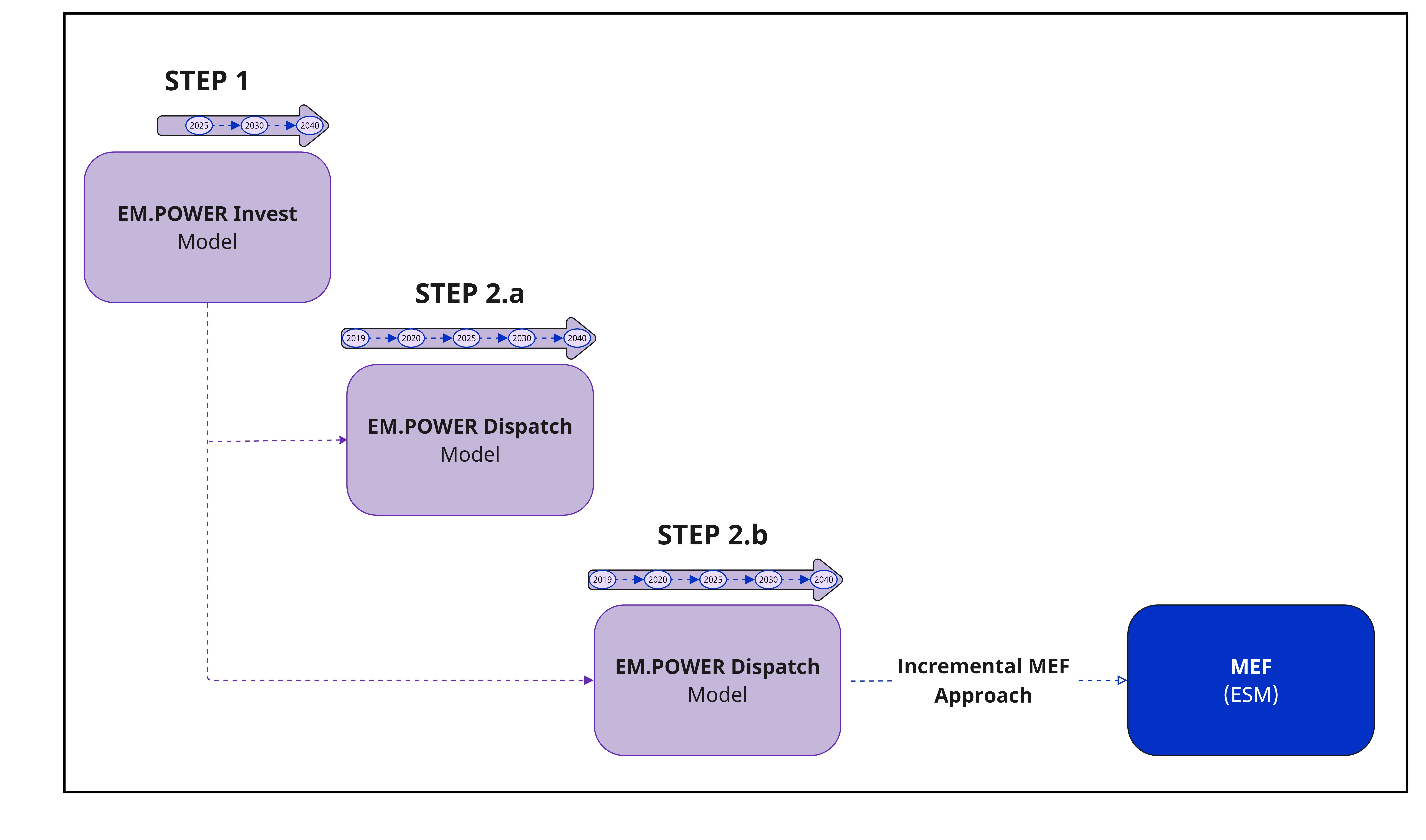}
  \caption{ESM Methodology for estimating the MEF}
  \label{ESM_method}
\end{figure}

The ESM methodology adopted to estimate the incremental MEF is illustrated in Figure \ref{ESM_method}. This methodology consists of two steps:\\

In the first step, an ESM focused on investment decisions (EM.POWER Invest) optimizes installed capacities for the years 2025, 2030, and 2040. The investment model is needed to determine long-term changes in the system's power plants resulting from changes in fuel prices, CO$_2$ prices, and the technical lifetimes of power plants.\\
In the second step, an ESM specialized in dispatch decisions (EM.POWER Dispatch) is used. This model is run successively in two sub-steps: In Step 2.a, the EM.POWER Dispatch model is used in a rolling horizon approach 365 times per year to optimize 24 hours of each day simultaneously.\footnote{The ESM optimizes three days with 72 hours total in each run, but only the 24 hours of the middle day are kept as results in each run.} These model runs provide the hourly "baseline emissions." \\
In addition to these 365 model runs per year, we need further runs of the dispatch model to calculate the change in emissions resulting from a change in load. In Step 2.b, we perform these additional model runs, where demand is increased (plus one MWh) separately in each of the 8760 hours in a year. Note that this requires one additional model run for each hour, i.e. 8760 additional model runs per year. The increase in emissions, calculated as the difference in emissions between the benchmark model run for that hour (Step 2.a) and the increased demand model run for the hour (Step 2.b), is defined as the MEF for that particular hour. We will refer to this methodological approach as the "incremental MEF” approach. 

Calculating hourly MEFs for one year thus requires more than 9000 separate model runs,\footnote{8760 + 365 = 9125 for a normal year with 8760 hours.} making the calculations computationally demanding. For this reason, we calculate two historic years (2019, 2020) and three future years (2025, 2030, and 2040) only. For analytical reasons, the resulting MEFs serve as benchmarks for the statistical model. Note however, that most of the complexity results from Step 2.b. At the end of Step 2, incremental MEFs for five benchmark years are available in hourly resolution.

The ESMs were developed in the Chair of Energy Economics at BTU Cottbus-Senftenberg. We offer an overview of the model parametrization in Section \ref{datasection} and Section \ref{datasectionF}, and the complete parametrization of the model can be found in the supplementary material offered on GitHub.\footnote{https://github.com/BTU-EnerEcon/Modeling-Hourly-Marginal-CO2-Factors}

\subsubsection{EM.POWER Invest}

This section describes the EM.POWER Invest model used in the first step of Figure \ref{ESM_method}. The model endogenously quantifies investments in terms of generation capacity for the years 2025, 2030, and 2040.\footnote{These investments are used in the second step to parameterize the EM.POWER Dispatch model.} The optimization determines a (partial) equilibrium in the electricity market and is formulated as a linear problem. The objective function is to minimize the total system costs. The model determines the investment in conventional capacities for the European level, while the renewable and storage capacities are introduced as exogenous parameters. The coverage of the power system includes EU-27 countries, also including Norway, the United Kingdom, and Switzerland. Power plants are grouped based on both their age and technology. The complete mathematical formulation of the model is presented in \customcite{sgarciu2023co2}.

In the following, we will describe the structure of the model. The sets, parameters, and variables are described in the nomenclature provided in Appendix Section \ref{nomenclatureapp}.

The aggregated discounted system costs are minimized by the objective function of the model:

\begin{equation}
   \textit{min} \operatorname{COST}=\textit{min}\left(\sum_{n, y}\left(\operatorname{COST}_{n, y}^{G e n}+\operatorname{COST}_{n, y}^{F i x}+\operatorname{COST}_{n, y}^{I n v}\right)  d f_y\right)
    \label{mincost}
\end{equation}

The generation costs are the sum of three components, with the details explained in the following three equations: Eq. \ref{costgen}, \ref{costfix}, and \ref{costinvest}.

\begin{equation}
\begin{aligned}
\text{COST}_{(n,y)}^{\text{Gen}} = &\sum_{(p,t)} \Bigl[ \text{cvar}_{(n,p,y)}^{\text{Full}}  \text{GEN}_{(n,p,y,t)}^{\text{Full}} + \text{cvar}_{(n,p,y)}^{\text{Min}}  \text{GEN}_{(n,p,y,t)}^{\text{Min}} \\
&+ \text{cramp}_{(n,p,y)}  \left(\text{CAP}_{(n,p,y,t)}^{\text{Up}} + \text{CAP}_{(n,p,y,t)}^{\text{Down}}\right) \Bigr] \\
&+ \sum_{(r,t)} \text{cvar}_{(n,r,y)}  \text{GEN}_{(n,r,y,t)} \\
&+ \sum_{(\text{curt},t)} \text{ccurt}_{(n,r^\text{curt},y)}^{\text{Gen}}  \text{CURT}_{(n,r^\text{curt},y,t)}^{\text{Gen}} \\
&+ \sum_t \text{ccurt}_{y}^{\text{Load}}  \text{CURT}_{(n,y,t)}^{\text{Load}},  \hspace{3cm}  \forall n, y
\end{aligned}
\label{costgen}
\end{equation}

\begin{equation}
\operatorname{COST}_{n, y}^{Fix} = \sum_p c fix_{p, y}  CAP_{n, p, y}^{\text {Install }} \quad \hspace{4cm}  \forall n, y
\label{costfix}
\end{equation}

\begin{equation}
    \operatorname{COST}_{n, y}^{\text{Inv}} = \sum_p \sum_{y' \mid y' \leq y \leq y' + \text{lifetime}_{p}^{\text{eco}}} \operatorname{cinv}_{p, y'} \operatorname{CAP}_{n, p, y'}^{\text{Add}} \quad \forall n, y
    \label{costinvest}
\end{equation}

The key constraint in our research question is the demand constraint. This market clearing condition makes sure that each node's electrical demand is constantly met. This requirement can be fulfilled by producing energy using conventional or RES technologies, operating a single node of storage, as well as through imports, exports, and demand-side management.
\begin{equation}
\begin{aligned}
\operatorname{load}_{n, y, t} = & \sum_p \text{GEN}_{n, p, y, t} + \sum_r \text{GEN}_{n, r, y, t} \\
& + \sum_s \text{DISCHARGE}_{n, s, y, t} - \text{CHARGE}_{n, s, y, t}  \\
& + \text{CURT}_{n, y, t}^{\text{Load}} + \sum_{nn}\left(1 - \frac{\text{gridloss}}{2}\right) \text{FLOW}_{nn, n, y, t} \\
& - \sum_{nn}\left(1 + \frac{\text{gridloss}}{2}\right) \text{FLOW}_{n, nn, y, t} \quad \forall n, y, t
\end{aligned}
\label{load}
\end{equation}

Note that the EM.POWER Invest model has numerous other constraints, which can be found in \customcite{sgarciu2023co2}. 

\subsubsection{EM.POWER Dispatch}
\label{dispatchmodelsection}

\begin{figure}[H]
  \centering
  \includegraphics[width=0.7\textwidth]{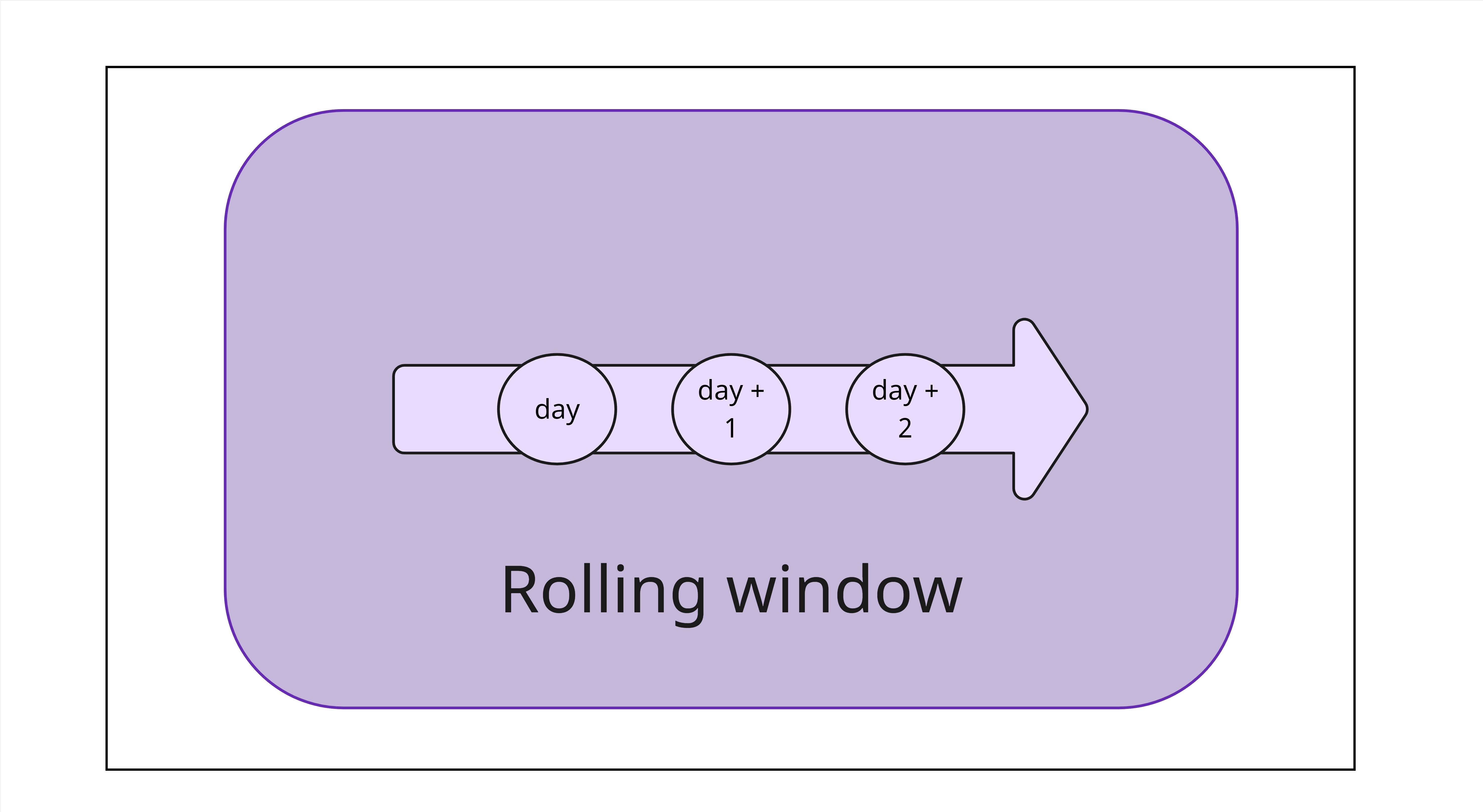}
  \caption{Rolling window used in EM.POWER Dispatch model (based on \customcite{mobius2023enhancing})}
  \label{rolling}
\end{figure}

This section describes the EM.POWER Dispatch model. The model is similar to the investment model in terms of its operations research set-up. The objective is to minimize the operational costs and is formulated as a linear problem. Key differences in the model formulation occur due to the different foci of the dispatch and investment models. While the investment model needs to optimize several years simultaneously to capture the interdependency with respect to the investment, the dispatch model can focus on a much shorter time period. This leaves computational resources in the dispatch model for an hourly time resolution and numerous important market factors on the supply and demand sides, including international power trade, must-run feed-in (due to heat constraints from CHP or for reasons of system stability), and inter-temporal effects (start-up decisions and storage optimization).\\ 
The EM.POWER Dispatch model has a focus on making predictions for the wholesale day-ahead energy market and therefore uses only information accessible to market participants one day prior to the delivery date. This is well suited to estimate MEFs, as it reflects market participants' level of information at the time they determine the emissions. The model is set up using a daily rolling window. An illustrative example of the rolling window is depicted in Figure \ref{rolling}. 
Note that such a daily rolling window matches the structure of electricity markets well: For example, in the European Power Exchange (EPEX) spot market, 24 hourly day-ahead prices are established at 12 p.m. the day before delivery (d).

The objective function described in Equation \ref{mincost1} aims to minimize the overall system costs, encompassing all the expenses encountered by generation units over the short term. It contains inter-temporal dynamic elements such as start-up decisions, demand-side management, renewable energy sources curtailment, and long-term storage (Equation \ref{mincost1}).
    
\begin{equation}
\begin{aligned}
\textit{min} \operatorname{COST} = & \sum_{i, n, t} \left(GEN_{i, n, t} + \operatorname{cvar}_{i, n, t}^{Full}\right) \\
& + \sum_{i, n, t} \left(CAP_{i, n, t}^{Up}  \operatorname{cramp}_{i, n, t}\right) \\
& + \sum_{i, n, t} \left(CAP_{i, n, t}^{Started}  GEN_{i, n, t}  \frac{GEN_{i, n, t}^{Min}}{1 - GEN_{i, n, t}^{Min}}\right) \\
& - \sum_{stl, n, t} CL_{i, n, t}  wv_{i, n, t} \\
& + \sum_{n, t} \operatorname{SHED}_{n, t}  ccurt^{\text{load}} + \sum_{n, t} CURT_{res, n, t}  ccurt_{n, res}^{Gen}, \quad \forall i, n, t
\end{aligned}
\label{mincost1}
\end{equation}

\

\begin{equation}
\begin{aligned}
& \operatorname{load}_{n, t}=\sum_i G E N_{i, n, t}+\sum_{s t m \subset I} C M_{s t m, n, t} \\
&-\sum_{s t l \subset I} C L_{s t l, n, t}+S H E D_{n, t} \\
&+\sum_{n n} F L O W_{n n, n, t}-\text{FLOW}_{n, n n, t}
\quad \hspace{5cm} \forall n, t
\end{aligned}
\label{load1}
\end{equation}
Since we compute our results with an hourly resolution and focus on the day-ahead market, we classify the storage activity of pumped hydropower plants as mid-term storage. In contrast, short-term storage operates at an intra-hourly resolution, which is closer to the time of delivery.
The market clearing condition included in Equation \ref{load1} ensures that demand is satisfied by supply in any given hour. For every hour within the specified rolling window, the demand should be equivalent to the sum of generation, load shedding, and electricity imports. This sum is then adjusted by subtracting the electricity consumption of mid-term energy storage, long-term energy storage, and electricity exports. For the market clearing condition, the dynamic elements included are cross-border trading and short-term storage. 

\begin{equation}
    G E N_{i, n, t} \leq P_{i, n, t}^{O N}-P C R_{i, n, b p}-S C R_{i, n, b s}^{P O S}, \\
\begin{aligned}
& \quad \hspace{1cm} \forall b p \in B P, \text { bs } \in B S,\\
& \quad \hspace{1cm}  i \in I, \quad t \in T 
\end{aligned}
\label{upperlimit}
\end{equation}

The electricity generation within each capacity cluster is constrained by both upper and lower limits. 
Equation \ref{upperlimit} defines the upper limit, ensuring that the electricity generation does not surpass the running capacity of the cluster. Furthermore, the feasible electricity generation from the running capacity is additionally constrained by the reserve allocated for positive control power provision.

\begin{equation}
    P_{i, n, t}^{O N}  g^{\min }+P C R_{i, n, b p}+S C R_{i, n, b s}^{N E G} \leq G E N_{i, n, t} \\
    \begin{aligned}
& \quad \hspace{1cm} \forall b p \in B P, \text { bs } \in B S,\\
& \quad \hspace{1cm}  i \in I, t \in T,  n \in N
\end{aligned}
\label{lowerlimit}
\end{equation}

Equation \ref{lowerlimit} establishes the lower limit of the generation, stipulating that running capacities must operate at a minimum power level, including the capacity reserved for negative control power provision. 
We must clarify that, in the German market, primary control power is synchronously provided, meaning that a unit must deliver both positive and negative primary control power. Distinct products for positive and negative control power have been introduced for secondary control power.

\begin{equation}
    P_{i, n, t}^{O N} \leq C A P_{i, n, t} \cdot a f_{i, n, t}-o u t_{i, n, t}\\
    \begin{aligned}
& \quad \hspace{2cm} \forall i \in I, t \in T,  n \in N
\end{aligned}
\label{power}
\end{equation}

The operational capacity of a power system is constrained by the installed capacity, which interacts with either the availability factor or power plant outages, as depicted in Equation \ref{power}. Thermal generation capacities employ hourly power plant outage data, while renewable energy sources are assigned an hourly availability factor. Hydroelectric units, on the other hand, operate based on a monthly availability factor.

\begin{equation}
    P_{i, n, t}^{O N}- P_{i, n, t-1}^{O N}  \leq  S U_{i, n, t}\\
      \begin{aligned}
& \quad \hspace{2cm} \forall i \in I, t \in T,  n \in N
\end{aligned}
\label{startup}
\end{equation}

Equation \ref{startup} monitors start-up activities, which represent instances where the running capacity is elevated from one hour to the next. Adhering to the non-negativity condition, start-ups are constrained to be either positive or zero.

\begin{equation}
    \operatorname{cap}_{r e s, n, t}  a f_{r e s, n, t}= G E N_{r e s, n, t}+ C U R T_{r e s, n, t}\\
     \begin{aligned}
& \hspace{1cm}  \forall res \in I, t \in T,  n \in N
\end{aligned}
\label{curtailment}
\end{equation}

In Equation \ref{curtailment}, the difference between the available feed-in from intermittent renewables and their actual generation determines the curtailment of renewable energy.

\begin{equation}
    S L_{s t m, n, t}=S L_{s t m, n, t-1}-G E N_{r e s, n, t}+C M_{s t m, n, t}  \eta_{s t m}, \\
\begin{aligned}
& \hspace{0.5cm}  \forall stm \in I, t \in T,  n \in N
\end{aligned}
\label{storagelevel}
\end{equation}

Equation \ref{storagelevel} outlines the status of the storage level in a mid-term storage system. The storage level experiences an increase through generation ($GEN_{stm,n,t}$) and a decrease through consumption during the charging phase ($ST_{in stm,n,t}$). The efficiency of the entire storage cycle ($\eta_{stm}$) is applied to the charging process.

\begin{equation}
    S L_{s t m, n, t} \leq \operatorname{cap}_{s t m, n, t} \\
    \begin{aligned}
& \hspace{3cm}  \forall stm \in I, t \in T,  n \in N
\end{aligned}
\label{maxstorage}
\end{equation}

The maximum energy storage capacity of mid-term storage is determined by the product of the maximum installed turbine capacity and an energy-power factor, as indicated in Equation \ref{maxstorage}.

\begin{equation}
    G_{s t m, n, t}+1.1  C M_{s t m, n, t} \leq  \operatorname{cap}_{s t m, n, t} \\
    \begin{aligned}
& \hspace{2cm}  \forall stm \in I, t \in T,  n \in N
\end{aligned}
\label{constraints} 
\end{equation}

Equation \ref{constraints} imposes constraints on both the turbine and pumping capacity, with the assumption that the pumping capacity is lower than the turbine capacity.

\begin{equation}
    G_{s t l, n, t}+C L_{s t l, n, t} \leq \operatorname{cap}_{s t l, n, t}\\
    \begin{aligned}
& \hspace{2cm}  \forall stl \in I, t \in T,  n \in N
\end{aligned}
\label{constraints2} 
\end{equation}

Long-term storage does not incorporate a storage mechanism. Nevertheless, the electricity generation and consumption of long-term storage units are constrained by the installed capacity of long-term storage, as indicated in Equation \ref{constraints2}.

\begin{equation}
\begin{aligned}
    C L_{s t l, n, t}, C M_{s t m, n, t}, G E N_{i, n, t}, C U R T_{r e s, n, t}, F L O W_{n, n n, t},\\
    P_{i, n, t}^{O N}, S C R_{i, n, b s}^{P O S}, S C R_{i, n, b s}^{N E G}, S U_{i, n, t}, S L_{s t m, n, t}, S H E D_{n, t} \geq 0\\
    \end{aligned}
 \hspace{1cm}  \forall stl \in I, t \in T,  n \in N
\label{constraints3} 
\end{equation}

Equation \ref{constraints3} represents the non-negativity constraint.

\subsubsection{The Incremental Marginal Emission Factor Approach}
\label{incrementalmef}

We use the incremental marginal emission factor approach to derive the change in emissions once we increase the demand marginally (by 1 MW).
This step is conducted for the years 2019, 2020, 2025, 2030, and 2040 and is computationally intensive.
The time series of MEFs is used as a benchmark for those obtained from the statistical model.
 The incremental marginal emission factor approach contains two phases: 
 \begin{itemize}
     \item a) In phase 1, we conduct model runs to obtain a time series of total carbon emissions for an entire year (8760 time steps). 
     \item b) In phase 2, the model runs are conducted again with a 1 MWh increase in demand to calculate the incremental MEF, defined as the difference in emissions between the baseline and increased-demand scenarios. 

The formula for determining the incremental MEF is shown in Equation \ref{incrementalmefequation}
 \end{itemize}

\begin{equation}
\begin{aligned}
\text{MEF}_t = \frac{\partial \text{emissions}}{\partial \text{load}_t} \approx \frac{\sum_{i,n,t} \left(\frac{{\text{GEN}_{i,n,t}^{*,\text{inc}}  \text{carb\_cont}_i}}{\eta_{i,n}}\right)  - \sum_{i,n,t} \left(\frac{{\text{GEN}_{i,n,t}^{*}  \text{carb\_cont}_i}}{\eta_{i,n}}\right)}{{\text{load}_t^{\text{inc}} - \text{load}_t}}
\end{aligned}
   \label{incrementalmefequation}
\end{equation}
Where $\text{load}_t^{\text{inc}} - \text{load}_t=1 MWh$\\
Demand increase translating the conditional since target day is the 2nd in the rolling window.

\begin{equation}
    \operatorname{load}_{D E, t}=\operatorname{load}_{D E, t}+X_{D E, t} * 1 M W
    \label{loaddiff}
\end{equation}
e.g. increasing the demand on $day+1$, hour$ 1$
\begin{equation}
    X_{D E, t}=
    \begin{cases}
        \begin{aligned}
            &1 &\text{if } \operatorname{ord}(t)&=1 \text{ and ord}(\text{ days})=2 \\
            &0 &\text{otherwise}
        \end{aligned}
    \end{cases}
\end{equation}

We compute the difference in total carbon emissions between the model run where demand is increased marginally and the initial model run.

\subsection{The Statistical Models}
\label{stat mod}

\begin{figure}[H]
  \centering
  \includegraphics[width=1\textwidth]{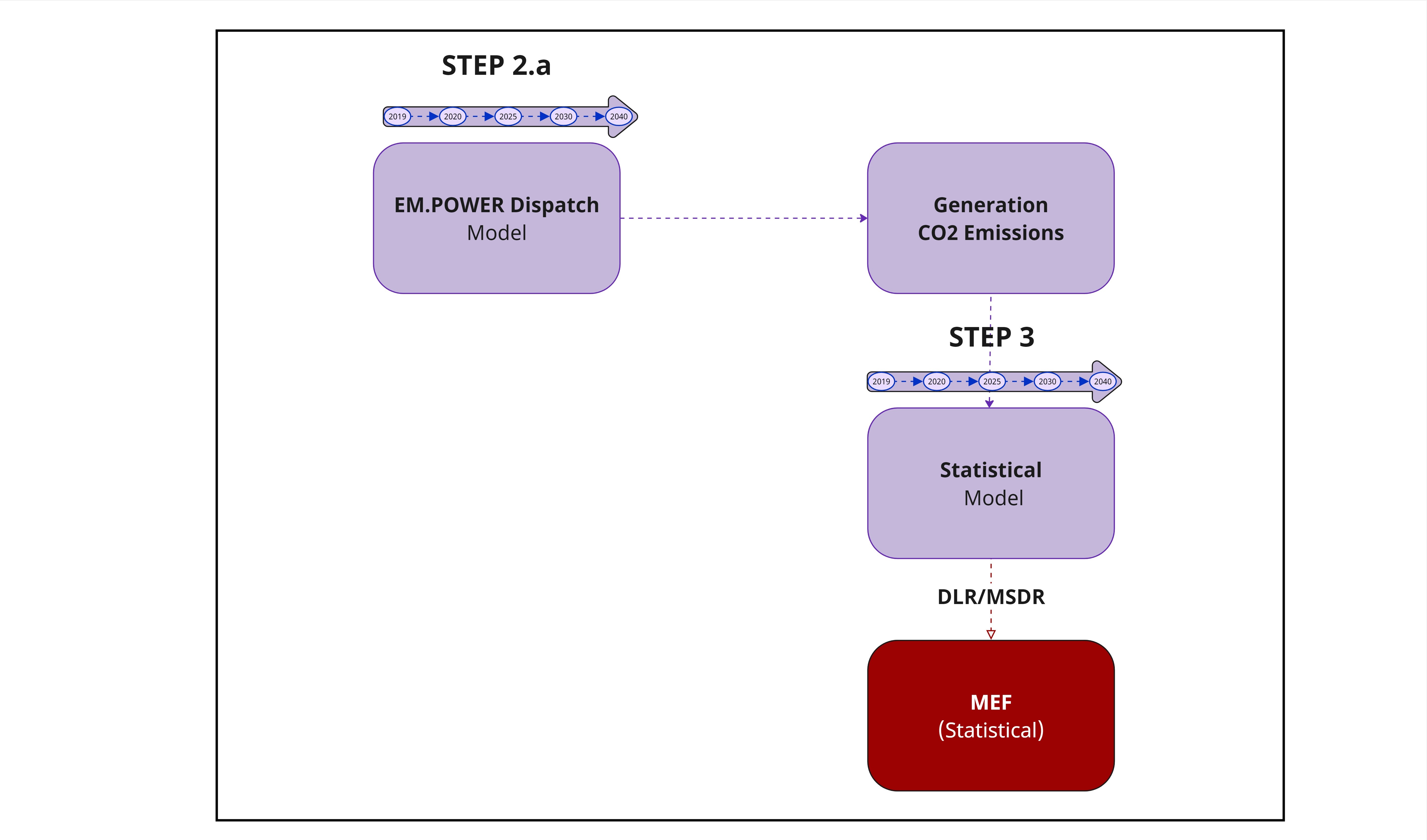}
  \caption{Statistical methodology for estimating the MEF}
  \label{stat_method}
\end{figure}

Statistical models are easy to implement and can reduce calculation time and computational resources significantly in comparison to the ESM. Our aim is to offer a simpler and faster alternative for replacing the second step (more precisely, Step 2.b in Figure \ref{ESM_method}) of the proposed ESM (Section \ref{incrementalmef}) and evaluating its accuracy against the incremental MEF (our benchmark).\\
For that purpose, we propose the Markov Switching Dynamic Regression model (MSDR). Furthermore, for comparison purposes, we estimate a Dynamic Linear Regression model. The latter is mostly used in preview studies estimating average MEFs (\customcite{hawkes2010estimating} and \customcite{hawkes2014long}). However, we consider estimating hourly MEFs, instead of average MEFs, which can provide more accurate and up-to-date MEF estimates.\\
The proposed statistical models use historical data from Agora Energiewende to estimate the empirical MEFs for 2019 and 2020. For the future year, the statistical model incorporates information from the first-step dispatch model (Step 2.a in Figure \ref{ESM_method}), in particular the CO$_2$ emissions and the electricity generation, to estimate the MEFs for 2025, 2030, and 2040. Previous studies estimating MEFs with statistical models relied on observed dispatch data, so they could only estimate historical MEFs. By utilizing the ESM model output, we provide results for future years. As the analysis of future years requires the results from the ESM (Steps 1 and 2.a), the statistical approach represents the third step of our proposed methodology. As illustrated in Figure \ref{stat_method}, the statistical model receives input data (CO$_2$ emissions and electricity generation) from the ESM model (Step 2.a) to estimate the MEF time series.

\subsubsection{The Markov Switching Dynamic Regression MSDR Model}

Statistical models most often employed for MEFs were linear models, specifically the linear regression model (\customcite{hawkes2010estimating}, \customcite{hawkes2014long}, \customcite{seckinger2021dynamic}, \customcite{huber2021carbon}). However, linear models have limitations; although adequate to estimate linear equations, they cannot capture non-linear dynamics between variables.  Additionally, models cannot be estimated for variables that move between significantly different states. In the present case, the CO$_2$ emission levels shift between regimes, resulting in  a change in merit order due to the change in generation technology.\\
To address these shortcomings, we propose a non-linear model called the Markov Switching Dynamic Regression model (MSDR). The MSDR model, first developed by (\customcite{ hamilton1989new}), allows for switching between states; it can contain more than one equation that describes time series behavior in different states. In the MSDR, the switch mechanism is controlled by an unobservable state variable that follows a Markov chain. A certain state only persists for a random period, and then "switches" to another state. \\
The shifting behavior of energy prices has been the focus of many empirical studies in the last few years, which motivated the use of regime-switching models in our context as well. As an example of this, \customcite{galyfianakis2016modeling} employ a Markov Switching Dynamic Regression (MSDR) model to examine the behavior of five energy price series for a period before and after the financial crisis of 2008.  \customcite{chen2022renewable} examine the dynamic causal relationship between renewable energy prices and economic growth using the Markov Switching Vector Autoregression (MS-VAR) model. \customcite{moutinho2022examining} examined the relationships between energy commodity and carbon prices and electricity prices using the Markov Switching Regression model. \\
In the current research, the MSDR model should capture the intricate relationship between carbon emissions and electricity generation due to its nuanced handling of various generation technologies within the merit order. More precisely, a shift in regimes can result from a change in generation technology in the merit order. During the low regime, when a given technology is in operation, emissions remain relatively stable. However, the model adeptly recognizes and accounts for the high regime, marked by a switch in generation technology within the merit order. This transition leads to a consequential change in carbon emissions, as each technology contributes differently to overall emissions. 
Incorporating the MSDR model into our analysis provides us with a nuanced understanding of these transitions, providing a more realistic representation of the energy market's behavior, thereby enabling more precise and comprehensive estimates of MEFs. \\
We parameterize the MSDR using a rolling window of 168 hours (one week) and implement the model using the \textit{"statsmodels"} library in Python.\footnote{Details and python code are provided in the GitHub link: https://github.com/BTU-EnerEcon/Modeling-Hourly-Marginal-CO2-Factors} \\
According to \cite{hamilton1989new}, the development of a given variable can be explained by states (or regimes) where a $k$-regime Markov switching regression model is given as follows:\\

\begin{equation}
\Delta E_t=\beta_{0,k,t}+\beta_{1, k,t} \Delta G_t+\theta_t 
\label{regime1-1}
\end{equation}
\begin{equation}
\theta_t=\phi_k+\theta_{t-1}+\omega_t
\label{regime1-2}
\end{equation}
where $\omega_t \sim N(0, W)$.\\

$\beta_{0,k,t}$ represents the intercept for regime $k$ at time $t$. $\beta_{1, k,t}$ represents the time-varying coefficients for electricity generation ($\Delta G_t$) for regime $k$ at time $t$; this coefficient represents the MEF. $\theta$ is a regime-specific parameter. Despite the fact that noise $\omega_t \sim N(0, W)$ is common in the $k$ regimes with constant variance $W$, there is heterogeneity in the variance of the latent parameter $\theta$.
For each regime, Eq. \ref{regime1-1} is called a measurement or observation. Analogously, Eq. \ref{regime1-2} is called a dynamic equation. 
In our model, the maximum number of regimes $k$ can be either $2$ or $3$, i.e., $k=\{2,3\}$. The model determines the appropriate number of regimes for each year under study.
\customcite{rosen2023intra} has applied a similar model for photovoltaic forecasting.\

The model also incorporates a regime-switching mechanism, allowing for transitions between different regimes over time. The k-state Markov process itself is governed by the following state transition matrix $P$:

\begin{equation}
P\left(S_t=s_t \mid S_{t-1}=s_{t-1}\right)=\left(\begin{array}{cccc}
P_{11} & P_{12} & \cdots & P_{1 k} \\
P_{21} & P_{22} & \cdots & P_{2 k} \\
\vdots & \vdots & \ddots & \vdots \\
P_{k 1} & P_{k 2} & \cdots & P_{k k}
\end{array}\right)
\end{equation}
where $S_t\in \{1,2,3\}$, $p_{ii}$ is the probability of being in regime $i$, and $p_{ij}$ is the probability of transition from regime $i$ at time $t-1$ to regime $j$ at time $t$.\\

To estimate the model, the joint distribution of the observed data $\left\{\Delta E_t, \Delta G_t\right\}$ and the unobservable regime indicator ${S_t}$ is given by the product of the likelihood functions for each time point:
\begin{equation}
    L(\Theta)=\prod_{t=1}^T\left[P_{S_{t-1}}  f\left(\Delta E_t, \Delta G_t \mid \Theta_{S_t}\right)\right]
\end{equation}

Here, $\Theta$ represents the set of all model parameters, $P_{S_{t-1}}$ is the probability of being in regime $S_{t-1}$ at time $t-1$, and $f\left(\Delta E_t, \Delta G_t \mid \Theta_{S_t}\right)$
is the conditional density function of the observed data given the regime $S_t$.\\
In order to determine the parameters of the MSDR, the maximum likelihood estimator is used. Taking the logarithm of the likelihood function, 
\begin{equation}
    \log L(\Theta)=\sum_{t=1}^T\left[\log P_{S_{t-1}}+\log f\left(\Delta E_t, \Delta G_t \mid \Theta_{S_t}\right)\right]
\end{equation}

the maximum likelihood estimator involves maximizing the log-likelihood function with respect to the parameters $\Theta$. This is typically done using optimization algorithms:
\begin{equation}
    \hat{\Theta}=\arg \max _{\Theta} \mathcal{L}(\Theta)
\end{equation}
The model parameters $\Theta$ are estimated using available data within rolling windows of 168 hours. The set of parameters $\Theta$, including the intercepts $\beta_{0,k,t}$ of each regime $k$, as well as the coefficients  $\beta_{1,k,t}$, provide insights into the impact of energy generation on CO$_2$ emissions within each regime. The time-varying nature of the coefficients reflects potential shifts in the relationship over time.\\
Each regime in MSDR has its own set of parameters, including different coefficients ($\beta_k$) for the exogenous variable of electricity generation, $\Delta G$. Hence, we calculate the regime-weighted coefficient, which reflects the weighted sum of these regime-specific coefficients, with the weights representing the probability of being in each regime at a given time $t$. 
The resulting regime-weighted coefficient reflects the time-varying nature of the regime dynamics and provides a single coefficient for the change in electricity generation, $\Delta G$, that can be used for interpretation. As a result, this regime-weighted coefficient for $\Delta G$ at time $t$, denoted as $MEF_t$, is computed as follows:

\begin{equation}
    MEF_t=\sum_{k=1}^K \beta_k^{\left(S_t\right)} \cdot P\left(S_t=k \mid y_{1: t}\right)
\end{equation}

where $K$ is the total number of regimes. 

\subsubsection{The Dynamic Linear Regression Model}
\label{dlrmodel}
In the literature, the commonly used statistical model for MEF estimation is the linear regression model (\customcite{hawkes2010estimating}, \customcite{hawkes2014long}, \customcite{seckinger2021dynamic}, \customcite{huber2021carbon}).
Unlike the above-mentioned studies that estimated average MEFs, we propose to estimate hourly MEFs with a Dynamic Linear Regression DLR model using a rolling window of 168 hours. The model is given by the following equation:

\begin{equation}
    \Delta E_t=\alpha_{0,t}+\alpha_{1,t} \Delta G_t+\eta_t
    \label{linear}
\end{equation}
\begin{equation}
    \eta_t=\psi_t+\eta_{t-1}+\omega_t
\end{equation}

where $\Delta E_t$ and $\Delta G_t$  are the variations in emissions and generation between two successive hours. $\alpha_{0,t}$ and $\alpha_{1,t}$ are the model parameters. More precisely, $\alpha_{1,t}$ represents the time-varying coefficient of the electricity generation  $\Delta G$, which also represents the MEF. $\eta$ denotes the time-varying state component or error term that evolves dynamically, and $\psi$ is a time-varying drift term that controls the evolution of $\eta$. 

\subsection{CO$_2$ Footprint of Electric Vehicle Charging as an Application Using MEFs}
\label{smartchargingmodel}
The MEF results can be used for a variety of applications. One intuitive use would involve consumers who want to reduce the emission footprint of the electricity they consume. To demonstrate this idea, we analyze the example of electric vehicle charging. Note that we purposefully introduce a very simple and straightforward model; our aim is to demonstrate the intuitive use case and provide an estimate of the impact.\\
We thus present a model where a consumer charges an electric vehicle at home overnight, i.e. between 20:00 and 06:00. We assume that the vehicle must be charged four hours to reach full capacity. To simplify the application, all other parameters (e.g., the capacity of the battery, charging power, etc.) are neglected. Note that we assume the car needs to be charged overnight, essentially preventing the use of excess solar power during the day. Hence, our estimate of the achievable reduction in the CO$_2$ footprint is likely to be a conservative estimate.\\
To assess the impact of minimizing the CO$_2$ footprint, we compare two scenarios. The first scenario is "Normal Charging," where the vehicle charges during the first 4 hours (20:00 to 00:00). The second scenario is "Emission-Minimized Charging," where the vehicle charges for the 4 hours during which the MEF is the lowest within the planned charging time (20:00 to 06:00) to reach a combination of charging hours that minimizes the MEF. \

Let MEF$_{h,d}$ represent the rate of CO$_2$ emissions per unit of electricity generated, expressed at hour $h$ of day $d$, with the hours being numbered from 1 to 10 and the days numbered from 1 to 365. We aim to quantify the total amount of CO$_2$ emissions each day and aggregate them over a year.

\subsubsection{Normal Charging}\

In this scenario, the vehicle charges during the first 4 hours of the charging period (from 20:00 to 06:00), i.e. from 20:00 to 00:00. For each day $d$, the total emissions $E1_d$ are computed by summing up the MEF values for these 4 hours $h$:
 \begin{equation}
     E1_d=\sum_{h=1}^4 \operatorname{MEF_{h,d}}
     \label{e1perday}
 \end{equation}

The cumulative emissions for the entire year $E1^{\mathrm{total}}$ are the sum of daily emissions:
\begin{equation}
    E1^{\mathrm{total}}=\sum_{d=1}^{365} E1_d
    \label{cume1peryear}
\end{equation}

\subsubsection{Emission-Minimized Charging}\

In the Emission-Minimized Charging scenario, the objective is to shift the charging hours to the 4 hours within the given 10-hour window (20:00 to 06:00) such that the total emissions are minimized.\\
The resulting optimization problem on a given day thus minimizes $E2_d$:
\begin{equation}
 E2_d = \min \sum_{h=1}^{10} I_{h,d} \operatorname{MEF_{h,d}}  
 \label{e2perday}
\end{equation}
where $I_{d,d}$ is an integer variable that is either zero (no charging) or one (charging). 

To make sure that the vehicle is charged, we introduce the constraint:
\begin{equation}
\sum_{h=1}^{10} I_{d,d} = 4,  \: \: \: \: \: \forall d
 \label{e2const}
\end{equation}

Again, the cumulative emissions for the entire year $E2^{\mathrm{total}}$ are the sum of daily emissions:
\begin{equation}
    E2^{\text {total }}=\sum_{d=1}^{365} E2_d
    \label{e2peryear}
\end{equation}

\subsubsection{Emission Savings}\

The potential emission savings from emission-minimized charging are calculated as:
\begin{equation}
 \Delta E=E1^{\text {total}}-E2^{\text {total}}   
 \label{emissionsaving}
\end{equation}
where $\Delta E$ represents the total reduction in CO$_2$ emissions due to optimized charging based on the MEF values.

\section{Data}
\label{data}
Since we used two different methodologies to estimate the MEF, in this section we describe the data used for the ESM model and the data used in the statistical models, distinguishing between historical year data and future year data. 

\subsection{Data Used in the ESM | Historical Years}
\label{datasection}

To compute marginal emission factors with a high level of accuracy, a detailed parametrization of the electricity market model is necessary. This section presents the input data used to develop and calibrate the EM.POWER Dispatch market model. The model is parametrized for historical and future years and includes comprehensive datasets for both the demand and supply sides. For historical years, the data covers the European electricity market from 1 January 2019 until 31 December 2020. The required dataset will be described first. The analysis of future years requires additional input, which is based on the European Transmission System Operators' ten-year network development plan (TYNDP), and will be described second.\\

Demand in hourly resolution from ENTSO-E Transparency serves as the foundation for modeling the demand side. The supply shortages are considered through load-shedding costs of €3,000/MWh. On the supply side, the EM.POWER Dispatch model includes renewable and thermal generating technologies, as well as energy storage systems, and cross-border trade. The model distinguishes between ten thermal generation technologies grouped into clusters based on commissioning years. Each cluster is parameterized with its corresponding efficiency, minimum generation, and partial-load efficiency loss, as well as CO$_2$ content and CO$_2$ price. Renewable energy sources, such as onshore wind, offshore wind, and photovoltaic are modeled using hourly availability factors from Renewable Ninja. \\
An overview of the sources used to parameterize the EM.POWER Dispatch model is given in Table \ref{ESM_parameters}.

On the \emph{supply side}, a set of technologies is generating and storing electricity. For a better depiction of the generating technologies, the installed capacity of power plants is divided into classes depending on the year in which they are commissioned.
Data reflects the empirical status quo of each generation unit, including conventional plants (such as nuclear, coal, gas) and renewable energy sources (such as solar, wind, pumped storage hydropower plants, reservoirs, run-of-river). We include techno-economic properties for each power plant class, such as fuel costs, CO$_2$ factor and CO$_2$ cost, efficiency, plant lifetime, etc. High-resolution data is implied to create solar and wind profiles for accurately modeling variable renewable energy systems. This data reflects the intermittent nature of renewables, which depends on each node (country) and hour.
To model hydropower systems, data on water inflows, reservoir capacities, and turbine power is required to ensure the accurate depiction of these technologies in the model.

\begin{table}[htbp]
\centering
\caption{Data used in the energy system models}
\begin{tabular}{ll}
\hline
			\multicolumn{1}{l}{\textbf{Parameter}} &
			\multicolumn{1}{l}{\textbf{Source}} \\
            \hline

Demand &  \customcite{EntsoeTPe} \\
\hline
CO\textsubscript{2} prices	& \customcite{Sandbag2021}		\\
\hline
Control power procurement & \customcite{Regelleistung}	\\
\hline 
Curtailment costs for RES	& own assumption: 20 EUR/MWh		\\
\hline
Renewable energy installed capacity & \customcite{EntsoeTPe} \\ \hline
Renewable energy availability factor & \customcite{RenewablesNinja} \\ \hline
Efficiency of generation capacities	&	\customcite{schroder2013current}, 	\\
 & \customcite{OPSDa}	\\
\hline
Efficiency losses at partial load	&	\customcite{schroder2013current} \\
\hline
Electricity demand \\(original day-ahead forecast)	& \customcite{EntsoeTPe}		\\
\hline
Energy-power factor (for storages) &	own assumption: 9	\\
\hline
Fuel prices	&	\customcite{Destatis2020},	\\
(Lignite, nuclear, coal, gas, oil)	&	\customcite{EEX2021}, \customcite{EntsoS} \\
\hline
Generation and storage capacity	&	\customcite{BNetzA2021}, \customcite{UBA2020}, \customcite{EBC2021},  \\
& \customcite{EntsoeTPa}, \\
&  \customcite{OPSDa}   \\
\hline
Generation by CHP units	&	\customcite{EC2021}	\\
\hline
Historic electricity generation	&	\customcite{EntsoeTPd}	\\
\hline
Load-shedding costs	&	own assumption: 3,000 EUR/MWh	\\
\hline
Minimum output levels  &  \customcite{schroder2013current}	\\
\hline
Imports and exports	&	\customcite{EntsoeTPf}, \\

\hline
Variable O\&M costs  &  \customcite{schroder2013current}	\\
\hline
Power plant outages	&	\customcite{EntsoeTPb}	\\
\hline

Start-up costs	&	\customcite{schroder2013current}	\\
\hline
Seasonal availability of hydropower &   \customcite{EntsoeTPd}	\\
\hline
Temperature (daily mean)	&	\customcite{OPSDb}	\\
\hline
Water value 	&  \customcite{EntsoeTPd}, 	\\
& \customcite{EntsoeTPg}	\\
\hline
			\hline
\end{tabular}
\label{ESM_parameters}
\end{table}

The data provided in the table above is available on the GitHub repository.\footnote{https://github.com/BTU-EnerEcon/Modeling-Hourly-Marginal-CO2-Factors}\\

\subsection{Data Used in the ESM | Future Years}
\label{datasectionF}

While modeling historical years requires gathering extensive empirical data, future year data requires a dataset of projections from a reliable source. We base our future assumptions on the "National Trends" scenario used in the Ten-Year Network Development Plan, developed by the European National Transmission System Operators, to parameterize the energy system model for the years 2025, 2030, and 2040. This scenario represents the implementation of existing national energy and climate policies, including National Energy and Climate Plans. The National Trends scenario is based on current policy frameworks and the rate of change to cleaner energy sources allowed by these policies. 
The parameterization of EM.POWER Invest uses the following parameters from the National Trends scenario: electricity demand, planned capacities for wind, solar, and biomass, aligned with national renewable targets, existing and planned capacities for pumped hydropower storage facilities, projected capacities for cross-border interconnections, and forecasts for fuel costs.\\
Furthermore, to determine the renewables' availability in future years, we use availability factors corresponding to the representative weather year 2012 from \customcite{RenewablesNinja} and projected installed capacity from the National Trends scenario. 
The parametrization of EM.POWER Dispatch consists of the same abovementioned parameters available from the National Trends scenario provided by \customcite{TYNDP2022}. The installed capacity for renewable and storage systems is parametrized with data provided by the National Trends scenario, but installed capacity for conventional technologies is parametrized with the output provided by computations from EM.POWER Invest, since both models include the same power plant classes based on commissioning years.

\subsection{Data Used in the Statistical Analysis}

The two statistical models use the same dataset: hourly electricity generation data, namely aggregated conventional generation, and hourly CO$_2$ emissions.\footnote{Note that hourly CO$_2$ emissions as input in the statistical model denote aggregated CO$_2$ emissions in that hour, i.e. the sum over all emissions from all technologies producing electricity in that hour.} While demand and electricity production are highly correlated, we chose conventional generation as our input variable to estimate CO$_2$ emissions, specifically from fossil fuel-based sources like coal, natural gas, and oil, for two reasons: First, the inclusion of renewable energy sources such as wind, solar, and hydropower, which do not emit CO$_2$, will not provide meaningful insight for this purpose. Second, the relationship between production and CO$_2$ emissions would be blurred when the overall demand for electricity is used, as it includes both conventional and renewable energy sources. For a clearer understanding of the relationship between electricity production and CO$_2$ emissions, we direct our attention to conventional generation.\\
The data required for statistical analysis can be divided into two categories: historical data and future projections. For historical years, we can use real-world observations or publicly available estimates for both electricity generation and CO$_2$ emissions data. 
For that purpose, we use real data from Agora Energiewende\footnote{Agorameter is a tool developed by Agora Energiewende for data visualization and analysis that offers historical and real-time insights into Germany's energy generation and related carbon emissions. It provides an all-encompassing perspective of the power system, allowing users to examine the composition of the different energy-generating sources which contribute to carbon emissions, as well as the overall variations in carbon emissions over time. Due to easy data access and documentation, it is a valuable tool that we can use for validating the statistical approach employed in our study.\\} for 2019 (\cite{agorameter2019}) and 2020 (\cite{agorameter2020}).
\footnote{Other sources which determine CO$_2$ emissions and electricity generation are also available: CO$_2$-Monitor (\customcite{co2monitor}), eCO$_2$grid (\customcite{eco2grid}), CO$_2$Map ( \customcite{co2map}), Electricity Maps (\customcite{co2map}), STROMDAO (\customcite{stromdao}).}
Alternatively, we can use historical year data from the ESM model as described in Section \ref{emissionandgeerationdata}.\\
For future years, we will derive input data from ESM model estimates since they are otherwise unavailable, enabling us to overcome the limitations of the statistical model, which can only estimate empirical MEFs based on historical data. 

\section{Results and Discussion}
\label{results}
We present and interpret the results for the ESM model first. Second, we analyze and discuss the results from the statistical models, including the results from the model parameters estimation and the MEF estimation, which we also compare to the incremental MEF from the ESM model. 

\subsection{Energy System Model Results}
\label{esmresultsinterpretation}

We start the discussion of the ESM results with the key result: the incremental MEF estimates in the following subsection. Afterwards, we also present the CO$_2$ emissions and the generation of electricity from the EM.POWER Dispatch model, which are used as input in the statistical models to estimate MEFs for future years.

\subsubsection{Incremental MEF Results}

\begin{figure}[H]
  \centering
  \includegraphics[width=1\textwidth]{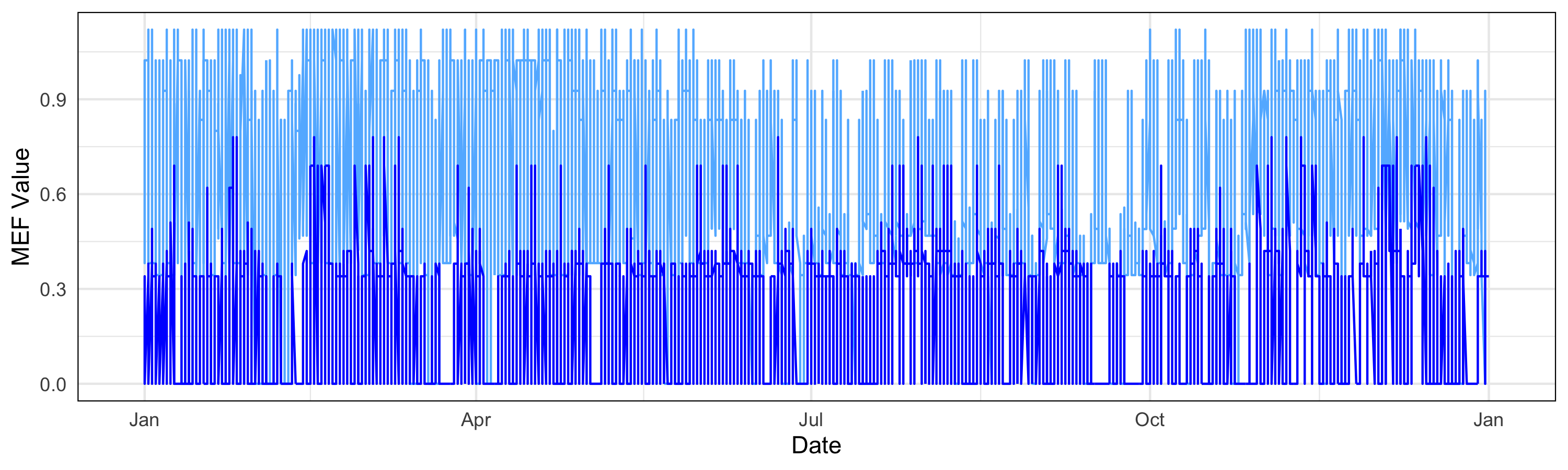}
  \caption{\textcolor{SteelBlue1}{Incremental MEFs for 2020} and \textcolor{blue}{incremental MEFs for 2040} in \textit{kg CO$_2$ eq./kWh}}
  \label{meffund20vs40}
\end{figure}

Figure \ref{meffund20vs40} shows the time series of marginal emission factors for the years 2020 (light blue) and 2040 (dark blue) under the assumptions discussed in the data section. Results for the other years (2019, 2025, and 2030) can be found in Appendix Section \ref{otherfund_mefs}.\\
The incremental MEF for 2020 shows a high level of variation throughout the year, varying dynamically between 0 and approximately 1 ton CO$_2$/MWh. This is particularly relevant because it shows that the \emph{when} of consumption matters: Consuming $1$ additional MWh of electricity in an hour $x$ can be much more harmful to the climate than consuming $1$ MWh in hour $y$.\\
For example, higher MEFs are observed in the winter and early spring months, which could be attributed to several factors, such as comparably high demand and low RES feed-in. We note that due to the COVID-19 outbreak, 2020 is not a representative year for the energy sector and the lower MEFs could have occurred due to changes in the energy demand patterns and the limited use of industrial processes.\\
MEFs in 2040 are significantly lower on average. This results from the significant increase in renewable capacities and the phase-out of coal generation.\footnote{Based on our assumption, the planned German coal phase-out for both lignite and hard coal will be finished by 2035.} Nonetheless, the majority of hours in 2040 still have strictly positive MEFs, reflecting that the system is not fully transitioned to a low-carbon state and that non-renewable sources still contribute to the energy mix, albeit at a reduced level. Nonetheless, the year 2040 already exhibits several hours with a MEF of zero, meaning that a marginal load increase in these hours would not lead to any additional CO$_2$ emissions. Overall, seasonal variation in 2040 is less pronounced compared to the year 2020. 

\subsubsection{CO$_2$ Emissions and Electricity Generation Data}
\label{emissionandgeerationdata}
In addition to calculating incremental MEFs, which is time- and resource-intensive due to its high complexity (in particular, requiring Step 2.b of the ESM model), the EM.POWER Dispatch model generates several other insightful outputs already in Step 2.a, including hourly CO$_2$ emissions and electricity generation data. These outputs will be fed into the statistical model (see Figure \ref{stat_method}), enabling us not only to estimate empirical MEFs based on historical data but also to project MEFs for future scenarios.

\begin{figure}[H]
  \begin{subfigure}{0.5\textwidth}
    \centering
    \includegraphics[width=1\linewidth]{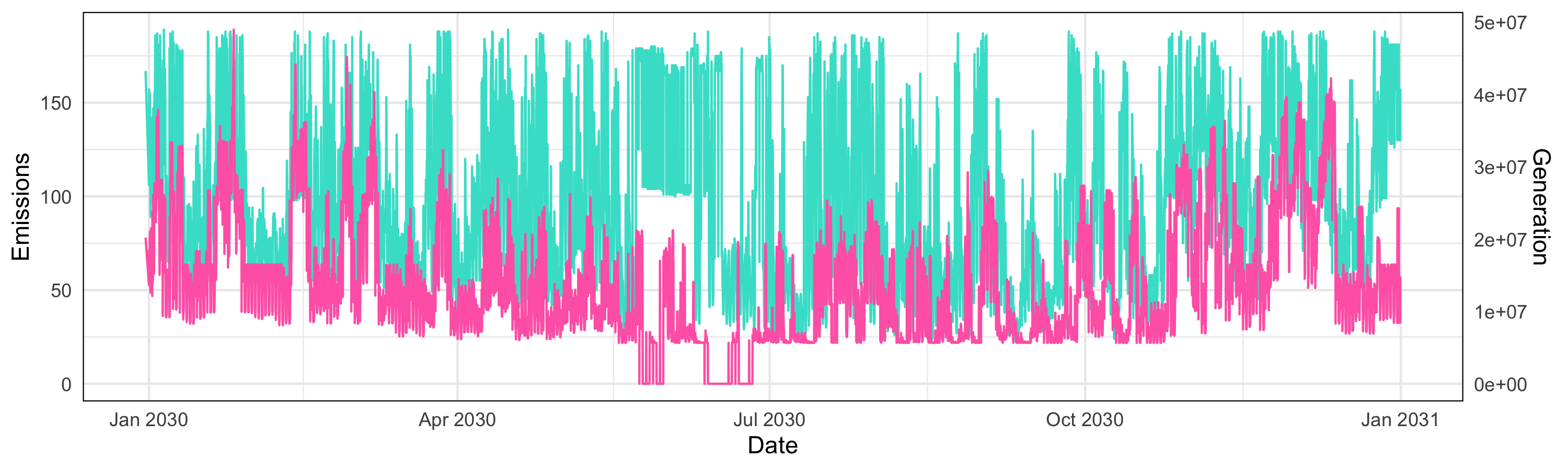}
    \caption{}
  \end{subfigure}%
  \begin{subfigure}{0.5\textwidth}
    \centering
    \includegraphics[width=1\linewidth]{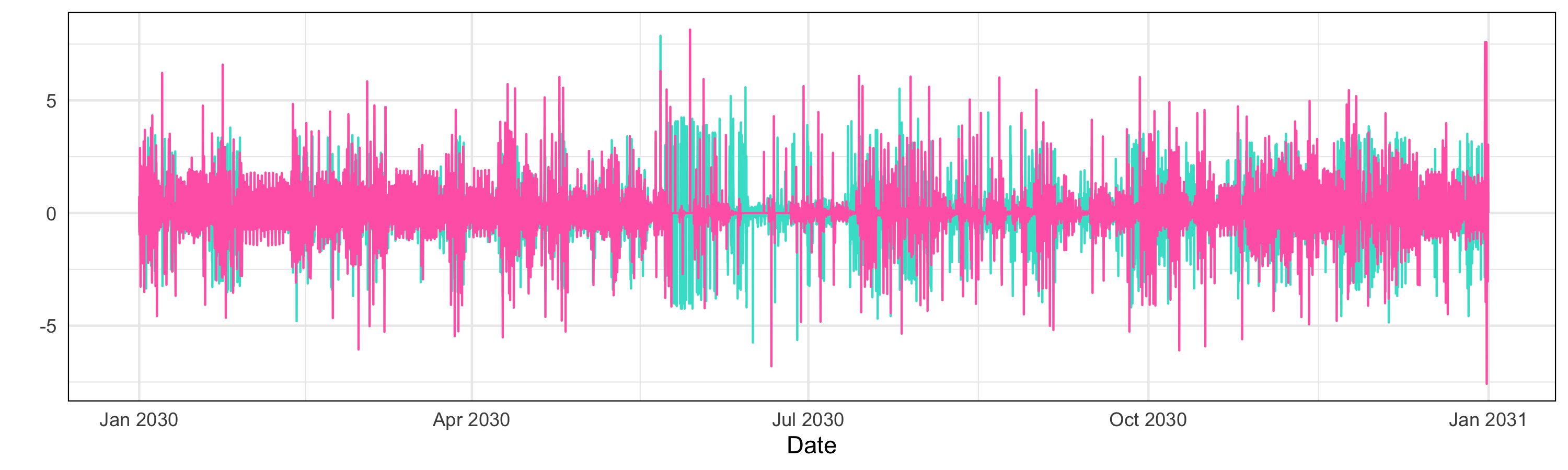}
    \caption{}
  \end{subfigure}
  \caption{\textcolor{bluegreen}{CO$_2$ emissions} and \textcolor{hotpink}{electricity generation} data and (b) their variations $\Delta$ in 2040}
  \label{input2040}
\end{figure}

Figure \ref{input2040} plots the CO$_2$ emissions and electricity generation for 2040. We observe a clear dependency between the two time series, showing the same increase and decrease patterns in the data.\footnote{Note that this is good news in our setting, as emissions are the dependent variable in the statistical models.} The right plot illustrates the variations in CO$_2$ emissions and electricity generation for successive hours. The remaining plots for the other years (2019, 2020, 2025, and 2030) are included in Appendix Section \ref{emissionsandgenerations_plots}.\\
Given that CO$_2$ emissions and electricity generation data from the ESM model will be used as input for the statistical model, a preliminary statistical analysis is necessary to investigate the characteristics of the time series. Appendix Section \ref{emissionsandgenerations_tests} provides  Table \ref{statistics2040}, which gives a comprehensive overview of the descriptive statistics and test results for CO$_2$ emissions ($E$), electricity generation ($G$), and their respective variations or initial differences ($\Delta E$) and ($\Delta G$)\footnote{($\Delta E$) and ($\Delta G$) have been standardized using "\textit{StandardScaler}" from the "\textit{sklearn}" Python package, to ensure that both time series, originally on different scales, are normalized to a common scale. This allows for more accurate comparisons and analyses by adjusting both series to have a mean of zero and a standard deviation of one.} in 2040. Appendix Section \ref{emissionsandgenerations_tests} also provides various additional statistical tests. 

\subsubsection{ESM Model Validation}

We have already pointed out in the introduction that MEFs are unobservable, even in retrospect. At the same time, we have also pointed out that \customcite{elenes2022well} compared approximation methods in an analytical setting and found that an estimate of MEFs based on complex energy systems modeling was the most accurate. Nevertheless, we provide an additional -- even though indirect -- way to benchmark our ESM: We validate our model with a comparison of our ESM's price estimator, calculated as the shadow price of the demand constraint, and historic electricity price realizations. \\
Electricity prices are often used to validate energy system models, and are a key instrument reflecting market dynamics. Electricity prices are sensitive to a wide range of factors including supply and demand, policy implications, operation constraints, weather conditions, etc.
By validating against historical wholesale prices, we show to what extent the model emulates market conditions.\\
We compare the hourly results for the historic sample years of 2019 and 2020, presenting the mean square error (MSE), mean absolute error (MAE), and root mean square error (RMSE). The corresponding formulas for these metrics are provided in Appendix Section \ref{evaluationmetricdef}.\\
According to Table \ref{tab:error_metrics_2019_2020}, the MAE for 2019 was 9.55 and the RMSE 12.47. The results are broadly in line with the literature analyzing ESM to forecast hourly prices. As an example,
\customcite{qussous2022understanding} obtained an MAE of 6.69\euro / MWh and an RMSE of 10.91\euro / MWh using a non-equilibrium oriented techno-economic market model aimed at reproducing the day-ahead electricity prices for the German bidding zone for the year 2019.

\begin{table}[H]
\centering
\caption{Evaluation metrics for 2019 and 2020 for the EM.POWER Dispatch model for electricity price estimation} 
\begin{tabularx}{\textwidth}{l *{2}{>{\centering\arraybackslash}X}}
  \hline
Metric & Year\_2019 & Year\_2020 \\ 
  \hline
MAE & 9.55 & 12.47 \\  
RMSE & 14.78 & 18.54 \\ 
   \hline
\end{tabularx} 
\label{tab:error_metrics_2019_2020}
\end{table}

\subsection{Statistical Model Results}
\label{statmodelresults}

In this section, we first focus on estimating key parameters of our models (Section \ref{statparest}), with particular emphasis on the MEF (coefficient of electricity generation). Afterward, the resulting statistical MEFs and their accuracy of estimation are discussed in the following Section \ref{mefresults}.\\ 
We applied the model to historical years (2019, 2020) using real data for CO$_2$ emissions and electricity generation from Agora Energiewende. Data for future years (2025, 2030, and 2040) were derived from the ESM model, as described in Section \ref{emissionandgeerationdata}.\\

\subsubsection{Results of the Model Parameters Estimation}\
\label{statparest}
This section will first present the parameter estimates for the MSDR, followed by the DLR, allowing us to understand the statistical relationship between CO$_2$ emissions and electricity generation (the MEF).

\paragraph{The Markov Switching Dynamic Regression MSDR Parameters Estimation} \leavevmode\\

Table \ref{markowestimation2040} provides the estimation results of the Markov Switching Dynamic Regression (MSDR) model for 2040 (for simplicity, we interpret only the 2040 results).\\
Note that our MSDR code is designed to select between 2 and 3 regimes based on the data provided. The model identifies three distinct regimes, for all studied years, that show distinct relationships between CO$_2$ emissions and electricity generation, which can be linked to real-world phenomena and the merit order effect.

\begin{table}[htbp]
\centering
\caption{Markov switching model results for 2040}
\label{tab:markov_switching_model_2040}
\begin{tabularx}{\textwidth}{@{} l *{6}{>{\centering\arraybackslash}X} @{}}
\toprule
\multicolumn{5}{l}{Dep. variable: Emissions} & \multicolumn{2}{r}{AIC: 9645.559} \\
\multicolumn{5}{l}{No. observations: 8759} & \multicolumn{2}{r}{BIC: 9751.727} \\
\multicolumn{5}{l}{Log likelihood: -4807.780} & \multicolumn{2}{r}{HQIC: 9681.734} \\
\midrule
\multicolumn{7}{c}{\textbf{Regime 1 parameters}} \\
\textbf{Variable} & \textbf{coef} & \textbf{std err} & \textbf{z} & \textbf{P>|z|} & \textbf{[0.025} & \textbf{0.975]} \\
\midrule
const  & 0.0026 & 0.002 & 1.279 & 0.201 & -0.001 & 0.007 \\
x1     & 0.0065 & 0.003 & 2.507 & 0.012 & 0.001 & 0.012 \\
sigma2 & 0.0102 & 0.001 & 17.241 & 0.000 & 0.009 & 0.011 \\
\midrule
\multicolumn{7}{c}{\textbf{Regime 2 parameters}} \\
\textbf{Variable} & \textbf{coef} & \textbf{std err} & \textbf{z} & \textbf{P>|z|} & \textbf{[0.025} & \textbf{0.975]} \\
\midrule
const  & 0.0043 & 0.008 & 0.533 & 0.594 & -0.011 & 0.020 \\
x1     & 0.4948 & 0.014 & 35.452 & 0.000 & 0.467 & 0.522 \\
sigma2 & 0.1114 & 0.007 & 16.026 & 0.000 & 0.098 & 0.125 \\
\midrule
\multicolumn{7}{c}{\textbf{Regime 3 parameters}} \\
\textbf{Variable} & \textbf{coef} & \textbf{std err} & \textbf{z} & \textbf{P>|z|} & \textbf{[0.025} & \textbf{0.975]} \\
\midrule
const  & -0.0356 & 0.047 & -0.761 & 0.447 & -0.127 & 0.056 \\
x1     & 0.3478 & 0.033 & 10.473 & 0.000 & 0.283 & 0.413 \\
sigma2 & 3.9333 & 0.161 & 24.376 & 0.000 & 3.617 & 4.250 \\
\midrule
\multicolumn{7}{c}{\textbf{Regime transition parameters}} \\
\textbf{Transition} & \textbf{coef} & \textbf{std err} & \textbf{z} & \textbf{P>|z|} & \textbf{[0.025} & \textbf{0.975]} \\
\midrule
p[1->1] & 0.8603 & 0.009 & 97.925 & 0.000 & 0.843 & 0.878 \\
p[2->1] & 0.0979 & 0.011 & 9.212 & 0.000 & 0.077 & 0.119 \\
p[3->1] & 0.1203 & 0.014 & 8.867 & 0.000 & 0.094 & 0.147 \\
p[1->2] & 0.0832 & 0.010 & 8.580 & 0.000 & 0.064 & 0.102 \\
p[2->2] & 0.7532 & 0.014 & 54.083 & 0.000 & 0.726 & 0.780 \\
p[3->2] & 0.2453 & 0.018 & 13.339 & 0.000 & 0.209 & 0.281 \\
\bottomrule
\end{tabularx}
\label{markowestimation2040}
\end{table}

As indicated in Table \ref{markowestimation2040}, \textit{in Regime 1}, the coefficient for electricity generation is 0.0065, the average MEF, (p-value = 0.012) with a variance of 0.010. This regime reflects a stable low-emission environment where increases in electricity generation have a modest impact on emissions. It likely corresponds to periods when the energy mix is dominated by low-carbon or renewable sources, resulting in minimal emissions increases despite increased generation.

\textit{Regime 2} shows a much stronger coefficient of 0.4948 (p-value = 0.000) and a variance of 0.111. This regime represents a high-emission environment in which electricity generation significantly increases CO$_2$ emissions. It aligns with scenarios where fossil fuels are heavily used, leading to a pronounced impact on emissions as generation increases. This period might also reflect high demand conditions, where less efficient and higher-emission power plants are dispatched, amplifying emissions.\\
\textit{In Regime 3}, the coefficient is 0.3478 (p-value= 0.0001), with a high variance of 3.933. This regime indicates a moderate positive relationship between electricity generation and CO$_2$ emissions, but with considerable variability. It might represent periods of transition or instability, such as shifts in the energy mix, or fluctuating market conditions affecting the merit order. The high variability suggests unpredictable emission outcomes due to these external factors.\\
Transition probabilities reveal that the model is more likely to remain in Regimes 1 or 2, with less frequent transitions involving Regime 3, which appears to represent less common or exceptional conditions. This stability in Regimes 1 and 2 aligns with expected energy generation patterns, while Regime 3’s variability highlights periods of greater uncertainty or change in the energy system.\\

\begin{figure}[H]
  \centering
  \includegraphics[width=1\textwidth]{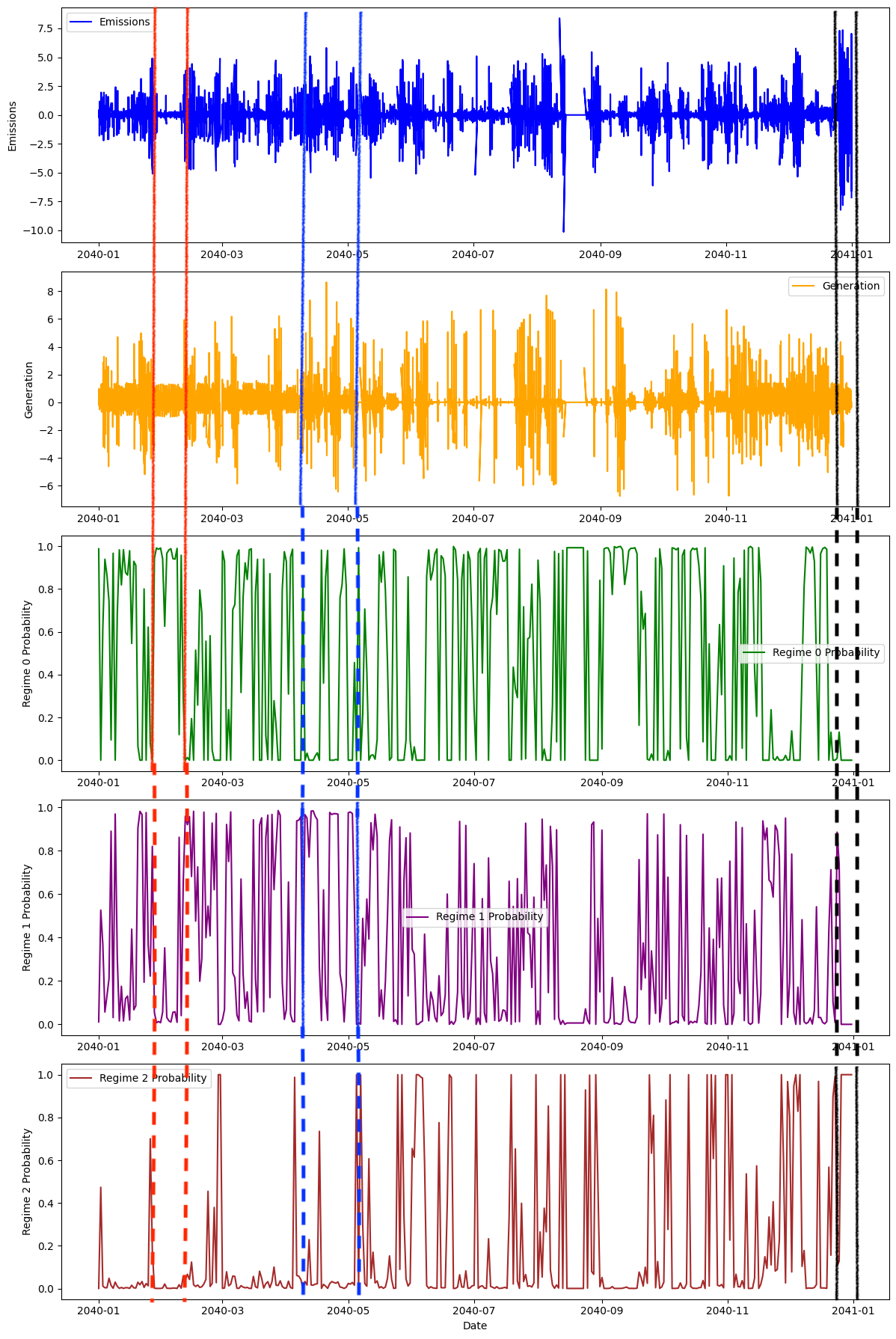}
  \caption{MSDR regime probabilities for 2040, with vertical boundaries indicating examples of \textcolor{red}{Regime 1}, \textcolor{blue}{Regime 2}, and \textbf{Regime 3}}
  \label{proba}
\end{figure}

To gain deeper insights into this phenomenon, we examine the smoothed probabilities of \textcolor{red}{Regime 1}, \textcolor{blue}{Regime 2}, and \textbf{Regime 3} in conjunction with the time series of changes in CO$_2$ emissions and electricity generation, as shown in Figure \ref{proba}.
Based on this figure, the decrease in the generation and CO$_2$ emissions is associated with \textcolor{red}{Regime 1} (low regime). However, an increase in generation is followed by an increase in CO$_2$ emissions, leading to a transition from \textcolor{red}{Regime 1} to \textcolor{blue}{Regime 2}. This transition can be attributed to a shift in generation technology, enabling the model to account for the dynamic merit order, consequently leading to a more accurate estimation of the inherent relationship between the studied variables. During periods associated with \textbf{Regime 3}, the higher variation in CO$_2$ emissions relative to electricity generation likely reflects the use of more carbon-intensive generation technologies during this period.

\paragraph{The Dynamic Linear Regression Parameters Estimation}\leavevmode\\

The Dynamic Linear Regression (DLR) model results are presented in Appendix Section \ref{linearresults}, in particular Figure \ref{lineareg}). Results show a generally positive relationship between changes in electricity generation and CO$_2$ emissions. However, a significant dispersion of data points around the fitted regression line suggests that the model may not fully capture the complexity of this relationship, indicating the potential influence of non-linear factors.


\subsubsection{Assessing the Accuracy of Statistical Models for MEF Estimations}
\label{mefresults}
\begin{figure}[H]
  \centering
  \includegraphics[width=1\textwidth]{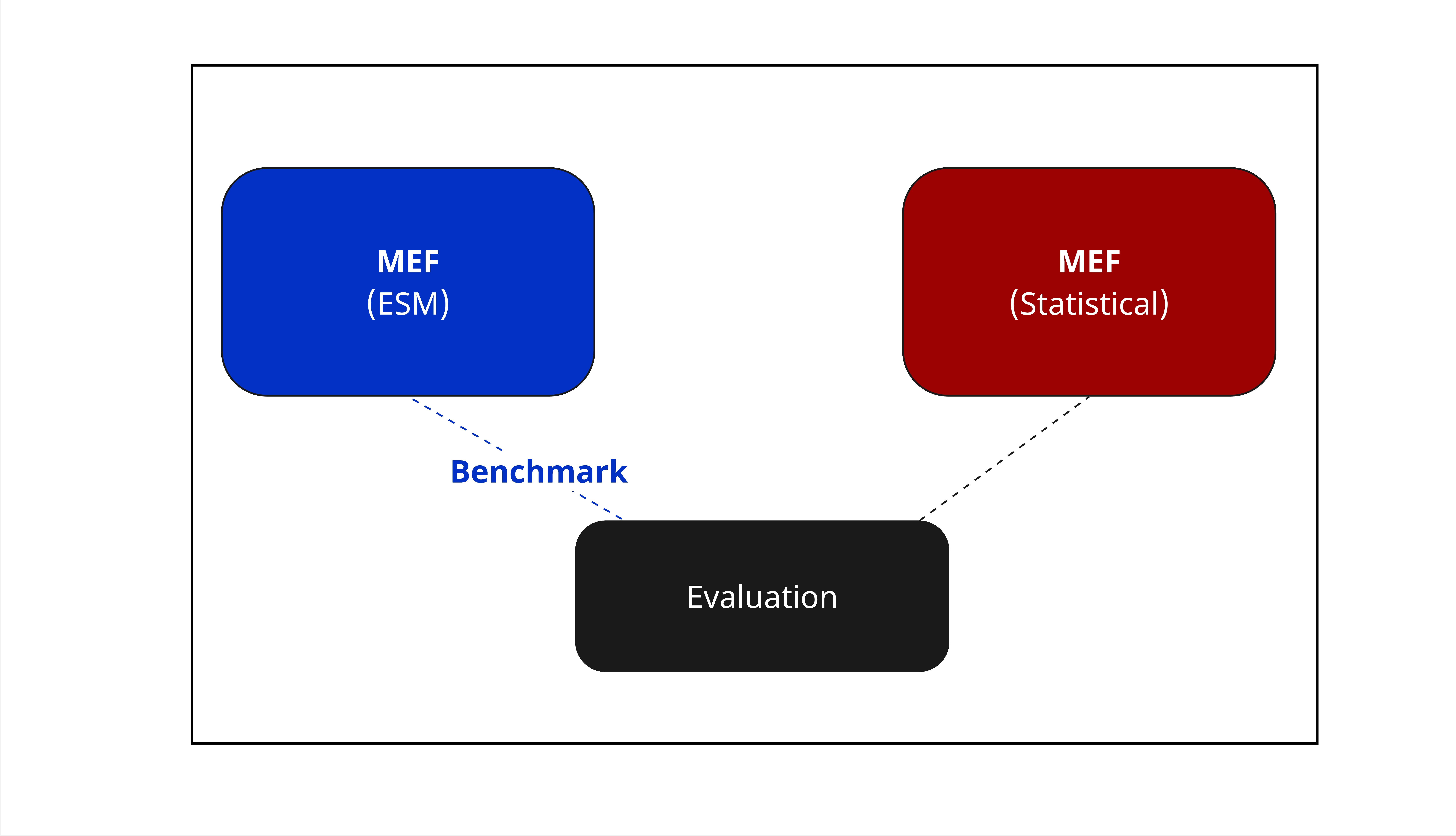}
  \caption{MEF evaluation flowchart}
  \label{Fund_stat_mefs}
\end{figure}

This section evaluates the MEF estimates from the statistical models using the incremental MEF as a benchmark. Figure \ref{Fund_stat_mefs} illustrates the evaluation diagram.

Table \ref{mefevaluation} presents the evaluation metrics for the statistical models' MEF estimation results. We assess the performance using the mean square error (MSE), mean absolute error (MAE), and root mean square error (RMSE). The corresponding formulas for these metrics are provided in Appendix Section \ref{evaluationmetricdef}.
The results indicate that the MSDR model outperforms the DLR model for most years and consistently shows the lowest average values for all metrics.

\begin{table}[htbp]
\centering
\caption{Evaluation metrics for MEF estimations}
\renewcommand{\arraystretch}{1.2} 
\footnotesize 
\begin{tabularx}{\textwidth}{@{}l*{6}{>{\raggedleft\arraybackslash}X}@{}}
\toprule
\multicolumn{7}{c}{\textbf{Markov Switching Dynamic Regression Model}}                                                              \\ \midrule
     & \multicolumn{2}{c}{\textbf{Historical Years}} & \multicolumn{3}{c}{\textbf{Future Years}} & \textbf{Average} \\ \midrule
     & 2019  & 2020  & 2025  & 2030  & 2040  \\ \midrule
\textbf{MSE}  & \textbf{0.328} & \textbf{0.279} & 0.202 & \textbf{0.138} & \textbf{0.122} & \textbf{0.214} \\
\textbf{MAE}  & \textbf{0.482} & \textbf{0.446} & 0.354 & \textbf{0.281} & \textbf{0.267} & \textbf{0.366} \\
\textbf{RMSE} & \textbf{0.573} & \textbf{0.528} & 0.450 & \textbf{0.371} & \textbf{0.350} & \textbf{0.454} \\ \midrule
\multicolumn{7}{c}{\textbf{Dynamic Linear Regression Model}}                                                                       \\ \midrule
     & \multicolumn{2}{c}{\textbf{Historical Years}} & \multicolumn{3}{c}{\textbf{Future Years}} & \textbf{Average} \\ \midrule
     & 2019  & 2020  & 2025  & 2030  & 2040   \\ \midrule
\textbf{MSE}  & 0.480 & 0.450 & \textbf{0.180} & 0.303 & 0.227 & 0.328 \\
\textbf{MAE}  & 0.608 & 0.604 & \textbf{0.334} & 0.359 & 0.289 & 0.439 \\
\textbf{RMSE} & 0.692 & 0.670 & \textbf{0.424} & 0.550 & 0.476 & 0.562 \\ \bottomrule
\end{tabularx}
\label{mefevaluation}
\end{table}

\begin{figure}[H]
  \centering

  \begin{subfigure}{1\textwidth}
    \centering
    \includegraphics[width=\linewidth]{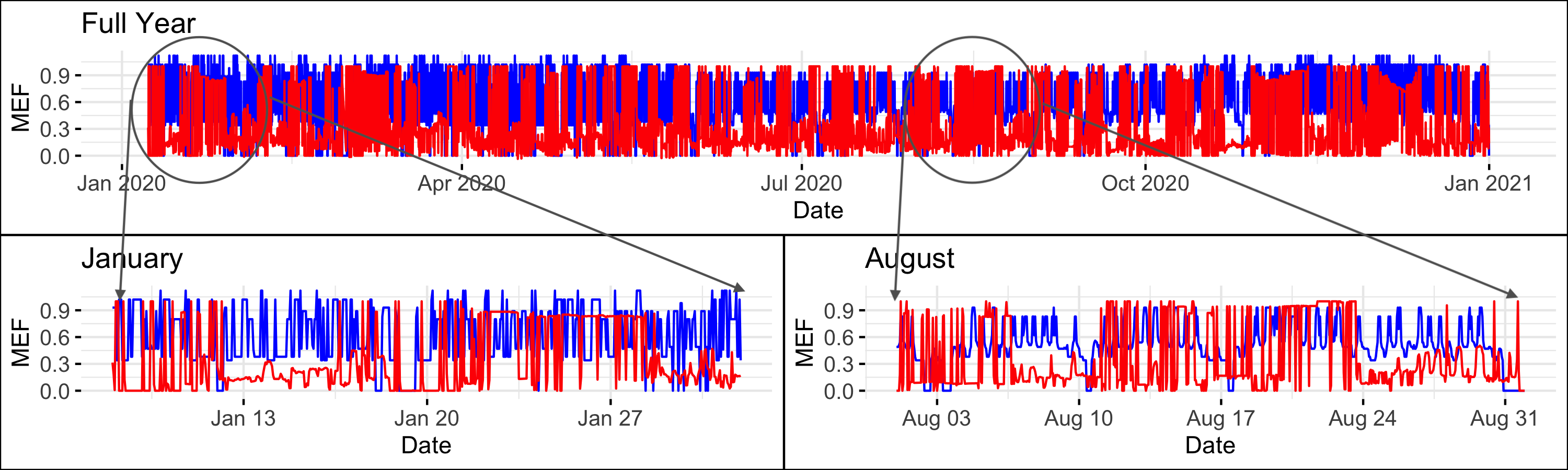}
    \caption{}
  \end{subfigure}
  \hfill
  \begin{subfigure}{1\textwidth}
    \centering
    \includegraphics[width=\linewidth]{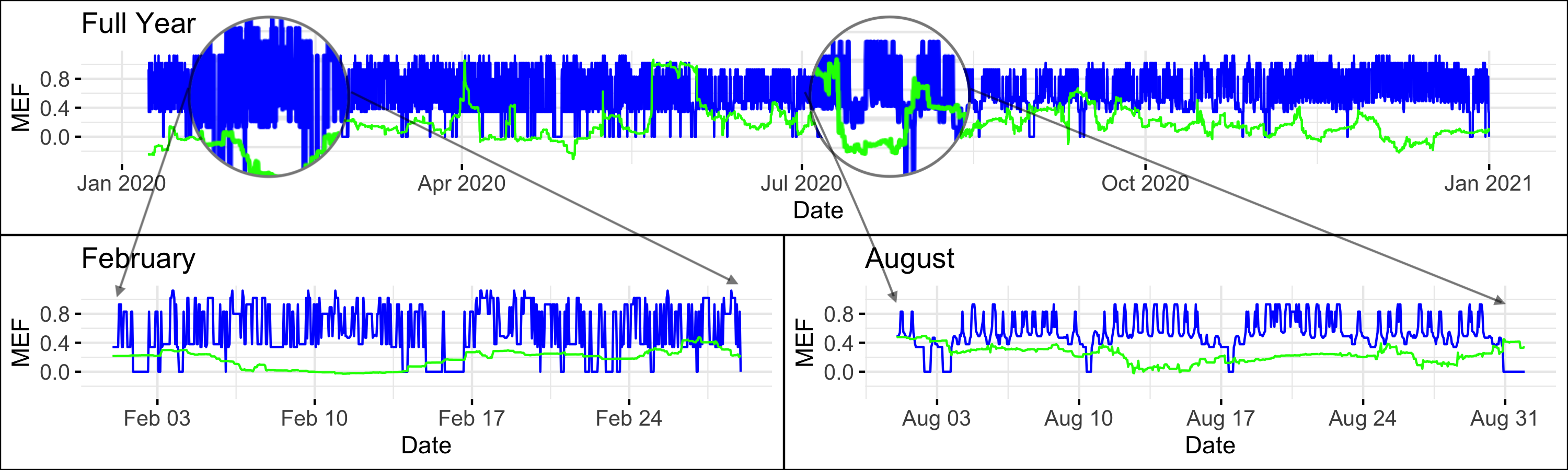}
    \caption{}
  \end{subfigure}
  \caption{\textcolor{blue}{Incremental MEF}, \textcolor{red}{MEF--MSDR}, and \textcolor{green}{MEF--DLR} time series from Agora Energiewende data in \textit{kg CO$_2$ eq./kWh} in 2020}
\label{compmef2020}
\end{figure}

\begin{figure}[H]
  \centering

  \begin{subfigure}{1\textwidth}
    \centering
    \includegraphics[width=\linewidth]{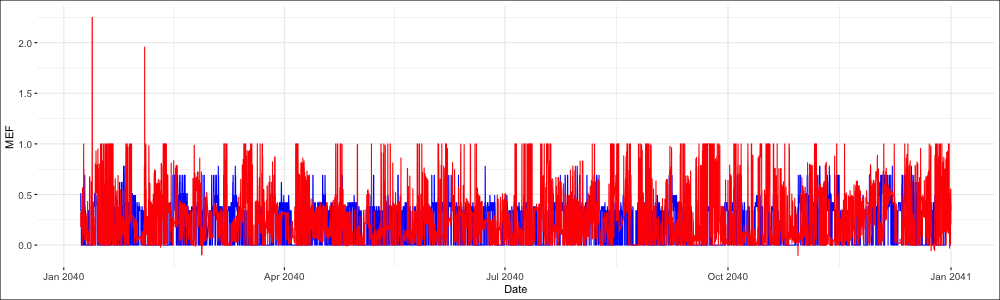}
    \caption{}
  \end{subfigure}
  \hfill
  \begin{subfigure}{1\textwidth}
    \centering
    \includegraphics[width=\linewidth]{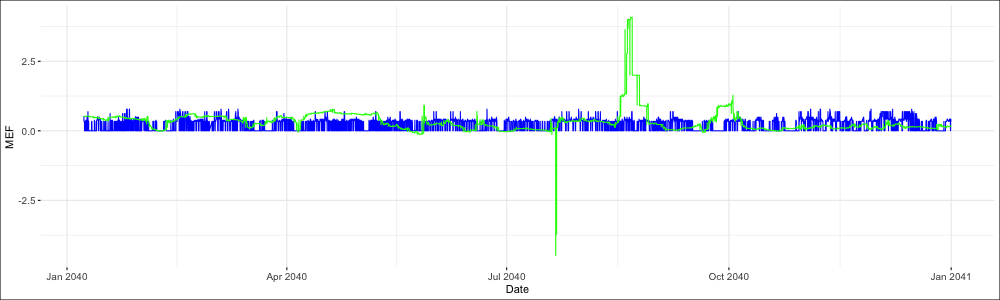}
    \caption{}
  \end{subfigure}
  \caption{\textcolor{blue}{Incremental MEF}, \textcolor{red}{MEF--MSDR}, and \textcolor{green}{MEF--DLR} time series in \textit{kg CO$_2$ eq./kWh} for ESM data in 2040}
\label{compmef2040}
\end{figure}

In fact, the previous results are further validated by Figures \ref{compmef2020} and \ref{compmef2040}, which show the hourly MEFs for the years 2020 and 2040. These figures compare the incremental MEF with the MEFs of the MSDR model (MEF--MSDR) and the DLR model (MEF--DLR). The plots demonstrate that the MEF values from the MSDR (MEF--MSDR) model closely align with those of the incremental MEF. 
To further explain the underlying estimates, we zoomed in on the January (winter month) and August (summer month) periods in 2020, as shown in Figure \ref{compmef2020}. The MEF--MSDR model closely follows the incremental MEF data during these months, effectively capturing the dynamic fluctuations and rapid changes in the time series. This shows the strong ability of the MSDR model to adapt to varying patterns and maintain accuracy.\\
In contrast, the MEF--DLR model struggles to capture the fast fluctuations evident in January and August. The MEF--DLR model's estimations appear more static, indicating that it does not respond as well to the dynamic variations present in the incremental MEF, making it less effective in scenarios requiring responsiveness to quick changes.\\
Additional comparison plots for 2019, 2025 and 2030 can be found in the Appendix Section \ref{mefscompa_app}.\\
Therefore, the close alignment of the estimated MEF of the MSDR model with the incremental MEF highlights its robustness and effectiveness in capturing the dynamics of CO$_2$ emissions and electricity generation. \

In fact, within the context of econometrics, the accuracy of the model is assessed by its ability to predict the dependent variable, the variation in CO$_2$ emissions in this case. The MEF, as a pivotal parameter derived from these models, plays a critical role in this process. Hence, the accuracy of MEF estimation is inherently linked to the accuracy of CO$_2$ emissions estimation.\\
Although we ultimately aim to compare the MEF time series obtained from these statistical models with the benchmark incremental MEF derived from the ESM model, this comparison requires careful consideration due to the distinctive design and underlying principles of both models (statistical vs. ESM). Rather than a direct comparison of MEFs, we first interpret the model parameters (mainly MEFs) and evaluate the accuracy of CO$_2$ emissions estimates. This preliminary step is crucial to understanding the ability of each model to reflect emission variations, setting the stage for a meaningful comparison with the incremental MEF. The detailed analysis of the estimation of CO$_2$ emissions and its implications for the accuracy of MEFs is provided in the Appendix (Section \ref{estimationaccuracyofstatmodel}). This additional information explores how well the statistical models estimate CO$_2$ emissions, providing insight into their overall effectiveness and the consistency of their MEF estimates compared to the incremental MEF benchmark.

In this framework, the consistency between the MEF estimation results and the CO$_2$ emission estimation results (in the Appendix Section \ref{estimationaccuracyofstatmodel}) suggests that an accurate statistical model -particularly in terms of estimating dependent variables - may be a preferable alternative to complex energy system models for estimating the MEF.

\section{Emission-Minimized Vehicle Charging Application and Environmental Benefits}
\label{smartchargingsection}

In this section, we present the results of the emission-minimized charging vehicle strategy described in Section \ref{smartchargingmodel}.
Table \ref{smartchargingtable} shows the cumulative MEF outcomes for a normal charging strategy and an emission-minimized charging strategy. 

\begin{table}[htbp]
\centering
\caption{Emission-Minimized Vehicle Charging Results}
\renewcommand{\arraystretch}{1.2} 
\footnotesize 
\begin{tabularx}{\textwidth}{@{}l*{6}{>{\raggedleft\arraybackslash}X}@{}}
\toprule
\multicolumn{7}{c}{\textbf{Scenario's MEF Results in \textit{kg CO$_2$ eq./kWh}}}                                                              \\ \midrule
     & \multicolumn{2}{c}{\textbf{Historical Years}} & \multicolumn{3}{c}{\textbf{Future Years}} & \textbf{Average} \\ \midrule
     & 2019  & 2020  & 2025  & 2030  & 2040  \\ \midrule
\textbf{Cumulative MEF $E1$} & 868.96 & 757.49 & 650.13 & 338.94 & 327.36 & \textbf{588.58} \\
\textbf{Cumulative MEF $E2$}  & 561.02 & 577.44 & 466.58 & 212.51 & 207.23  & \textbf{404.95} \\
\textbf{Emission Savings} & 307.93 & 180.06 & 183.55 & 126.43 & 120.13 & \textbf{183.62}\\ \midrule

\end{tabularx}
\label{smartchargingtable}
\end{table}

\begin{figure}[H]
  \centering

  \begin{subfigure}{0.45\textwidth}
    \centering
    \includegraphics[width=\linewidth]{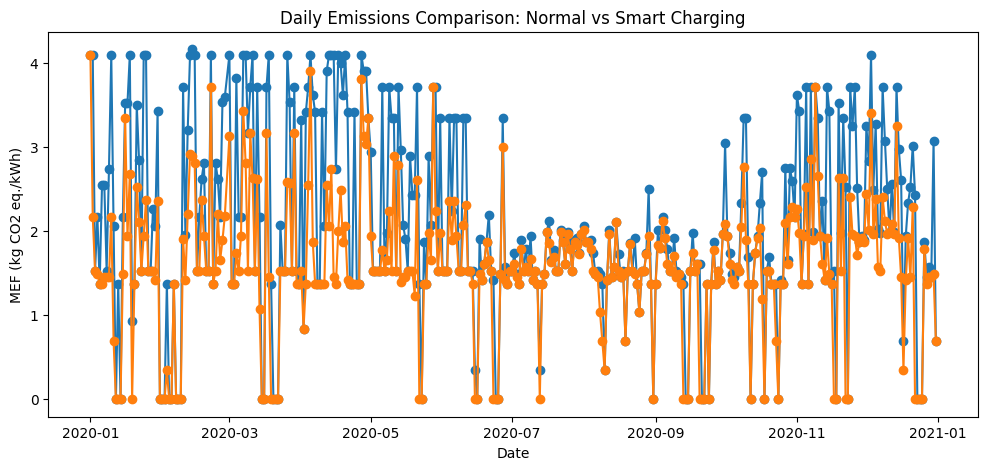}
    \caption{}
  \end{subfigure}
  \hfill  
  \begin{subfigure}{0.45\textwidth}
    \centering
    \includegraphics[width=\linewidth]{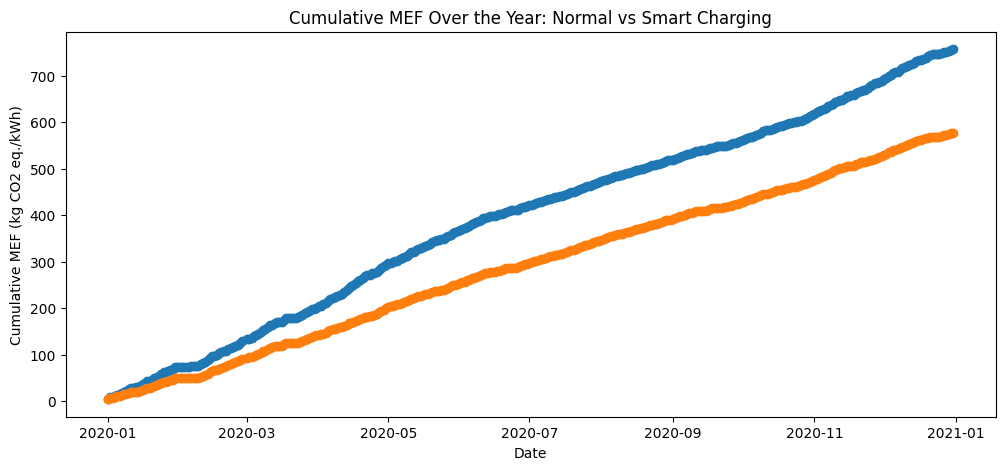}
    \caption{}
  \end{subfigure}

  \caption{ (a) Daily MEF and (b) cumulative MEF over the year for \textcolor{blue}{the normal charging scenario} and \textcolor{orange}{the emission-minimized charging strategy} in 2020}
\label{plotssmart2020}
\end{figure}

\begin{figure}[H]
  \centering

  \begin{subfigure}{0.45\textwidth}
    \centering
    \includegraphics[width=\linewidth]{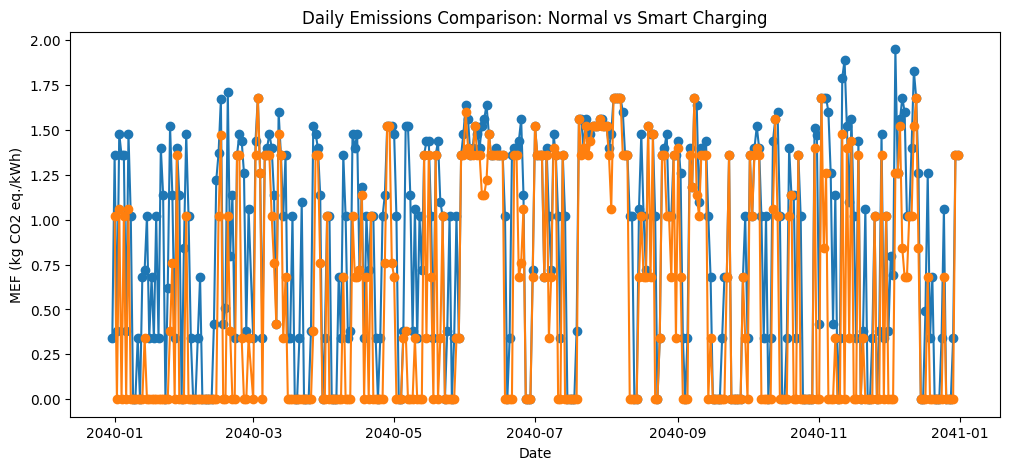}
    \caption{}
  \end{subfigure}
  \hfill  
  \begin{subfigure}{0.45\textwidth}
    \centering
    \includegraphics[width=\linewidth]{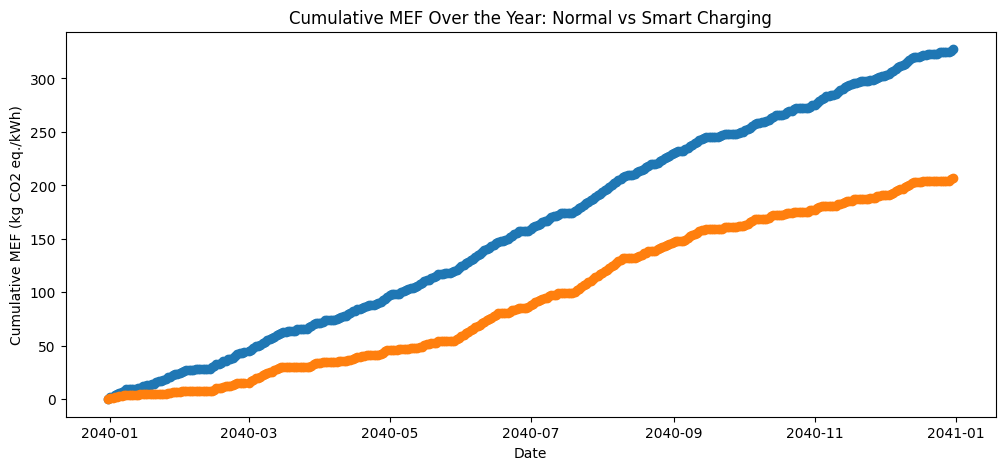}
    \caption{}
  \end{subfigure}

  \caption{ (a) Daily MEF and (b) cumulative MEF over the year for \textcolor{blue}{the normal charging strategy} and \textcolor{orange}{the emission-minimized charging strategy} in 2040}
\label{plotssmart2040}
\end{figure}

Table \ref{smartchargingtable} shows the cumulative MEF results for a normal charging strategy and for a vehicle charging strategy with a minimum emission. The results cover historic years (2019 and 2020) and future years (2025, 2030, and 2040). As a result of emission-minimized charging, we achieved an average MEF savings of 31\% over 5 years, with a decrease of 35\% in 2019 and a reduction of 36\% in 2040, compared to the normal charging strategy. In addition, for both scenarios, we noticed a decrease in MEF over the years, reflecting the increasing impact of climate mitigation efforts to achieve zero emissions by 2050.\\
To better understand the savings in MEF, Figures \ref{plotssmart2020} and \ref{plotssmart2040} illustrate the daily MEF and cumulative MEF in 2020 and 2040, to compare their levels in both charging scenarios. According to the graphs, the MEF levels are consistently lower for the emission-minimized charging strategy than for the normal charging scenario for both years. Therefore, emission-minimized charging reduces emissions by adjusting energy consumption during planned charging hours to the lowest MEF hours.
Figures for 2019, 2025, and 2030 are illustrated in Appendix Section \ref{smartchargappendixsection}.\\
Therefore, emission-minimized charging based on MEF data has several environmental benefits: First, it reduces the carbon footprint of electric vehicles by optimizing electricity use during times of lower grid emissions. Second, given that the grid shifts towards renewable energy sources, mixing cleaner energy with emission-minimized charging leads to more efficient energy consumption and hence a significant decrease in total greenhouse gas emissions. Finally, besides supporting the transition to zero emissions, this strategy reduces the environmental impact of increased electrification in the transport sector.

\section{Conclusions and Policy Recommendations}
\label{conclusionandpolicy}

Historical data for hourly marginal emission factors is not observable.
Determining marginal emission factors is essential to accurately assess regional impacts caused by changes in demand, particularly in the context of decolonization. Unlike average emission factors, marginal emission factors provide a more precise measure of the immediate impact of changes in electricity demand on CO$_2$ emissions. \\
In this paper, we determine the time series of marginal emission factors using an energy system model and statistical models. 
First, we use a complex energy system model to determine the marginal emission factor in high temporal resolution (hourly). The energy system model emulates the electricity market fundamentals, which results in computational complexity. We compute the marginal emission factor for selected historical and future years (2019, 2020, 2025, 2030, and 2040).
We use the marginal emission factors which result from the energy system model as a benchmark for testing the accuracy of the statistical methodologies. 
Second, we propose the Markov Switching Dynamic Regression and the Dynamic Linear Regression models as an alternative to the incremental MEF approach. 
The statistical models use information on CO$_2$ emissions and electricity generation from Agora Energiewende for historical years (2019 and 2020), and from the energy system model for future years (2025, 2030, and 2040) to estimate the MEF time series, using a rolling window with a length of 168 hours (one week). 
The MEFs from the MSDR model show the lowest error within the evaluation metrics.
The MSDR model could be adopted to estimate MEFs, especially when short-term MEF estimation is needed and only empirical data on CO$_2$ emissions and electricity generation is available.
We can confidently rely on the MSDR model and make informed decisions for sustainable energy practices based on its estimations.

Given its proven robustness, modelers and energy practitioners can use the energy system model to generate MEFs under a wide range of scenarios, including future projections, renewable energy integration, or shifts in electricity demand.\\
In particular, we present an example of demand-side flexibility to showcase the improvements achieved by using MEFs when the goal is to minimize GHG emissions. Determining marginal emission factors is crucial for optimizing strategies such as electric vehicle charging. To demonstrate this, we develop an application for EV charging under two scenarios: normal charging and emission-minimized charging.
The model makes use of the MEF time series and identifies periods with lower MEFs that can be targeted for electric vehicle charging. 
This optimization leads to substantial reductions in carbon emissions, as evidenced by the significant emission savings observed in the emission-minimized charging scenario compared to normal charging.
We observe that using information from the time series of MEFs leads to significant emission reductions, up to 307.93 kg CO$_2$ equivalent in 2019. Moreover, this application supports sustainable practices, and provides valuable insights for policy and technological advancements. As the energy system evolves, the role of MEFs in smart charging applications will become increasingly important, driving further innovations and improvements in grid management and environmental impact.
Determining time series of MEFs provides essential data for policymakers to design effective regulations and incentives that promote the use of smart charging technologies. This could include, but is not limited to, time-of-use tariffs that could encourage electric vehicle owners to charge during periods of low emissions.Future research could focus on determining marginal emission factors in the context of sector coupling, by linking the electricity sector with the heating and transport sectors. The ongoing electrification of various sectors, such as transportation and heating, will alter electricity demand patterns, which is crucial for MEFs. \\

For policymakers and industry practitioners, the statistical model offers a fast-computing, powerful resource for making informed decisions. By providing time series of MEFs, the model can help in designing and implementing effective demand response schemes, smart metering applications, and other environmental regulations aimed at reducing carbon emissions.
Practitioners and policymakers can use the statistical model to explore different future energy scenarios and their impact on MEFs. This includes analyzing the potential effects of increased renewable energy penetration, electrification of transport, or shifts in energy demand patterns. Such analyses will be crucial for planning a low-carbon energy transition.
The statistical model could also be coupled with real-time energy management systems, helping utilities and large energy consumers to make operational decisions based on the current and projected carbon intensity of the grid. By simulating various policy interventions, the model can help assess the potential impact of new climate policies on marginal emission factors. This allows for the refinement of policy measures to ensure that they generate the desired outcome.
Time series of hourly marginal emission factors are particularly valuable in several industrial applications. 
These include real-time energy management systems, where industries can make operational decisions based on the current carbon intensity of the grid, profits if CO$_2$ credits extend to other agents in the system, and smart charging for electric vehicles, where the charging sessions occur during lower emissions periods,
Additionally, these time series are essential for the accurate calculation of carbon footprints in processes that vary over time, such as demand-response programs. 
Precise determination of marginal emission factors is a powerful tool that would support both environmental goals and the operational efficiency of industries, making it a key component in the transition to a low-carbon economy.
Future research could also investigate how developments in decentralized grids influence MEFs. This implies studying the effects of micro-grids and peer-to-peer energy trading.
Additionally, investigating how MEFs are correlated with other elements in the energy market, such as weather data or energy prices, could be a valuable area of study.



\section*{Acknowledgements}
Souhir Ben Amor, Smaranda Sgarciu, and Taimyra Batz Lineiro gratefully acknowledge the support of the German Federal Ministry for Economic Affairs and Climate Action (BMWK), which funded this research through the CO2Calc4Quartier Project (Development of planning methods and tools for the  CO$_2$  assessment of neighborhood heating concepts; Award No. 03EN3040B). 
Felix Muesgens gratefully acknowledges financial support from the Federal Ministry of Education and Research, Award No. 19FS2032C, as well as the German Federal Government, the Federal Ministry of Education and Research, and the State of Brandenburg within the framework of the joint project EIZ: Energy Innovation Center (project numbers 85056897 and 03SF0693A) with funds from the Structural Development Act (Strukturstärkungsgesetz) for coal-mining regions.

\section*{Declaration of Competing Interests}
The authors declare that they have no known competing financial 
interests or personal relationships that could have appeared to influence 
the work reported in this paper.


\bibliography{bibliography.bib}

\appendix
\section{Appendix}
\label{appendixsection}

\subsection{Energy System Model Input}
\label{nomenclatureapp}
\begin{table}[H]
\centering
\scriptsize 
\caption{Nomenclature}
\begin{tabular}{|c|c|c|}
\hline
\textbf{Abbreviation} & \textbf{Unit} & \textbf{Description} \\ \hline
\textbf{Model sets} & & \\
$n, nn$ & & Node (country or region) \\
$i$ & & Technologies \\
$p$ & & Conventional technologies \\
$r$ & & Renewable technologies \\
$r^{\text{curt}}$ & & Curtailable renewables (onshore/offshore wind and solar) \\
$stm$ & & Mid-term storage \\
$stl$ & & Long-term storage \\
$t$ & & Time step \\
$y, yy$ & & Year \\ \hline
\textbf{Model parameters} & & \\
$\text{ccurt}_{n, r^{\text{curt}}, y}^{\text{Gen}}$ & \euro/MWh & Variable costs for renewables curtailment \\
$\text{ccurt}_y^{\text{Load}}$ & \euro/MWh & Variable costs for load curtailment \\
$\text{cramp}_{n, p, y}$ & \euro/MW & Ramping costs \\
$\text{cvar}_{n, r, y}$ & \euro/MWh & Average variable generation costs (incl. CO$_2$ costs) \\
$\text{cvar}_{n, p, y}^{\text{Full}}$ & \euro/MWh & Variable generation costs (incl. CO$_2$ costs) at full load \\
$\text{cvar}_{n, p, y}^{\text{Min}}$ & \euro/MWh & Variable generation costs (incl. CO$_2$ costs) at minimum load \\
$\text{cramp}_{n, p, y}$ & \euro/MW & Ramping costs \\
gridloss & \% & Grid loss \\
$wv$ & \euro/MWh & Water value for hydropower reservoirs and long-term storage\\
$ef_i$ & CO$_2$ t/MWh & Technology’s emissions factor \\
$\Delta f_y$ & & Discount factor \\ \hline
\textbf{Model variables} & & \\
$\text{CAP}_{n, p, y}^{\text{Add}}$ & MW & Capacity investment \\
$\text{CAP}_{n, p, y}^{\text{Install}}$ & MW & Installed capacity \\
$\text{CAP}_{n, p, y, t}^{\text{Started}}$ & MW & Operational generation capacity \\
$\text{CAP}_{n, p, y, t}^{\text{Up}}$ & MW & Increase of operational generation capacity \\
$\text{CAP}_{n, p, y, t}^{\text{Down}}$ & MW & Decrease of operational generation capacity \\
$\text{CHARGE}_{n, s, y, t}$ & MWh & Charge of storage \\
$\text{DISCHARGE}_{n, s, y, t}$ & MWh & Discharge of storage \\
COST & \euro & Total system costs (objective value) \\
$\text{COST}_{n, y}^{\text{Gen}}$ & \euro & Generation and curtailment costs \\
$\text{COST}_{n, y}^{\text{Fix}}$ & \euro & Fixed operation and maintenance costs \\
$\text{COST}_{n, y}^{\text{Inv}}$ & \euro & Investment in generation capacity \\
$\text{CURT}_{n, r, y, t}^{\text{Gen}}$ & MWh & Curtailment of renewable technologies \\
$\text{CURT}_{n, y, t}^{\text{Load}}$ & MWh & Curtailment of load \\
$\text{FLOW}_{n, nn, y, t}$ & MWh & Export flow \\
$\text{FLOW}_{nn, n, y, t}$ & MWh & Import flow \\
$\text{GEN}_{n, p, y, t}$ & MWh & Electricity generation from conventional power plants \\
$\text{GEN}_{n, r, y, t}$ & MWh & Electricity generation from renewable technologies \\
$\text{GEN}_{n, p, y, t}^{\text{Full}}$ & MWh & Electricity generation from power plants operating at full load \\
$\text{GEN}_{n, p, y, t}^{\text{Min}}$ & MWh & Electricity generation from power plants operating at minimum load \\
$\text{CL}_{i, n, t}$ & MWh/h & Charging activity for long-term storage \\
$\text{CM}_{i, n, t}$ & MWh/h & Charging activity for mid-term storage \\
$\text{SHED}_{n, t}$ & MWh/h & Load shedding \\
MEF$_t$ & CO$_2$ t/MWh & Marginal emissions factor \\
$\Delta CI_{T}$ & & Change in carbon emissions \\
$\Delta LOAD_{T}$ & & Change in electricity demand \\
$D_{n, t}$ & MWh & Technology’s demand in an hour \\ \hline
\end{tabular}
\label{nomenclature}
\label{Nomenclature}
\end{table}

\subsection{Energy System Model Output}

\subsubsection{Incremental MEF}
\label{otherfund_mefs}

\begin{figure}[H]
  \centering

  \begin{subfigure}{0.45\textwidth}
    \centering
    \includegraphics[width=\linewidth]{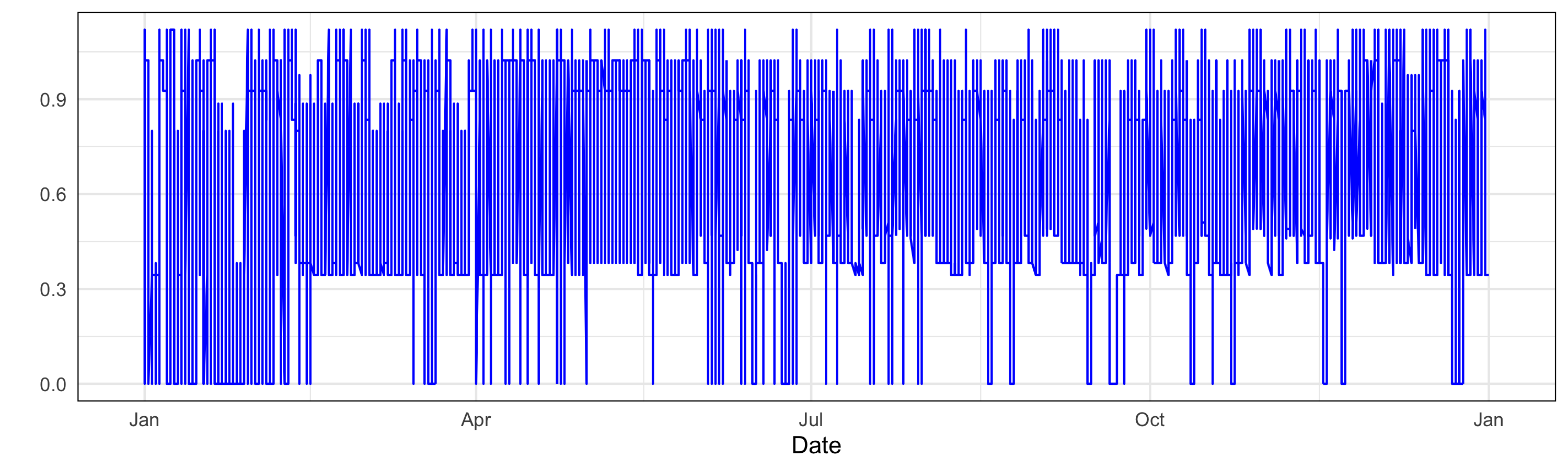}
    \caption{}
  \end{subfigure}
  \hfill
  \begin{subfigure}{0.45\textwidth}
    \centering
    \includegraphics[width=\linewidth]{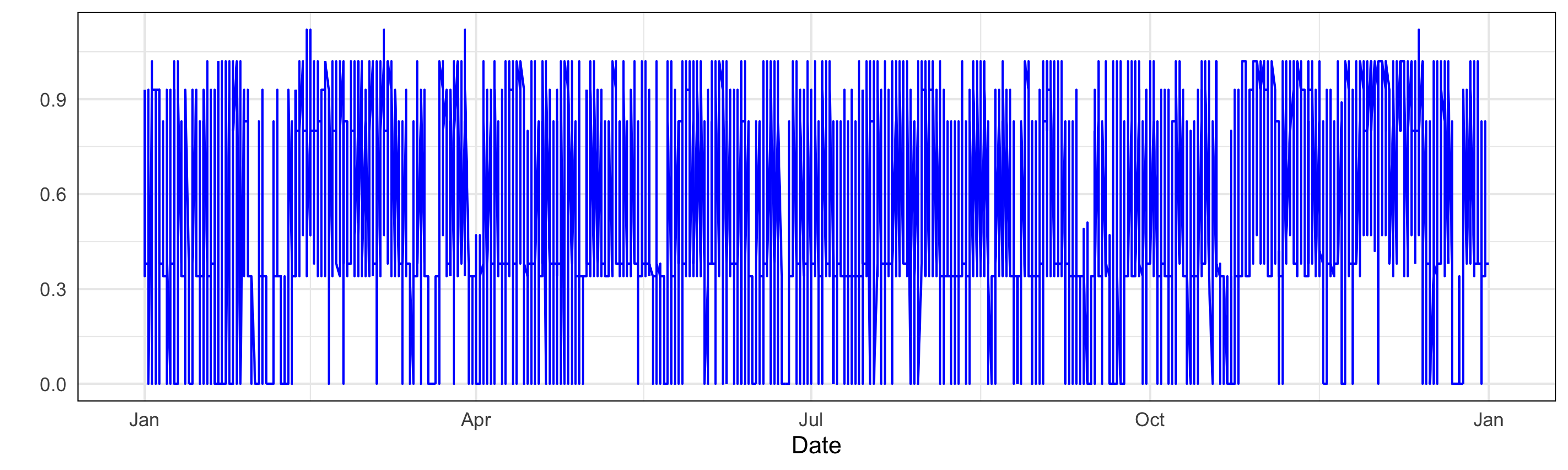}
    \caption{}
  \end{subfigure}

  \vspace{10pt} 

  \begin{subfigure}{0.45\textwidth}
    \centering
    \includegraphics[width=\linewidth]{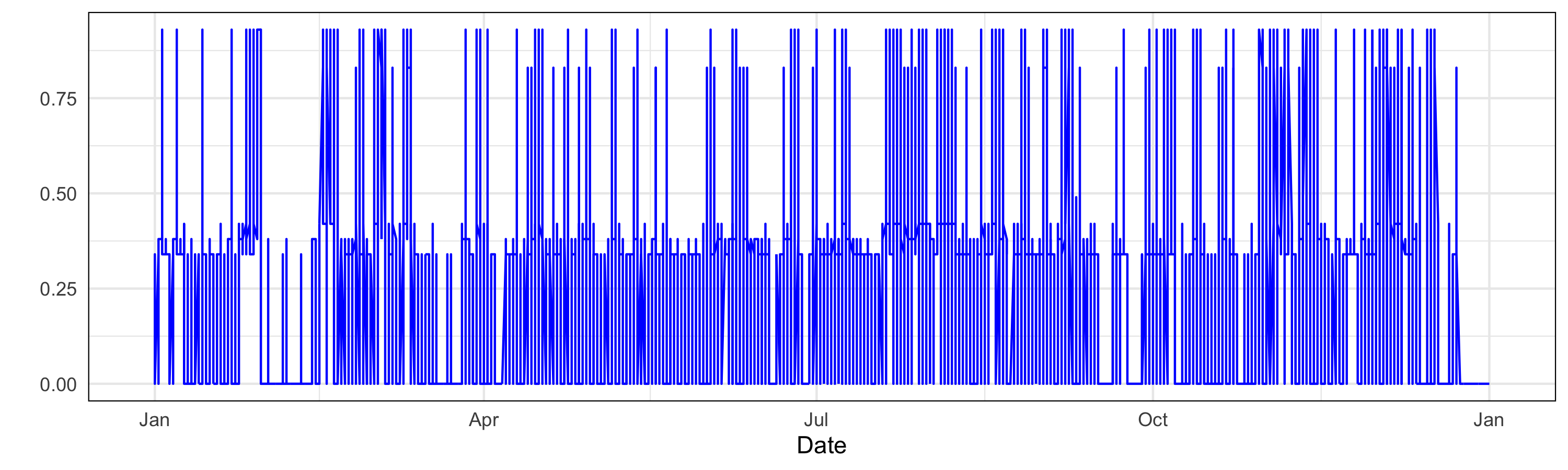}
    \caption{}
  \end{subfigure}

  \caption{Incremental MEF for (a) 2019, (b) 2025, and (c) 2030, in \textit{kg CO$_2$ eq./kWh}}
\label{otherMEFs}
\end{figure}

\subsubsection{CO$_2$ Emissions and Electricity Generation Data}
\label{emissionsandgenerations_plots}

\begin{figure}[H]
  \begin{subfigure}{0.5\textwidth}
    \centering
    \includegraphics[width=1\linewidth]{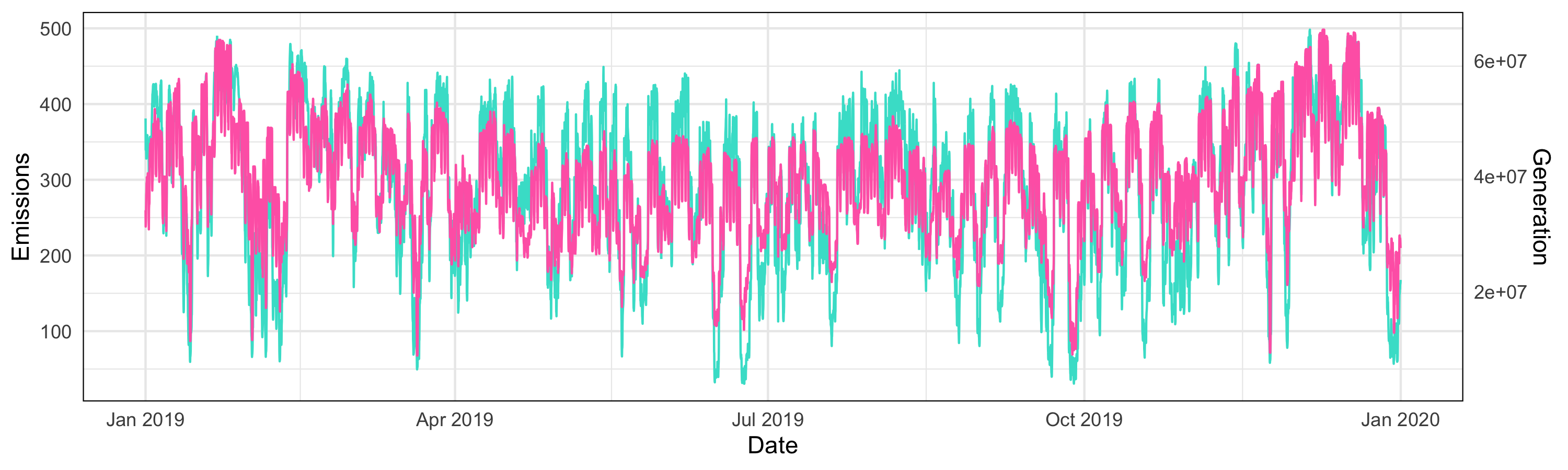}
    \caption{}
  \end{subfigure}%
  \begin{subfigure}{0.5\textwidth}
    \centering
    \includegraphics[width=1\linewidth]{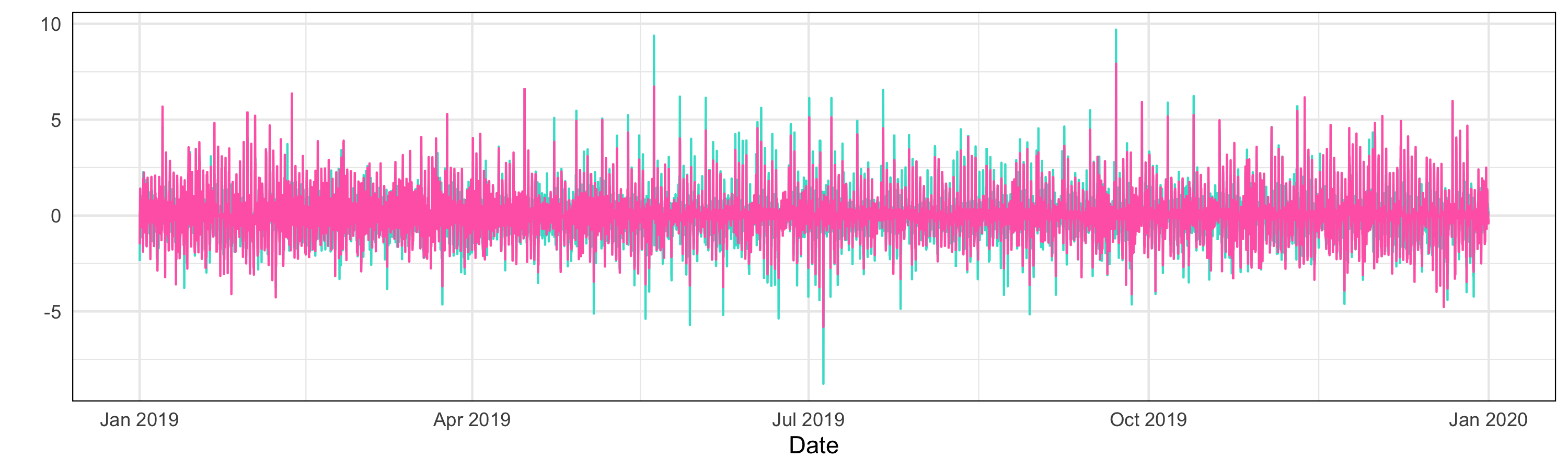}
    \caption{}
  \end{subfigure}
  \caption{\textcolor{bluegreen}{CO$_2$ emissions} and \textcolor{hotpink}{electricity generation} data and (b) their variations $\Delta$ in 2019}
  \label{input}
\end{figure}

\begin{figure}[H]
  \begin{subfigure}{0.5\textwidth}
    \centering
    \includegraphics[width=1\linewidth]{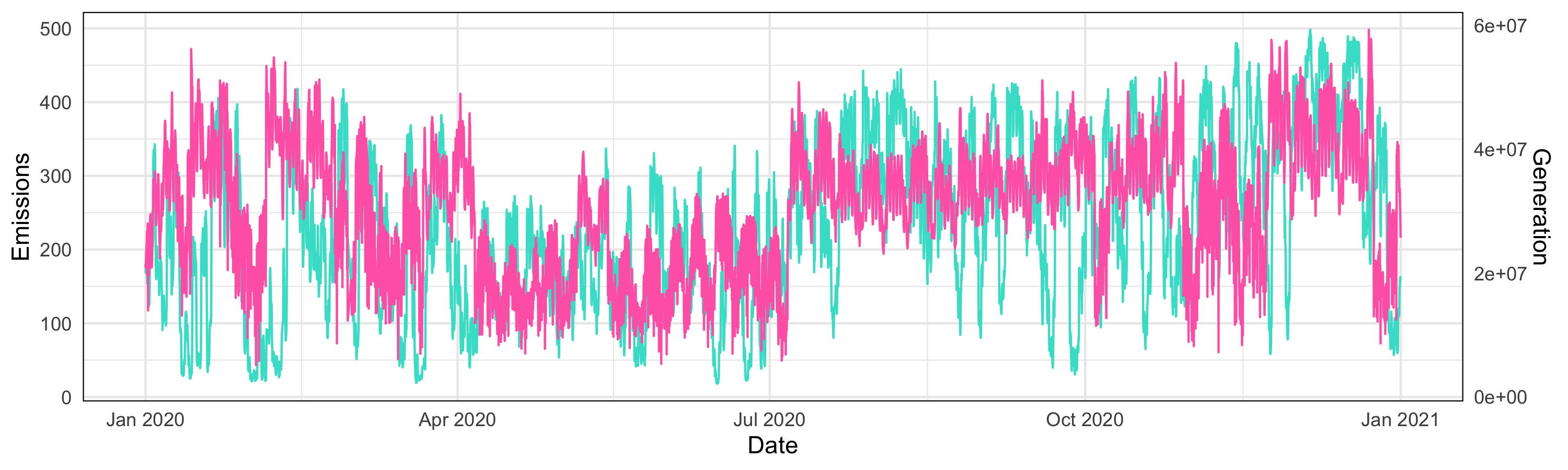}
    \caption{}
  \end{subfigure}%
  \begin{subfigure}{0.5\textwidth}
    \centering
    \includegraphics[width=1\linewidth]{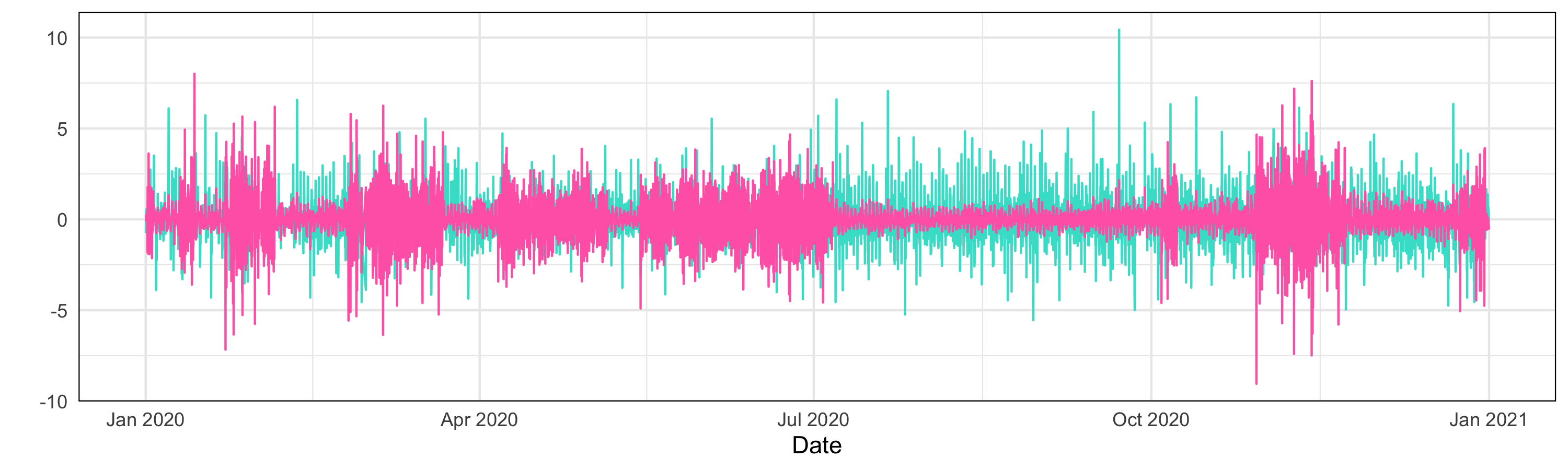}
    \caption{}
  \end{subfigure}
  \caption{\textcolor{bluegreen}{CO$_2$ emissions} and \textcolor{hotpink}{electricity generation} data and (b) their variations $\Delta$ in 2020}
  \label{input20}
\end{figure}

\begin{figure}[H]
  \begin{subfigure}{0.5\textwidth}
    \centering
    \includegraphics[width=1\linewidth]{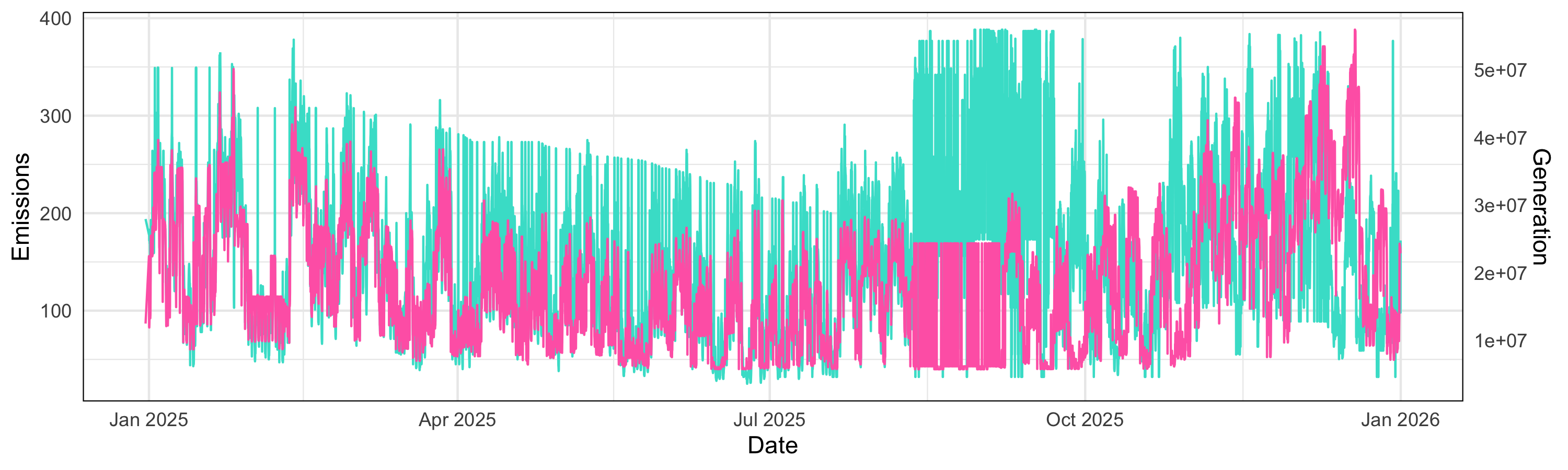}
    \caption{}
  \end{subfigure}%
  \begin{subfigure}{0.5\textwidth}
    \centering
    \includegraphics[width=1\linewidth]{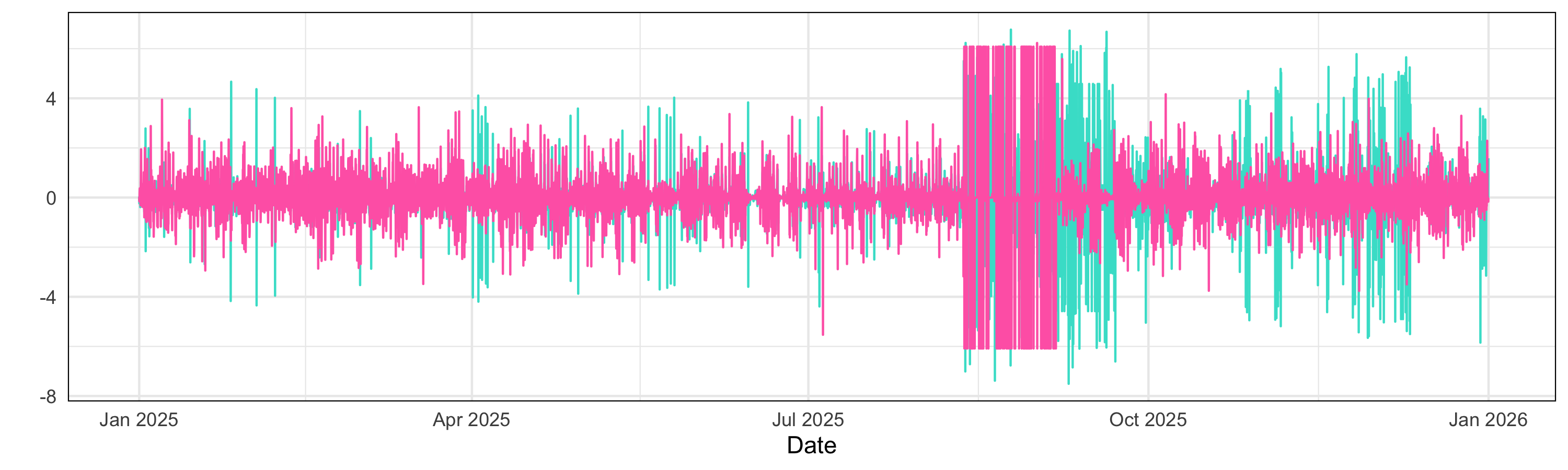}
    \caption{}
  \end{subfigure}
  \caption{\textcolor{bluegreen}{CO$_2$ emissions} and \textcolor{hotpink}{electricity generation} data and (b) their variations $\Delta$ in 2025}
  \label{input25}
\end{figure}
\begin{figure}[H]
  \begin{subfigure}{0.5\textwidth}
    \centering
    \includegraphics[width=1\linewidth]{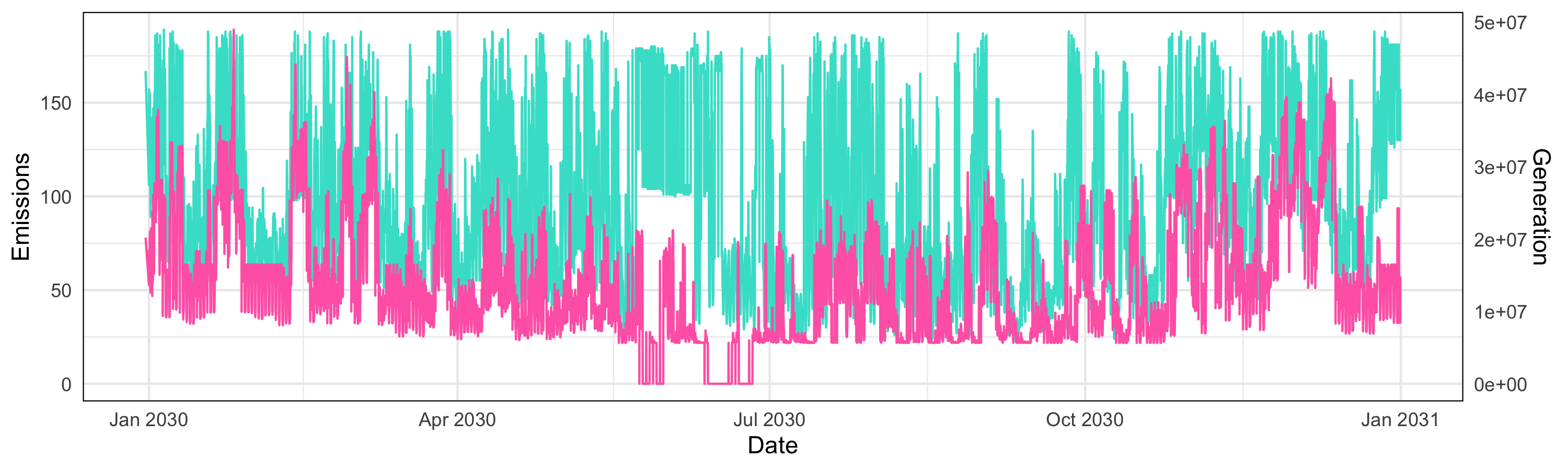}
    \caption{}
  \end{subfigure}%
  \begin{subfigure}{0.5\textwidth}
    \centering
    \includegraphics[width=1\linewidth]{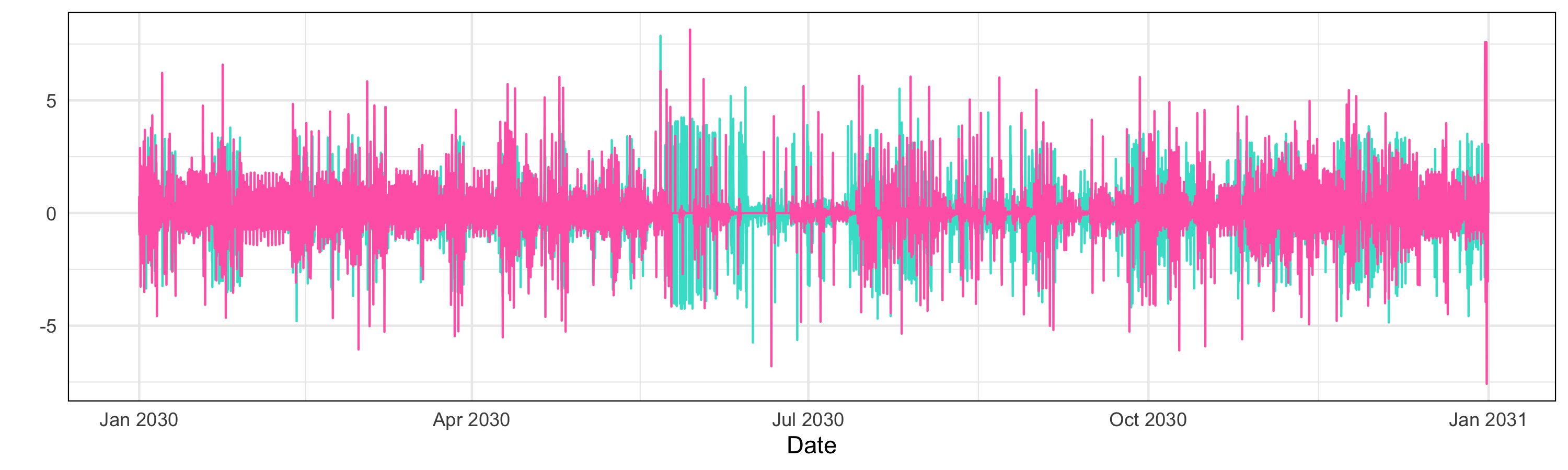}
    \caption{}
  \end{subfigure}
  \caption{\textcolor{bluegreen}{CO$_2$ emissions} and \textcolor{hotpink}{electricity generation} data and (b) their variations $\Delta$ in 2030}
  \label{input30}
\end{figure}

\subsubsection{Statistical Tests of ESM Results on Generation and CO$_2$ Emissions}
\label{emissionsandgenerations_tests}

\begin{table}[h]
\centering
\caption{Descriptive statistics and test results in 2040}
\begin{tabularx}{\textwidth}{l*{4}{r}}  
\toprule
 & \textbf{Emissions} & \textbf{Generation} & \textbf{$\Delta$ Emissions} & \textbf{$\Delta$ Generation} \\
\midrule
count & 8760 & 8760 & 8759 & 8759 \\
mean & 60.419 & 11710035 & 0.000 & 0.000 \\
std  & 50.668 & 8502200 & 1.000 & 1.000 \\
min  & 0.000 & 0.000 & -10.136 & -6.753 \\
25\% & 21.000 & 5643428 & -0.154 & -0.127 \\
50\% & 42.000 & 8472689 & 0.000 & 0.000 \\
75\% & 90.000 & 13657456 & 0.154 & 0.018 \\
max  & 200.000 & 53003389 & 8.387 & 8.642 \\
ADF statistic & -8.845 & -7.188 & -19.236 & -17.949 \\
ADF p-value & 0.000 & 0.000 & 0.000 & 0.000 \\
BDS statistic & 186.838 & 158.715 & 29.740 & 23.788 \\
BDS p-value & 0.000 & 0.000 & 0.000 & 0.000 \\
Jarque-Bera statistic & 1659.514 & 5467.878 & 64526.712 & 85422.512 \\
Jarque-Bera p-value & 0.000 & 0.000 & 0.000 & 0.000 \\
\bottomrule
\end{tabularx}
\label{statistics2040}
\end{table}

According to preliminary statistics, the Augmented Dickey-Fuller (ADF) test results indicate the stationarity of the CO$_2$ emissions and electricity generation and the variation of their times series, which ensures the reliability of various statistical models, including the linear regression model. The BDS test statistics suggest significant nonlinearity in the series, with p-values of zero across all datasets, highlighting the presence of complex dependencies and structure beyond linear models. The Jarque-Bera test results, with p-values of zero, indicate that the distributions of all series significantly deviate from normality, likely due to skewness and excess kurtosis. Note that the other years studied (2019, 2020, 2025, and 2030) exhibit similar statistical characteristics.


\begin{figure}[H]
  \begin{subfigure}{0.5\textwidth}
    \centering
    \includegraphics[width=1\linewidth]{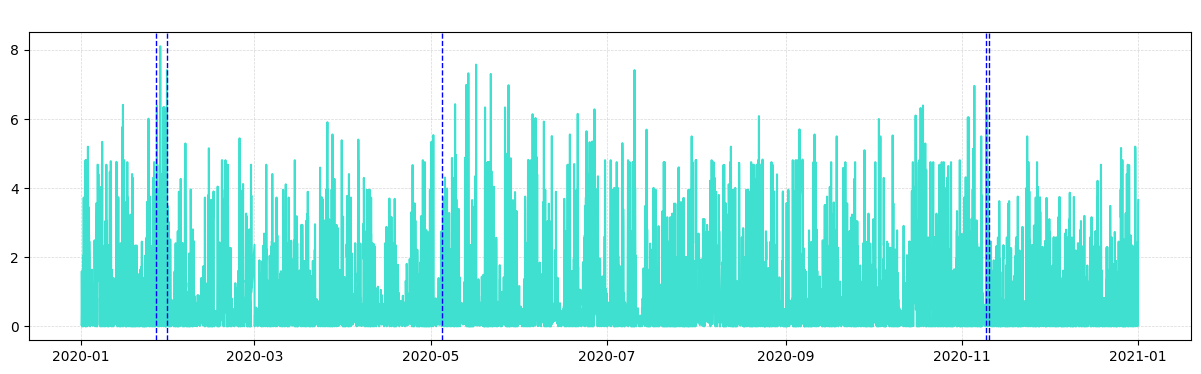}
    \caption{}
  \end{subfigure}%
  \begin{subfigure}{0.5\textwidth}
    \centering
    \includegraphics[width=1\linewidth]{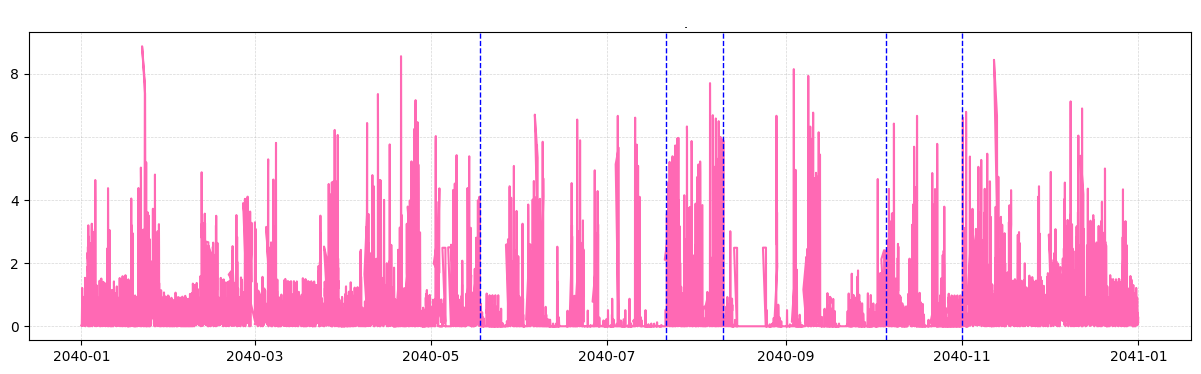}
    \caption{}
  \end{subfigure}
  \caption{\textcolor{bluegreen}{Changes in CO$_2$ emissions volatility} and (b) \textcolor{hotpink}{changes in electricity generation volatility} and their associated \textcolor{blue}{breakpoints} in 2040}
  \label{breakpoints2040}
\end{figure}

\begin{figure}[H]
  \begin{subfigure}{0.5\textwidth}
    \centering
    \includegraphics[width=1\linewidth]{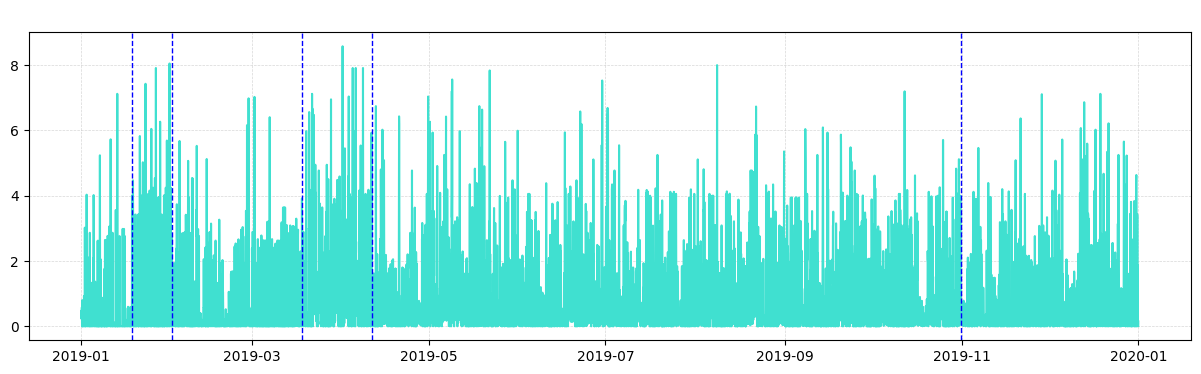}
    \caption{}
  \end{subfigure}%
  \begin{subfigure}{0.5\textwidth}
    \centering
    \includegraphics[width=1\linewidth]{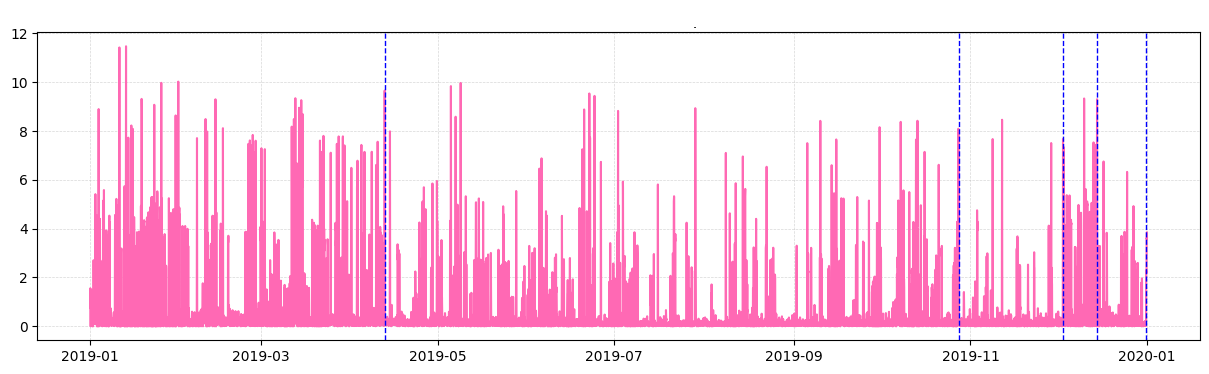}
    \caption{}
  \end{subfigure}
  \caption{\textcolor{bluegreen}{Changes in CO$_2$ emissions volatility} and (b) \textcolor{hotpink}{changes in electricity generation volatility} and their associated \textcolor{blue}{breakpoints} in 2019}
  \label{breakpoints2019}
\end{figure}

\begin{figure}[H]
  \begin{subfigure}{0.5\textwidth}
    \centering
    \includegraphics[width=1\linewidth]{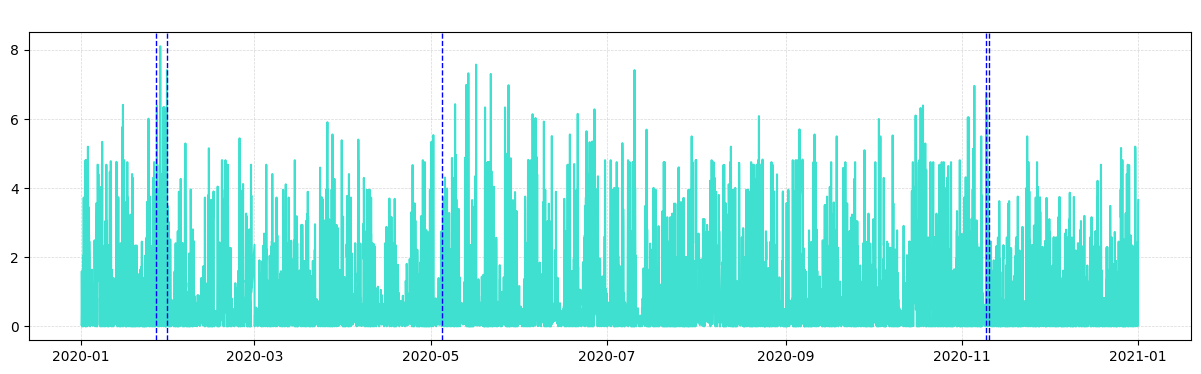}
    \caption{}
  \end{subfigure}%
  \begin{subfigure}{0.5\textwidth}
    \centering
    \includegraphics[width=1\linewidth]{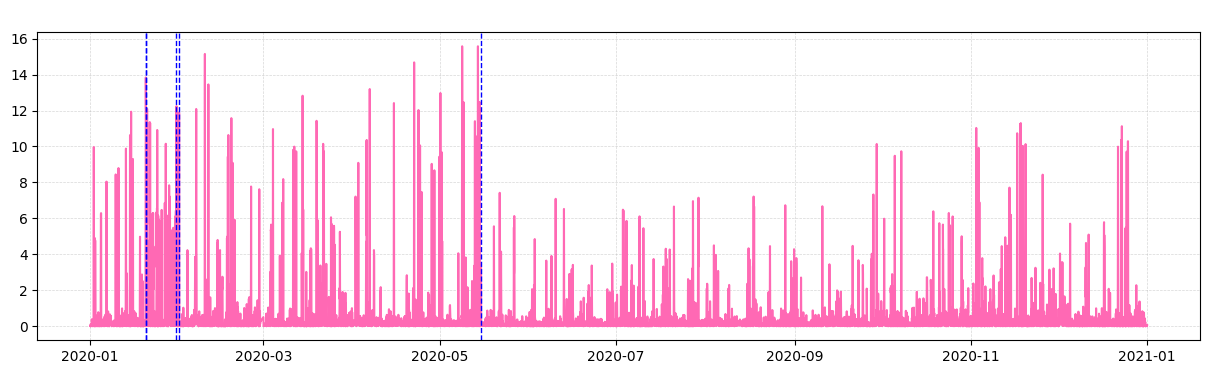}
    \caption{}
  \end{subfigure}
  \caption{\textcolor{bluegreen}{Changes in CO$_2$ emissions volatility} and (b) \textcolor{hotpink}{changes in electricity generation volatility} and their associated \textcolor{blue}{breakpoints} in 2020}
  \label{breakpoints2020}
\end{figure}

\begin{figure}[H]
  \begin{subfigure}{0.5\textwidth}
    \centering
    \includegraphics[width=1\linewidth]{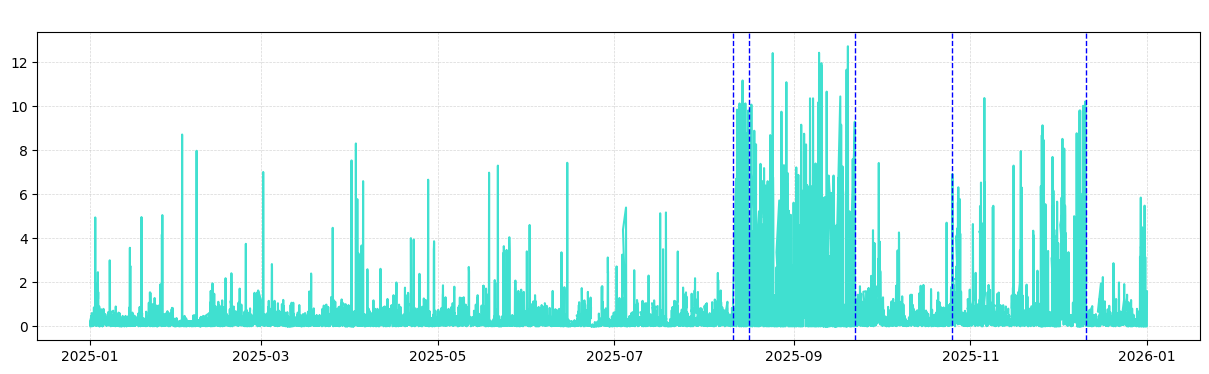}
    \caption{}
  \end{subfigure}%
  \begin{subfigure}{0.5\textwidth}
    \centering
    \includegraphics[width=1\linewidth]{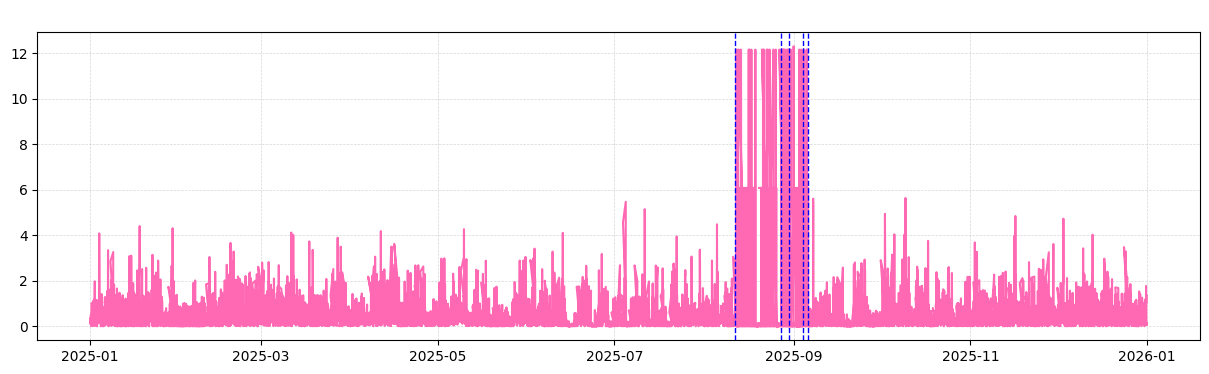}
    \caption{}
  \end{subfigure}
  \caption{\textcolor{bluegreen}{Changes in CO$_2$ emissions volatility} and (b) \textcolor{hotpink}{changes in electricity generation volatility} and their associated \textcolor{blue}{breakpoints} in 2025}
  \label{breakpoints2025}
\end{figure}

\begin{figure}[H]
  \begin{subfigure}{0.5\textwidth}
    \centering
    \includegraphics[width=1\linewidth]{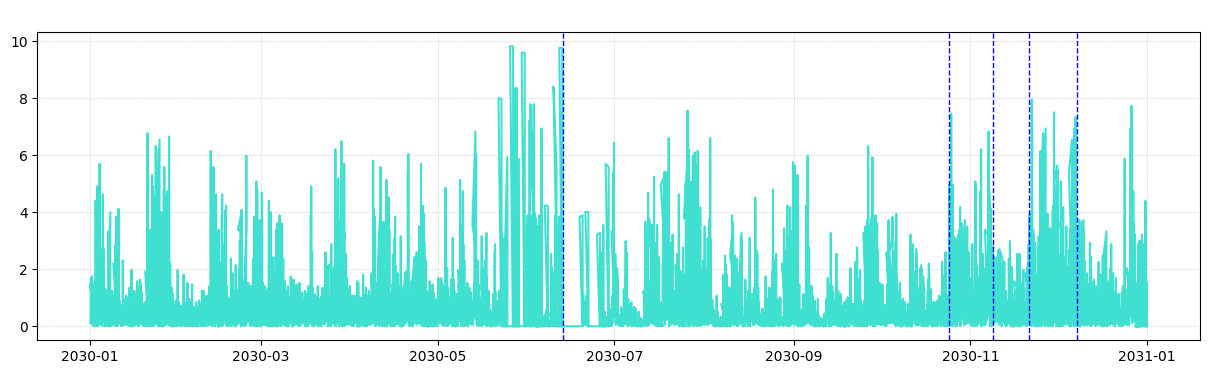}
    \caption{}
  \end{subfigure}%
  \begin{subfigure}{0.5\textwidth}
    \centering
    \includegraphics[width=1\linewidth]{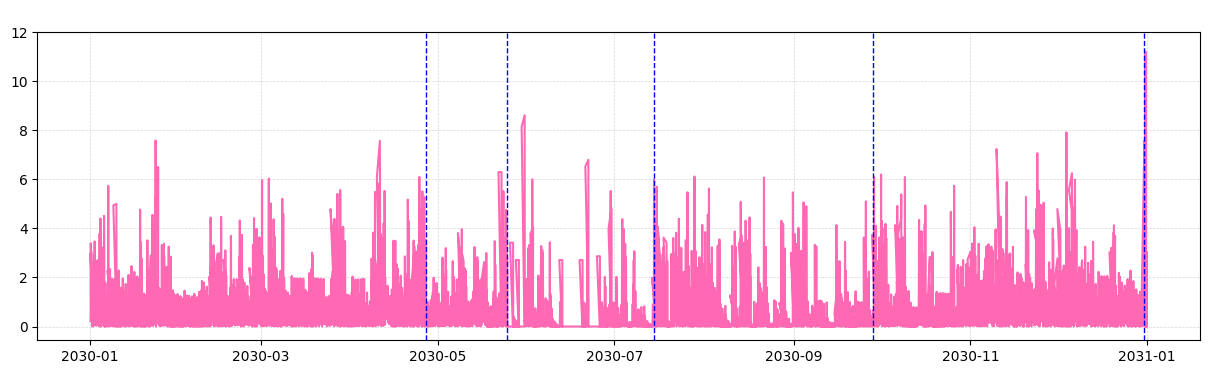}
    \caption{}
  \end{subfigure}
  \caption{\textcolor{bluegreen}{Changes in CO$_2$ emissions volatility} and (b) \textcolor{hotpink}{changes in electricity generation volatility} and their associated \textcolor{blue}{breakpoints} in 2030}
  \label{breakpoints2030}
\end{figure}

The \customcite{bai2003computation} test was applied independently to the CO$_2$ emissions and electricity generation data over the years studied. We used absolute returns as a proxy for volatility due to their stronger autocorrelation and reduced noise compared to squared returns (\customcite{andersen1997heterogeneous}, \customcite{taylor2008modelling}). The test results for 2020 (Figure \ref{breakpoints2020}) and 2040 (Figure \ref{breakpoints2040}) identified distinct structural breaks, indicating significant changes in volatility patterns in both series. For CO$_2$ emissions, these breaks suggest transitions between stable periods and phases of increased volatility, likely driven by factors such as regulatory changes or changes in electricity generation processes (mean order). Similarly, in the electricity generation series, the breakpoints correspond to periods of varying volatility, reflecting changes in energy demand, supply dynamics, or external shocks. The alignment of breakpoints across both datasets suggests a strong interdependence, where shifts in electricity generation directly impact emission volatility. These results were also observed for the years 2019, 2020, 2025, and 2030 (see Figure \ref{breakpoints2019}, \ref{breakpoints2020}, \ref{breakpoints2025}, and \ref{breakpoints2030}). These findings highlight the dynamic relationship between CO$_2$ emissions and generation, providing a basis for applying a Markov Switching Dynamic Regression model.

\subsubsection{Evaluation Metrics}
\label{evaluationmetricdef}

The most widely used metrics to evaluate the accuracy of point forecasts are mean absolute error (MAE), mean square error (MSE), and root mean square error (RMSE), given by the following equations:
\begin{equation}
\mathrm{MAE} = \frac{1}{N_{\mathrm{h}}} \sum_{h=1}^{N_{\mathrm{h}}} \left| p_h - \hat{p}_h \right|,
\end{equation}

\begin{equation}
\mathrm{MSE} = \frac{1}{N_{\mathrm{h}}} \sum_{h=1}^{N_{\mathrm{h}}} \left( p_h - \hat{p}_h \right)^2,
\end{equation}

\begin{equation}
\mathrm{RMSE} = \sqrt{\frac{1}{N_{\mathrm{h}}} \sum_{h=1}^{N_{\mathrm{h}}} \left( p_h - \hat{p}_h \right)^2},
\end{equation}

where $p_{h}$ and $\hat{p}_{h}$ denote the real and estimated time series for hour $h$, respectively. The total number of hours in our sample is denoted by $N_h$.\\

\subsection{Estimation Results}


\subsubsection{Dynamic Linear Regression Estimation Results}
\label{linearresults}

\begin{figure}[H]
  \centering

  \begin{subfigure}{0.45\textwidth}
    \centering
    \includegraphics[width=\linewidth]{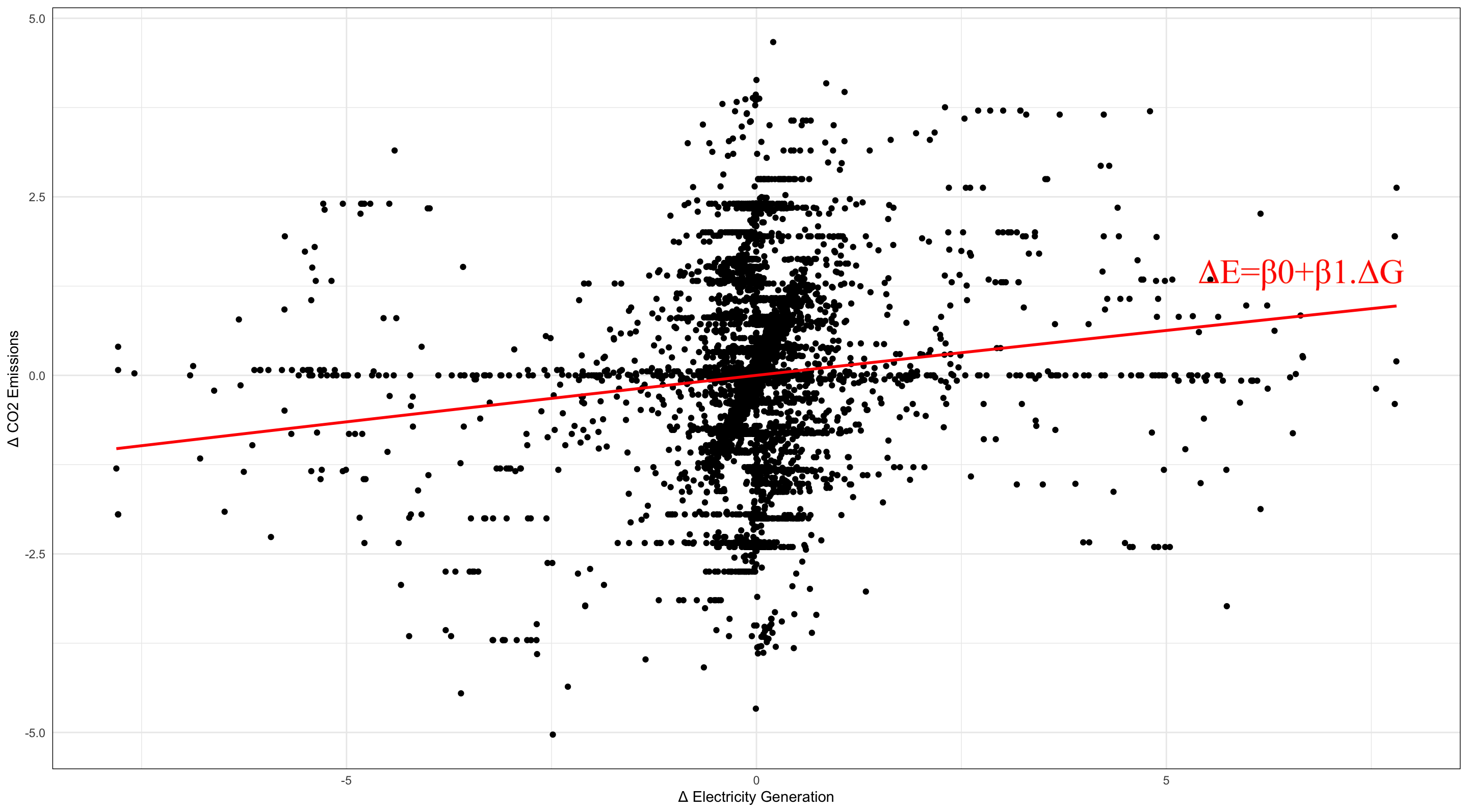}
    \caption{}
  \end{subfigure}
  \hfill
  \begin{subfigure}{0.45\textwidth}
    \centering
    \includegraphics[width=\linewidth]{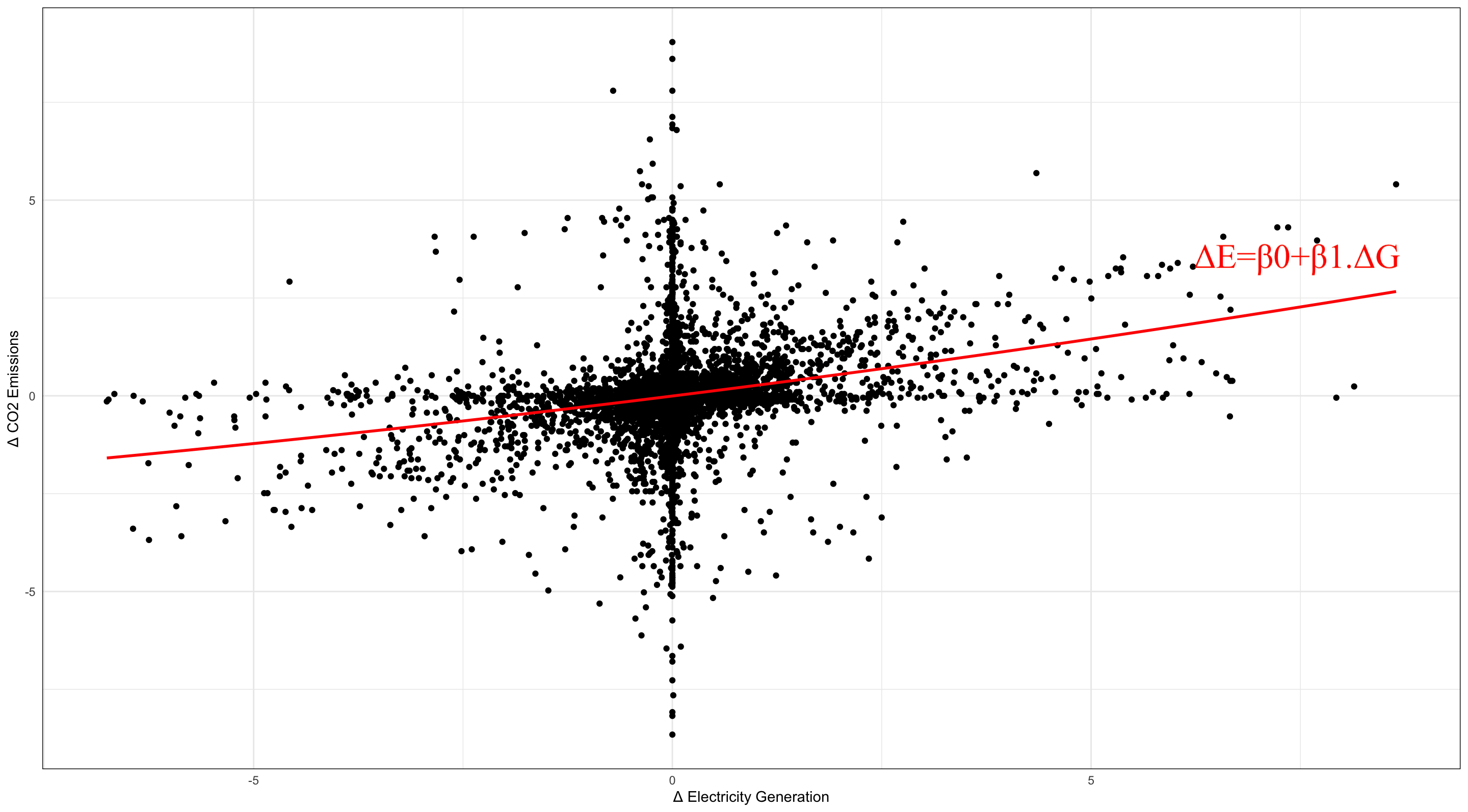}
    \caption{}
  \end{subfigure}
  \caption{Linear regression estimation results for (a) 2020 and (b) 2040}
\label{lineareg}
\end{figure}

In Figure \ref{lineareg} we observe the linear regression results from 2020 and 2040. The red line is a fitted linear regression line, and slope $\beta_1$
of this line is the coefficient of the $\Delta G_t$, indicating the MEF. 
The scatter plot illustrates a generally positive relationship between changes in electricity generation and CO$_2$ emissions, as indicated by the upward-sloping red regression line. However, the significant dispersion of data points around this line suggests a high level of variability, which means that the linear model does not fully capture the complexity of the relationship. The presence of outliers and the broad spread of data points imply that the relationship is not strictly linear, with other factors—such as shifts in the energy mix, varying generation technologies, or policy changes—likely contributing to the observed CO$_2$ emissions. While the red line indicates a positive trend, the overall pattern suggests a more complex, potentially non-linear relationship between these variables.


\subsubsection{MEF Estimation Results}
\label{mefscompa_app}

\begin{figure}[H]
  \centering

  \begin{subfigure}{0.45\textwidth}
    \centering
    \includegraphics[width=\linewidth]{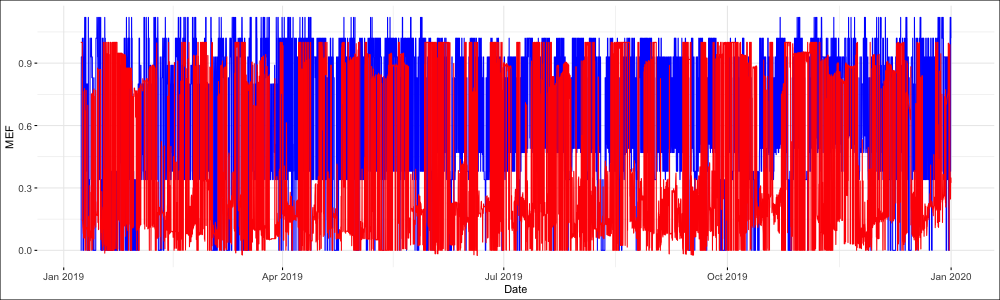}
    \caption{}
  \end{subfigure}
  \hfill
  \begin{subfigure}{0.45\textwidth}
    \centering
    \includegraphics[width=\linewidth]{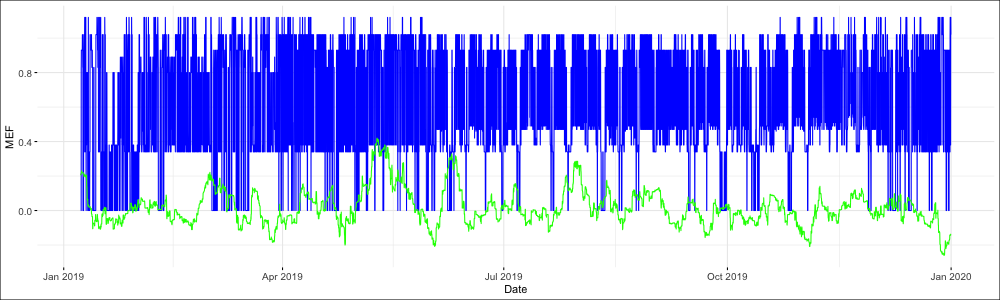}
    \caption{}
  \end{subfigure}
  \caption{\textcolor{blue}{Incremental MEF}, \textcolor{red}{MEF--MSDR}, and \textcolor{green}{MEF--DLR} time series from Agora Energiewende data in \textit{kg CO$_2$ eq./kWh} in 2019}
\label{compmefreal2019}
\end{figure}



\begin{figure}[H]
  \centering

  \begin{subfigure}{0.45\textwidth}
    \centering
    \includegraphics[width=\linewidth]{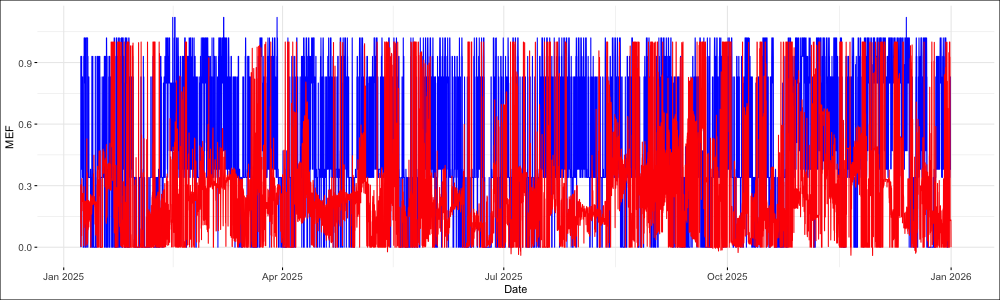}
    \caption{}
  \end{subfigure}
  \hfill
  \begin{subfigure}{0.45\textwidth}
    \centering
    \includegraphics[width=\linewidth]{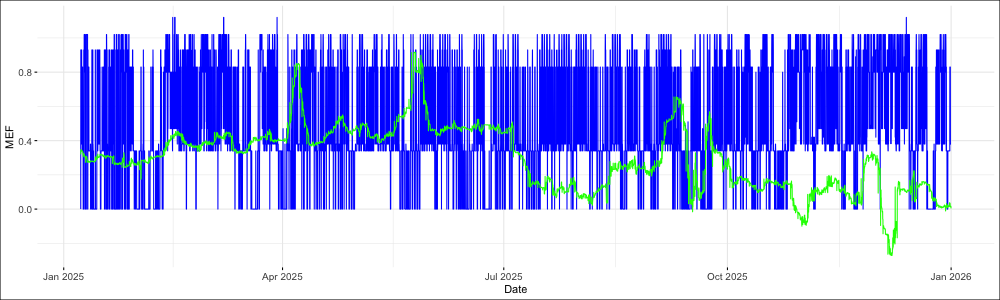}
    \caption{}
  \end{subfigure}
  \caption{\textcolor{blue}{Incremental MEF}, \textcolor{red}{MEF--MSDR}, and \textcolor{green}{MEF--DLR} time series in \textit{kg CO$_2$ eq./kWh} for 2025}
\label{compmef2025}
\end{figure}


\begin{figure}[H]
  \centering

  \begin{subfigure}{0.45\textwidth}
    \centering
    \includegraphics[width=\linewidth]{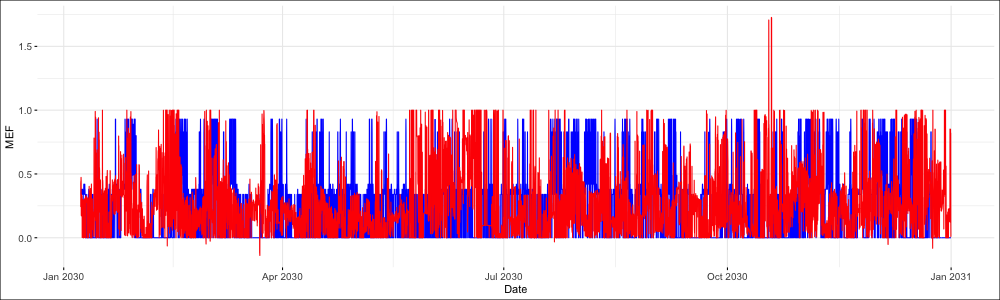}
    \caption{}
  \end{subfigure}
  \hfill
  \begin{subfigure}{0.45\textwidth}
    \centering
    \includegraphics[width=\linewidth]{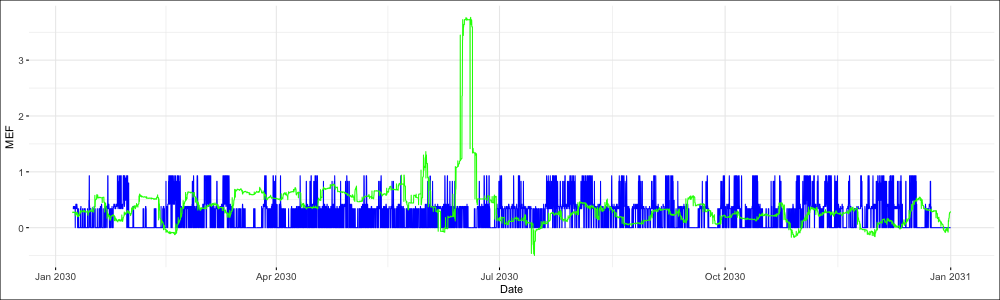}
    \caption{}
  \end{subfigure}
  \caption{\textcolor{blue}{Incremental MEF}, \textcolor{red}{MEF--MSDR}, and \textcolor{green}{MEF--DLR} time series in \textit{kg CO$_2$ eq./kWh} for 2030}
\label{compmef2030}
\end{figure}

\subsubsection{CO$_2$ Emission Estimation and the Accuracy of the Statistical Models}
\label{estimationaccuracyofstatmodel}

In this section, we assess the accuracy of CO$_2$ emissions estimates as a crucial informative step:\\
First, this emphasis on estimation accuracy is crucial for understanding the models' ability to capture variations in emissions, laying the groundwork for a meaningful comparison with the incremental MEF.\\
Second, the statistical model is designed to accurately estimate CO$_2$ emissions as a dependent or target variable. Considering that the MEF (the electricity generation coefficient) is a key parameter that allows the calculation of estimated CO$_2$ emissions, the precision of the statistical model in estimating CO$_2$ emissions is directly related to its ability to provide a good estimator of MEFs. Hence, by evaluating the accuracy of emissions estimates, we gain valuable insights into the models' overall effectiveness. This step proves instrumental in gauging the statistical abilities of each model to accurately capture the intricacies of emissions in the absence of fundamental data or a benchmark.\\
Finally, directly comparing the MEFs of the statistical models with the incremental MEF may not be reliable. This comparison hinges on the accuracy of aligning outputs with a fundamentally different model. However, by examining the accuracy of statistical models in estimating emissions, we can draw conclusions about the consistency of statistical and ESM results. This approach enables us to assert that a close alignment of the incremental MEF with the MEF from statistical models indicates a superior ability to estimate emissions when only electricity generation data is available.\\

In summary, we will examine the consistency of the statistical model when it comes to estimating CO$_2$ emissions and estimating the MEF in comparison to a benchmark (incremental MEF). Having confirmed this assumption, we can use the accuracy of CO$_2$ emission estimation results as a decision criterion when no benchmark is available.\\

The evaluation metrics for the CO$_2$ emissions estimation results are presented in Table \ref{estim_table}. We assess the performance using the mean square error (MSE), mean absolute error (MAE), and root mean square error (RMSE) (Section \ref{evaluationmetricdef}).

\begin{table}[htbp]
\centering
\caption{Evaluation metrics for the estimation of CO$_2$ emissions time series}
\renewcommand{\arraystretch}{1.2} 
\footnotesize 
\begin{tabularx}{\textwidth}{@{}l*{5}{>{\raggedleft\arraybackslash}X}@{}}
\toprule
\multicolumn{6}{c}{\textbf{Markov Switching Dynamic Regression Model}} \\ \midrule
     & \multicolumn{2}{c}{\textbf{Historical Years}} & \multicolumn{3}{c}{\textbf{Future Years}} \\ \midrule
     & 2019  & 2020  & 2025  & 2030  & 2040  \\ \midrule
\textbf{MSE}  & \textbf{0.296} & \textbf{0.373} & \textbf{0.433} & \textbf{0.409} & \textbf{0.373} \\
\textbf{MAE}  & \textbf{0.241} & \textbf{0.274} & \textbf{0.322} & \textbf{0.369} & \textbf{0.330} \\
\textbf{RMSE} & \textbf{0.544} & \textbf{0.610} & \textbf{0.658} & \textbf{0.640} & \textbf{0.611} \\ \midrule
\multicolumn{6}{c}{\textbf{Dynamic Linear Regression Model}} \\ \midrule
     & \multicolumn{2}{c}{\textbf{Historical Years}} & \multicolumn{3}{c}{\textbf{Future Years}} \\ \midrule
     & 2019  & 2020  & 2025  & 2030  & 2040  \\ \midrule
\textbf{MSE}  & 0.985 & 0.988 & 0.899 & 0.867 & 0.879 \\
\textbf{MAE}  & 0.489 & 0.484 & 0.418 & 0.514 & 0.450 \\
\textbf{RMSE} & 0.992 & 0.994 & 0.948 & 0.931 & 0.937 \\ \bottomrule
\end{tabularx}
\label{estim_table}
\end{table}

The results in Table \ref{estim_table} indicate that the MDSR model shows the minimum values of the MSE, MAE, and RMSE, compared to the DLR model over the years studied. \

\begin{figure}[H]
  \centering

  \begin{subfigure}{0.45\textwidth}
    \centering
    \includegraphics[width=\linewidth]{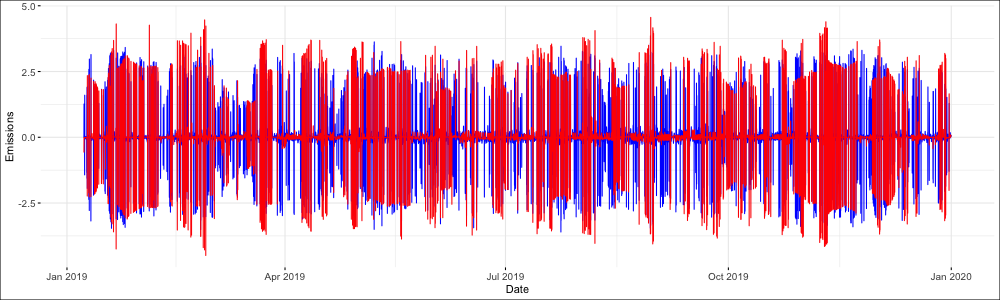}
    \caption{}
  \end{subfigure}
\hfill 
  \begin{subfigure}{0.45\textwidth}
    \centering
    \includegraphics[width=\linewidth]{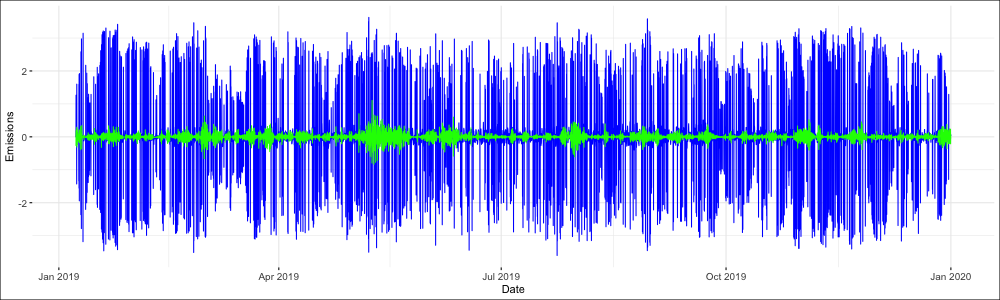}
    \caption{}
  \end{subfigure}

  \caption{Estimation results of \textcolor{red}{MSDR CO$_2$ emissions} and \textcolor{green}{DLR CO$_2$ emissions} compared to \textcolor{blue}{the Agora Energiewende real CO$_2$ emissions} data in 2019}
\label{compestimreal2019}
\end{figure}

\begin{figure}[H]
  \centering

  \begin{subfigure}{0.45\textwidth}
    \centering
    \includegraphics[width=\linewidth]{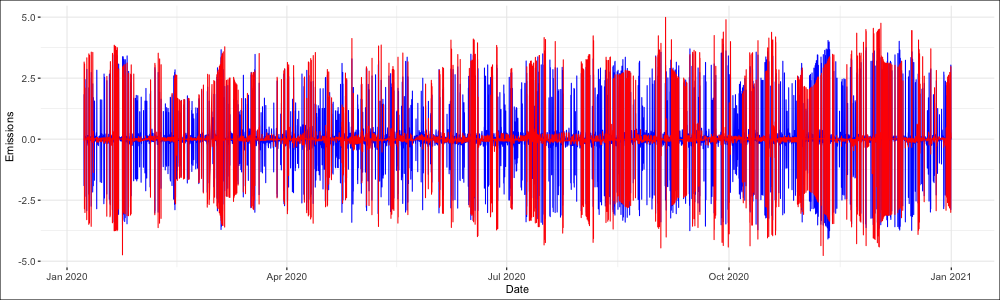}
    \caption{}
  \end{subfigure}
\hfill 
  \begin{subfigure}{0.45\textwidth}
    \centering
    \includegraphics[width=\linewidth]{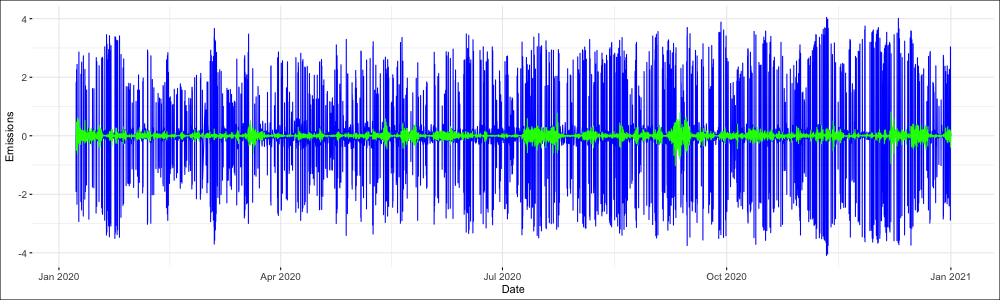}
    \caption{}
  \end{subfigure}
  \caption{Estimation results of \textcolor{red}{MSDR CO$_2$ emissions} and \textcolor{green}{DLR CO$_2$ emissions} compared to \textcolor{blue}{the Agora Energiewende real CO$_2$ emissions} data in 2020}
\label{compestimreal2020}
\end{figure}

\begin{figure}[H]
  \centering

  \begin{subfigure}{0.45\textwidth}
    \centering
    \includegraphics[width=\linewidth]{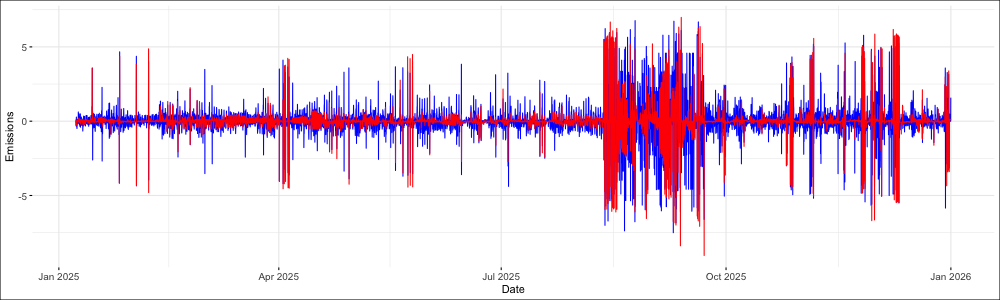}
    \caption{}
  \end{subfigure}
  \hfill  
  \begin{subfigure}{0.45\textwidth}
    \centering
    \includegraphics[width=\linewidth]{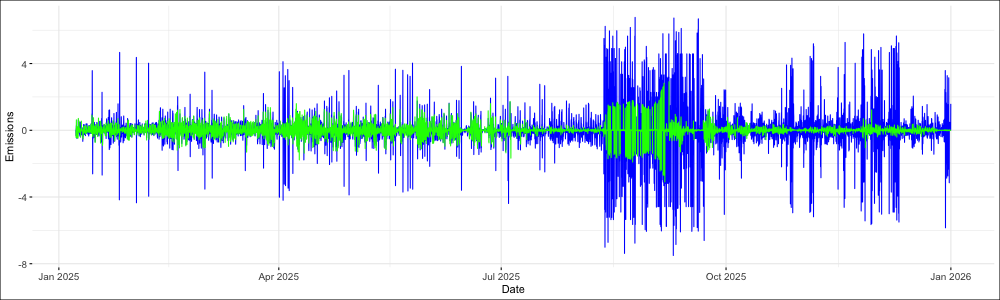}
    \caption{}
  \end{subfigure}

  \caption{Estimation results of \textcolor{red}{MSDR CO$_2$ emissions} and \textcolor{green}{DLR CO$_2$ emissions} compared to the observed \textcolor{blue}{ESM CO$_2$ emissions} in 2025}
\label{compestim2025}
\end{figure}

\begin{figure}[H]
  \centering

  \begin{subfigure}{0.45\textwidth}
    \centering
    \includegraphics[width=\linewidth]{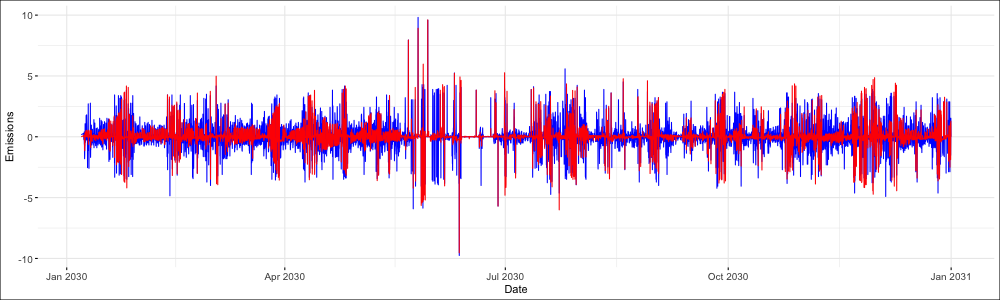}
    \caption{}
  \end{subfigure}
  \hfill  
  \begin{subfigure}{0.45\textwidth}
    \centering
    \includegraphics[width=\linewidth]{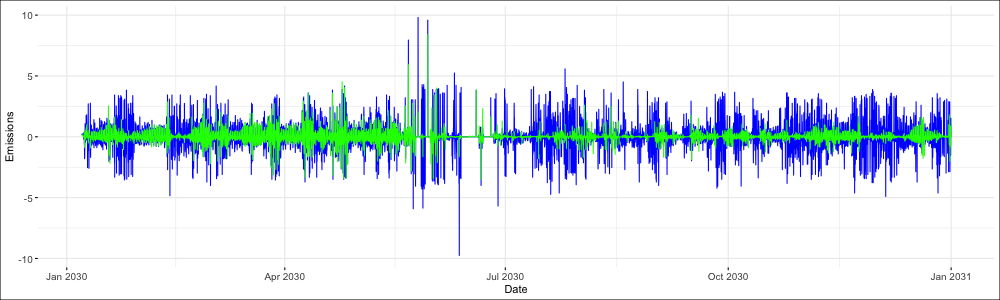}
    \caption{}
  \end{subfigure}

  \caption{Estimation results of \textcolor{red}{MSDR CO$_2$ emissions} and \textcolor{green}{DLR CO$_2$ emissions} compared to the observed \textcolor{blue}{ESM CO$_2$ emissions} in 2030}
\label{compestim2030}
\end{figure}

\begin{figure}[H]
  \centering

  \begin{subfigure}{0.45\textwidth}
    \centering
    \includegraphics[width=\linewidth]{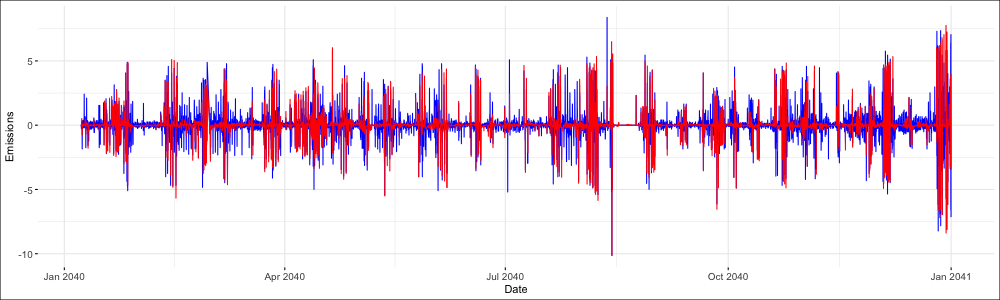}
    \caption{}
  \end{subfigure}
\hfill 
  \begin{subfigure}{0.45\textwidth}
    \centering
    \includegraphics[width=\linewidth]{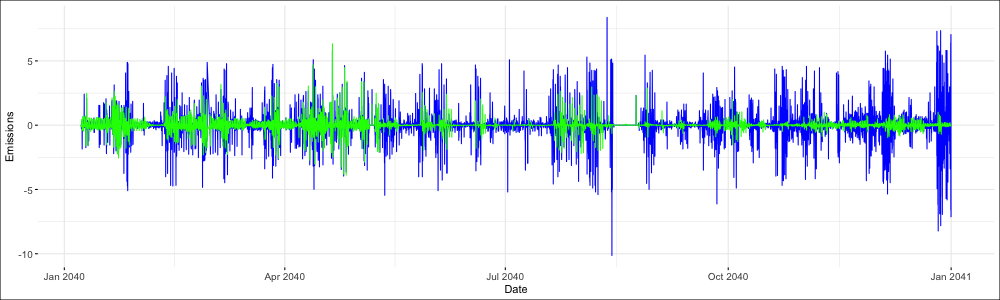}
    \caption{}
  \end{subfigure}

  \caption{Estimation results of \textcolor{red}{MSDR CO$_2$ emissions} and \textcolor{green}{DLR CO$_2$ emissions} compared to the \textcolor{blue}{ESM CO$_2$ emissions} in 2040}
\label{compestim2040}
\end{figure}

The MSDR model accuracy is further supported by Figures \ref{compestimreal2019},  \ref{compestimreal2020}, \ref{compestim2025}, \ref{compestim2030}, and \ref{compestim2040}, which compare the observed CO$_2$ emissions and the estimated CO$_2$ emissions from each statistical model. Notably, the MSDR model's estimated values closely align with the observed values, indicating its superior performance in capturing the complex dynamics between CO$_2$ emissions and electricity generation.


The favorable evaluation metrics and the CO$_2$ emission estimation plots indicate that the MSDR model is accurate in estimating CO$_2$ emissions. However, a key question remains: Does this model remain the most accurate when comparing MEF estimates from various statistical models to the incremental MEF, which is derived from an energy system model rather than the statistical model itself? Comparing parameters or time series from different models can be challenging. We will explore this issue in Section \ref{mefresults}.


\subsubsection{Results from the Smart Charging Application}
\label{smartchargappendixsection}
\begin{figure}[H]
  \centering

  \begin{subfigure}{0.45\textwidth}
    \centering
    \includegraphics[width=\linewidth]{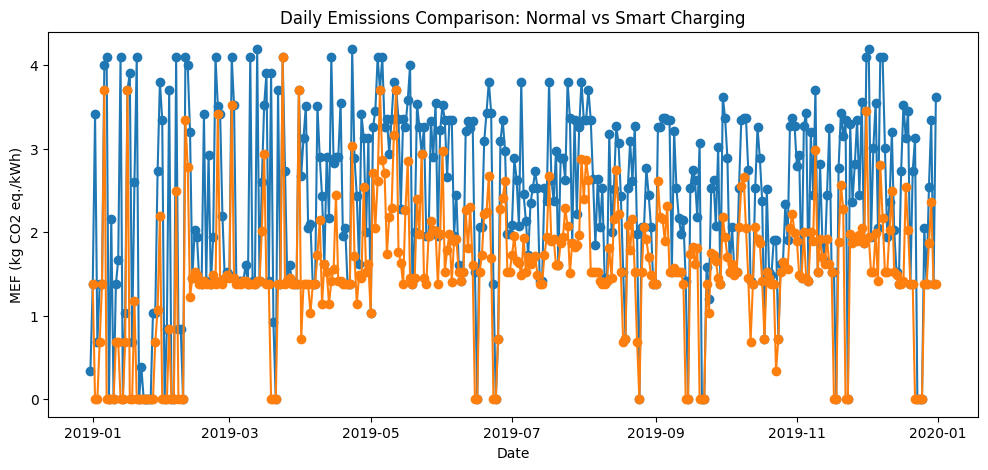}
    \caption{}
  \end{subfigure}
  \hfill  
  \begin{subfigure}{0.45\textwidth}
    \centering
    \includegraphics[width=\linewidth]{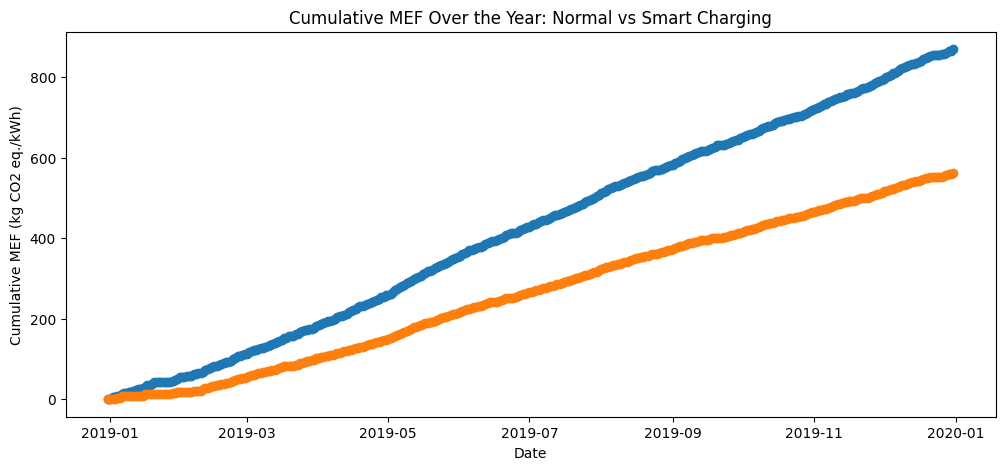}
    \caption{}
  \end{subfigure}

  \caption{ (a) Daily MEF and (b) cumulative MEF over the year for \textcolor{blue}{the normal charging scenario} and \textcolor{orange}{the smart charging scenario} in 2019}
\label{plotssmart2019}
\end{figure}

\begin{figure}[H]
  \centering

  \begin{subfigure}{0.45\textwidth}
    \centering
    \includegraphics[width=\linewidth]{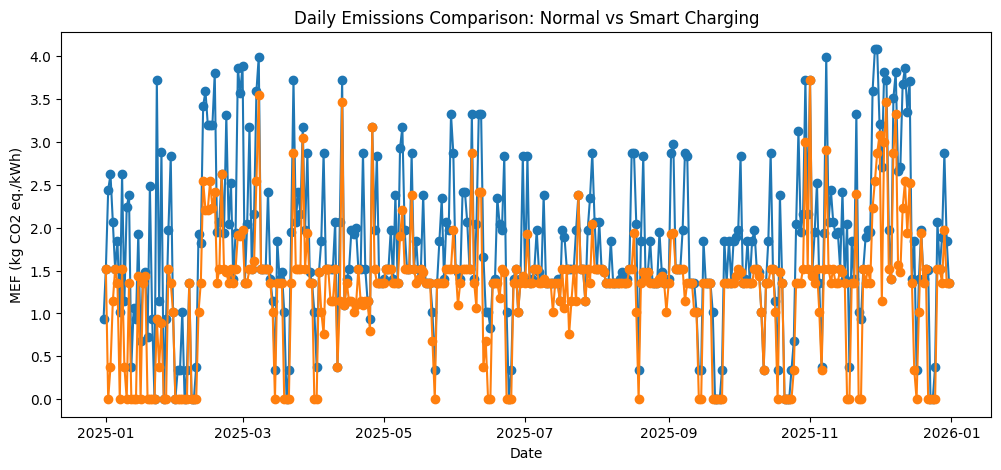}
    \caption{}
  \end{subfigure}
  \hfill  
  \begin{subfigure}{0.45\textwidth}
    \centering
    \includegraphics[width=\linewidth]{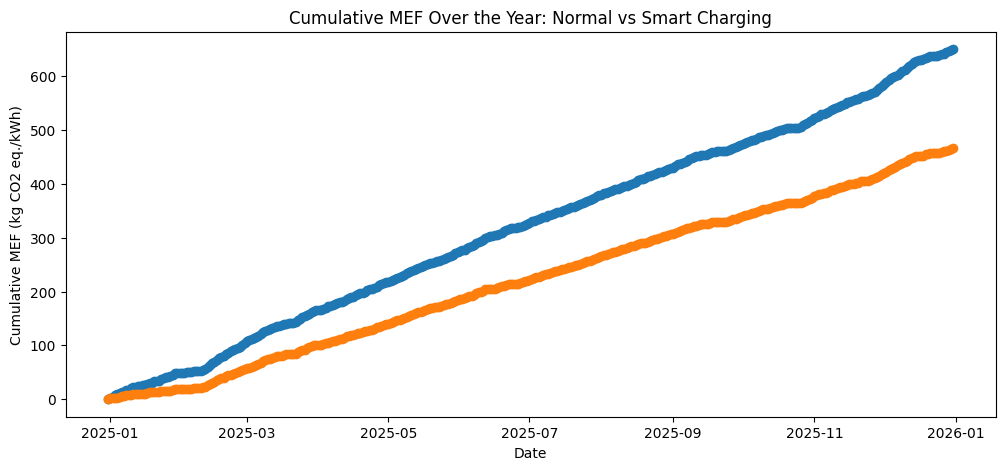}
    \caption{}
  \end{subfigure}

  \caption{ (a) Daily MEF and (b) cumulative MEF over the year for \textcolor{blue}{the normal charging scenario} and \textcolor{orange}{the smart charging scenario} in 2025}
\label{plotssmart2025}
\end{figure}

\begin{figure}[H]
  \centering

  \begin{subfigure}{0.45\textwidth}
    \centering
    \includegraphics[width=\linewidth]{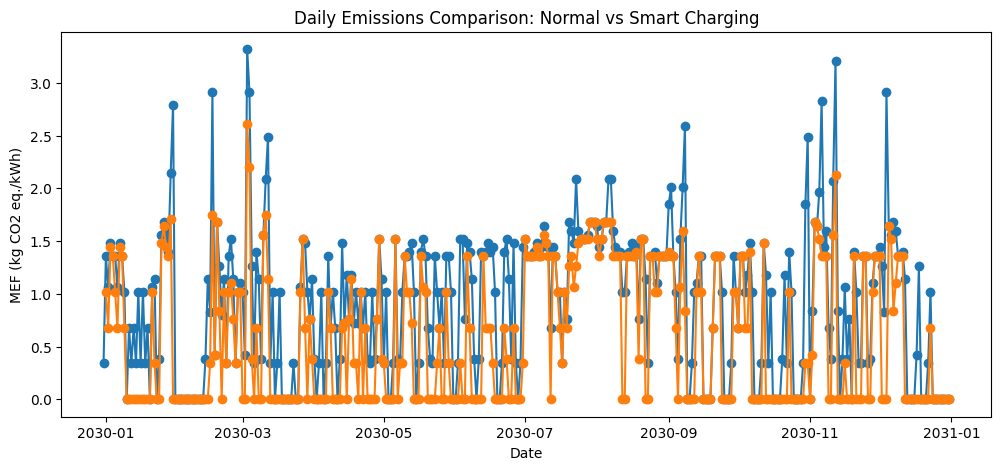}
    \caption{}
  \end{subfigure}
  \hfill  
  \begin{subfigure}{0.45\textwidth}
    \centering
    \includegraphics[width=\linewidth]{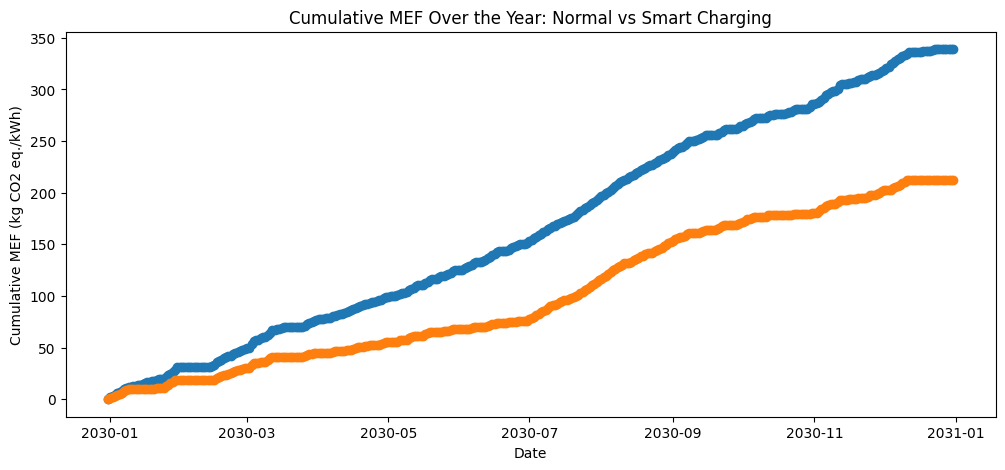}
    \caption{}
  \end{subfigure}

  \caption{ (a) Daily MEF and (b) cumulative MEF over the year for \textcolor{blue}{the normal charging scenario} and \textcolor{orange}{the smart charging scenario} in 2030}
\label{plotssmart2030}
\end{figure}

\end{document}